\documentclass[12pt]{article}
\usepackage{epsfig,latexsym,amsfonts,amsmath,amsthm,amssymb,amsbsy,multirow,slashed,wasysym,textcomp,comment,mathtools,cancel,cite,datetime,hyperref,stackrel,mathrsfs}

\usepackage{tikz}
\usetikzlibrary{backgrounds}
\usetikzlibrary{patterns}
\usetikzlibrary{calc}
\usetikzlibrary{arrows.meta}
\usetikzlibrary{shapes}
\tikzstyle{bag} = [align=center]
\usetikzlibrary{decorations.pathmorphing}
\usepackage[normalem]{ulem}
\usepackage{amsmath}

 \newcommand{\badat}{\begin{alignedat}}
 \newcommand{\eadat}{\end{alignedat}}
 \def\be{\begin{equation}}
\def\ee{\end{equation}}

\long\def\new#1\endnew{{\bf #1}}		
\long\def\del#1\enddel{}

\def\del{\partial}

\usepackage{color}
 
 \newcommand\scalemath[2]{\scalebox{#1}{\mbox{\ensuremath{\displaystyle #2}}}}

\newcommand{\pink}[1]{\textcolor{\pink}{#1}}

\definecolor{dblue}{rgb}{0.2,0.50,0.80} 

 \newcommand{\virg}{\hspace{1 mm}, \hspace{8 mm}}

\newcommand{\p}{\partial}

\def\O{\mathcal{O}}
\def\A{\mathcal{A}}

\def\M{\mathcal{M}}

\def\J{\mathcal{J}}
\def\F{\mathcal{F}}

\def\Q{\mathcal{Q}}

\def\bz{{\bar z}}
\def\bw{{\bar w}}

\def\bm{{\bar m}}

\def\tA{\widetilde A}

\def\th{\widetilde h}
\def\tJ{\widetilde J}

\def\tN{\widetilde N}

\def\tDelta{\widetilde \Delta}

\def\tW{\widetilde W}
\def\ty{\widetilde y}
\def\tY{\widetilde Y}

\def\th{\widetilde h}

\def\tDelta{\widetilde \Delta}

\def\tvarphi{\widetilde \varphi}

\def\n{k}

\def\scri{\mathscr{I}}

\numberwithin{equation}{section} % equation numbers follow sections

\begin{document}

 \begin{titlepage}
  \thispagestyle{empty}
  \begin{flushright}
  CPHT-RR004.012022
  \end{flushright}
  \bigskip
  \begin{center}

  \baselineskip=13pt {\LARGE 
   \scshape{Goldilocks Modes and 
   \\ \vspace{0em} the Three Scattering Bases}\\  

 } \vspace{1em} 
 
   \vskip1.5cm 

   \centerline{ 
   {Laura Donnay,}${}^\ddagger{}$ 
   {Sabrina Pasterski,}${}^\star{}$ 
   {and Andrea Puhm}${}^\lozenge{}$
   }

~\\~\\

\end{center}

\begin{abstract}
  \noindent

We consider massless scattering from the point of view of the position, momentum, and celestial bases. In these three languages different properties of physical processes become manifest or obscured. Within the soft sector, they highlight distinct aspects of the infrared triangle: quantum field theory soft theorems arise in the limit of vanishing energy $\omega$, memory effects are described via shifts of fields at the boundary along the null time coordinate $u$, and celestial symmetry algebras are realized via currents that appear at special values of the conformal dimension $\Delta$. We focus on the subleading soft theorems at $\Delta=1-s$ for gauge theory $(s=1)$  and gravity $(s=2)$  and explore how to translate the infrared triangle to the celestial basis.  We resolve an existing tension between proposed overleading gauge transformations as examined in the position basis and the `Goldstone-like' modes where we expect celestial symmetry generators to appear.  In the process we elucidate various order-of-limits issues implicit in the celestial formalism.  We then generalize our construction to the tower of $w_{1+\infty}$ generators in celestial CFT, which probe further subleading-in-$\omega$ soft behavior and are related to subleading-in-$r$ vacuum transitions that measure higher multipole moments of scatterers. In the end we see that the celestial basis is `just right' for identifying the symmetry structure.

\end{abstract}

 \bigskip \bigskip \bigskip \bigskip

\noindent{\em${}^\ddagger$ Institute for Theoretical Physics, Vienna University of Technology, A-1040 Vienna, Austria}
 
\noindent{\em ${}^\star$ Princeton Center for Theoretical Science, Princeton, NJ 08544, USA}

\noindent{\em${}^\lozenge$   CPHT, CNRS, Ecole polytechnique, IP Paris, F-91128 Palaiseau, France}
\end{titlepage}

\newpage

 \setcounter{tocdepth}{2}
\tableofcontents

\pagebreak
\section{Introduction}

The gravitational $\mathcal{S}$-matrix can be viewed as a hologram for quantum gravity in asymptotically flat spacetimes. Focusing on massless scattering we can prepare the initial and final states in terms of data at null infinity. This data can be presented in either the position, momentum, or celestial bases
\begin{equation}\label{udw}
    u \stackrel{\F}{\longleftrightarrow} \omega \stackrel{\M}{\longleftrightarrow} \Delta,
\end{equation}
related to one another by Fourier $(\mathcal{F})$ and Mellin $(\mathcal{M})$ transforms.  The goal of the flat space holographer is to present the $\mathcal{S}$-matrix in a manner that optimizes the number of symmetries that are manifest, so as to more clearly determine the constraints a consistent  $\mathcal{S}$-matrix  should obey. The claim of the celestial holographer is that the $\Delta$ basis is `just right' for this purpose.\footnote{Recent reviews and lecture notes on the celestial holography program include~\cite{Strominger:2017zoo,Pasterski:2019ceq,Raclariu:2021zjz,Pasterski:2021rjz,Pasterski:2021raf}.}

The fact that there are an infinite number of symmetries at our disposal is underscored by the observation by Strominger~\cite{Strominger:2017zoo} that Ward identities for the full asymptotic symmetry group are manifested non-trivially in $\mathcal{S}$-matrix elements as soft theorems. This fits into a larger universal structure called the {\it Infrared Triangle} (IR) connecting soft theorems,  asymptotic symmetries, and memory effects. 

Let us start by looking at the representations of memory effects in each basis. To have a well defined initial value problem, the data at null infinity must obey the constraint equations.  Integrals along retarded time $u$ on future null infinity of the constraints $\J_\mu$,
\begin{equation}\label{eq:memoryeffect}
   \int_{-\infty}^{+\infty} du\, n^\mu\J_\mu=   \int_{-\infty}^{+\infty} du\, n^\mu[\J^{soft}_\mu+\J^{hard}_\mu]=0\,,
\end{equation}
turn into relations between zero modes of the radiative field ($\J^{soft}_\mu$) and matter sources ($\J^{hard}_\mu$). When inserted into an $\mathcal{S}$-matrix element, we can view this in the usual momentum basis as a soft theorem
\be\label{eq:softtheorem}
\langle out|a_\pm(\omega)\mathcal{S}|in\rangle=\sum_{i}\sum_{k\in \{in,out\}} S_k^{(i)\pm}\langle out|\mathcal{S}|in\rangle\,,
\ee 
where the soft factors $S_k^{(i)\pm}$ scale with different powers of $\omega$. Upon a Mellin transform to the $\Delta$-representation 
\be\label{eq:celestialamplitude}
\langle \mathcal{O}^{\epsilon_1}_{\Delta_1,J_1}\dots\mathcal{O}^{\epsilon_n}_{\Delta_n,J_n}\rangle = \Bigl[ \, \prod_{i=1}^n \int_0^\infty\!\! d\omega_i\, \omega_i^{\Delta_i-1} \Bigr] 
\langle out|\mathcal{S}|in\rangle\,,
\ee
where $\epsilon_i=\pm$ indicates whether the $i^{th}$ particle is in the {\it out} or the {\it in} state, the energetically soft theorems~\eqref{eq:softtheorem} are encoded as simple poles at these special values of conformal dimension $\Delta=-n$ with $n=-1,0,1,...$, namely~\cite{Cheung:2016iub,Fan:2019emx,Fotopoulos:2019tpe,Pate:2019mfs,Adamo:2019ipt,Puhm:2019zbl,Guevara:2019ypd} 
\begin{equation}\label{eq:celestcurrent}
    \lim_{\Delta_i\to -n} (\Delta_i+n) \langle \mathcal{O}^{\epsilon_1}_{\Delta_1,J_1} \dots \O^\pm_{\Delta_i,J_i}\dots \mathcal{O}^{\epsilon_n}_{\Delta_n,J_n}\rangle = \sum_{k\in \{in,out\}}  \hat S_k^{(i)\pm} \langle \mathcal{O}^{\epsilon_1}_{\Delta_1,J_1} \dots \mathcal{O}^{\epsilon_n}_{\Delta_n,J_n}\rangle\,.
\end{equation}
The external scattering states in this basis are boost eigenstates and transform under the Lorentz group as primaries with definite weight $\Delta$ and spin $J=\ell$ matching the 4D helicity.

While the vacuum structure, the physicality of the vacuum-to-vacuum transitions, and their constrained nature (manifested as soft theorems/memory effects/celestial Ward identities) can be phrased in any scattering basis, the three scattering bases in~\eqref{udw} each naturally highlight different aspects of the IR triangle.  The memory effects provide the connection to experiment, the soft theorems highlight the fact that we are scattering undressed operators and draw attention to associated issues with IR divergences, while the celestial basis gives a natural way of seeing the hard + soft splitting as a manifestation of a conservation law. 

Meanwhile these bases can also obscure certain symmetries and features. For example, while translation symmetry is manifest in a momentum basis it is obscured in the conformal basis as a shift in the conformal dimensions. Similarly position space is well suited for examining and extending the proposed asymptotic symmetry group to more subleading soft theorems, whereas the conformal primary gauge fixing obscures this.  Somewhat paradoxically, the celestial basis obscures the identification the Goldstone modes and memory effects while at the same time highlighting the existence of symmetry generators.  Normally, from the canonical gauge theory side, we would look for residual gauge degrees of freedom and identify symmetries using Noether's theorem.  From the celestial perspective, however, we can let the operator weights point us to constraints on scattering. The `conformally soft' operators in~\eqref{eq:celestcurrent} have a natural interpretation as currents precisely because these values of $\Delta$ correspond to multiplets with primary descendants. If we can establish a radially quantized dictionary~\cite{Crawley:2021ivb,Pasterski:2022lsl,Pasterski:2022jzc}, we expect a decoupling of the primary descendants.

A crucial step towards understanding the celestial holographic dictionary is to identify the operator content that captures the universal sector of gauge theory and gravity in the infrared that encodes soft physics and memory.  In~\cite{Donnay:2020guq} we established how to analytically continue the bulk conformal primary wavefunctions $\Phi^\epsilon_{\Delta,J}$ off the principal series to capture conformally soft modes and demonstrated that the operators defined via the inner-product dictionary
\be\label{eq:2Dopintro}
\O^{\epsilon}_{\Delta,J}=i(\hat O^{s},(\Phi^{-\epsilon}_{\Delta,J})^*)\,
\ee
correspond to the anticipated soft charges in each instance of a known asymptotic symmetry (large $U(1)$, supertranslations, superrotations) in gauge theory and gravity $(s=1,2)$ at the expected integer values of $\Delta$. This was extended to the supergravity case in~\cite{Pano:2021ewd}.  The operators $\O_{\Delta,J}$ indeed become the canonical charge for a corresponding asymptotic symmetry whenever the wavefunctions $\Phi_{\Delta,J}$ is pure gauge~\cite{Donnay:2020guq}, which is the case for $1-s < \Delta \le 1 $, and we refer to them as Goldstone wavefunctions.  The remaining modes at $1 \le \Delta < 1+s $  in this range are celestial memory modes. These ranges overlap at $\Delta=1$ for the integer spin case precisely because one needs to keep track of the conformally soft modes studied in~\cite{Donnay:2018neh}. The celestial memory modes are canonically paired with the Goldstone modes.

The energetically soft expansion of amplitudes does not stop here. At the next subleading order, conformal analogs of the subleading soft theorem in gauge theory and the sub-subleading soft theorem at $\Delta=1-s$ imply (at least at tree level) that the other celestial operators will source this generalized current and that there should be a non-trivial conservation law. However, the corresponding wavefunctions $\Phi_{1-s,\pm s}$ are not pure gauge. The standard interpretation of the IR triangle thus breaks down: we have neither an obvious asymptotic symmetry nor a Goldstone mode but we do have a soft theorem, and hence a version of a memory effect. These Goldstone-like modes (affectionately shortened to `Goldilocks modes' in our title) will play a central role here. The lack of a Goldstone mode that satisfies the conformal primary condition does not mean that an asymptotic symmetry cannot be associated to these soft theorems in another gauge.  An asymptotic symmetry interpretation for this and the sub-subleading soft graviton were provided in~\cite{Campiglia:2016hvg,Campiglia:2016efb}, where the standard phase space was augmented with an overleading residual pure gauge mode in harmonic gauge.  Similar proposals for more subleading (but non-universal) soft theorems~\cite{Hamada:2018vrw} and memory effects~\cite{Seraj:2016jxi,Compere:2017wrj,Compere:2019odm} run into analogous issues despite natural dimension-matching to the $w_{1+\infty}$ symmetry in the self-dual sector~\cite{Guevara:2021abz,Strominger:2021lvk,Himwich:2021dau}.

The aim of this paper is to resolve these tensions and round out our investigations of the modes connected to conformally soft theorems~\cite{Donnay:2018neh,Donnay:2020guq,Pasterski:2021fjn,Pasterski:2021dqe,Donnay:2021wrk,Pano:2021ewd} by showing how to extend this picture to the remainder of the conformally soft modes that have primary descendants~\cite{Pasterski:2021fjn,Pasterski:2021dqe}.  These play an important role in the symmetry structure of celestial CFT (CCFT). We will focus on the $\Delta=1-s$ subleading %-most
soft modes in gauge theory and gravity since these distinctly highlight a variety of issues we encounter for the remainder of the tower, and also play an important role in constraining celestial OPE coefficients~\cite{Fan:2019emx,Fotopoulos:2019tpe,Pate:2019lpp,Fotopoulos:2019vac,Banerjee:2020kaa}, even giving rise to recursion relations that let us solve for them as an analytic function of $\Delta$.

Our main results are as follows. Intended as a direct extension of~\cite{Donnay:2018neh,Donnay:2020guq,Pano:2021ewd}, we show that the subleading soft charge in QED arises from the spin-1 primary wavefunction with conformal dimension $\Delta=0$ and its shadow with $\Delta=2$, while the spin-2 conformal primary with $\Delta=-1$ and its shadow with $\Delta=3$ give rise to the sub-subleading soft charge in gravity. We further identify the corresponding memory  modes and compute the relevant symplectic pairings.  The pairing between celestial memory and Goldstone was computed for large gauge symmetry and supertranslations in~\cite{Donnay:2018neh}. As a bonus we extend this analysis to superrotations here. We also resolve the apparent tension between the existence of a conformal soft theorem and the absence of conformal Goldstones by identifying the analogue of the large gauge parameters in conserved charges as encoded in the SL$(2,\mathbb{C})$ conformal primary wavefunctions $\Phi_{\Delta,J}$. 

\pagebreak

In order to get to this result we need to deal with two order-of-limits issues intrinsic to the conformally soft sector, both of which boil down to the light-ray supported objects we get in the extrapolate dictionary for celestial amplitudes.  First, there is the issue of $r$ versus (conformally) soft. Does the large-$r$ limit still localize on the celestial sphere when $\Delta \to 1,0,-1,...$?  This question is relevant already in the momentum space story where $\omega\to0$ en route to the `soft theorem = Ward identity' relation.  There we take large-$r$ first in order to use the saddle point approximation. Here we advocate for extending this to the celestial amplitude, which amounts to interpreting the Mellin transformed celestial amplitudes as an integral transform of radiative order data in contrast to literally scattering non-normalizable wavefunctions.  Our charge renormalization computations show how these two perspectives can be reconciled.  

Second, there is the question of taking the conformally soft limit before or after integrating along $\mathscr{I}^\pm$. Performing the $u$-integral of the saddle point result in~\eqref{eq:2Dopintro} before taking $\Delta$ conformally soft lands us on the expected form of the soft charges discussed in the asymptotic symmetry literature~\cite{He:2014laa,Kapec:2014opa,Lysov:2014csa}. For the Goldilocks modes the opposite order of limits yields a new logarithmic in~$u$ contribution to the soft charge. However, the analytic structure of celestial amplitudes suggests that only the first order of limits makes sense. From the work of Laddha and Sen~\cite{Laddha:2018myi} we know that there are logarithmic corrections to soft theorems in four spacetime dimensions that are related to infrared exchanges. In the momentum basis these are leading compared to the ordinary subleading and sub-subleading soft theorems~\cite{Laddha:2018rle,Laddha:2018myi,Sahoo:2018lxl,Ghosh:2021hsk}. Here we show that these logs give higher order poles at negative integer $\Delta$.  This extends the pole structure identified in~\cite{Arkani-Hamed:2020gyp} to statements about external states. In the end, we use insights from all three scattering bases~\eqref{udw} to define the Goldilocks modes in a manner that extends to the (semi-)infinite tower of conformally soft symmetry generators.

This paper is organized as follows. In section~\ref{sec:3Bases} we set the stage for discussing the IR sector in three different bases and expand on the extrapolate dictionary for celestial holography. This provides a bulk  explanation for the Carrollian/Celestial connection.  We make use of the three bases in section~\ref{sec:Soft} to highlight aspects of soft theorems, memory effects, and symmetries. In section~\ref{sec:CelestialTriangle} we discuss the celestial version of the IR triangle and how the conformal basis informs our understanding of quantum gravity in asymptotically flat spacetimes. This includes a brief review of asymptotic symmetries and their associated memory effects from the point of view of celestial CFT with a new discussion of conformal primary modes associated to spin memory. Our main focus in section~\ref{sec:Goldilocks} is on the subleading soft photon and the sub-subleading soft graviton, the construction of their canonically paired memory modes, the computation of the associated soft charges, and a proposal for mixed helicity Goldstone modes that give a natural interpretation of overleading-in-$r$ large gauge and diffeomorphism symmetries. We end in section~\ref{sec:Discussion} with a discussion of the general lessons we can extract for the tower of conformally soft modes.

\pagebreak

%%%%%%%%%%%%%%%%%%%%%%%%%%%%%%%%%%%%%%%%%%%%%%%%%%%%%%%%%%
\section{Three Bases for Scattering}
\label{sec:3Bases}
%%%%%%%%%%%%%%%%%%%%%%%%%%%%%%%%%%%%%%%%%%%%%%%%%%%%%%%%%%

In this section we set our notation for going between the position, momentum, and celestial bases for scattering which highlight different formulations of a flat space hologram.  We stick to the case of finite energy scattering for which the saddle point approximation is valid.

%%%%%%%%%%%%%%%%%%%%%%%%%%%%%%%%%%%%%%%%%%%%%%%%%%%%%
\subsection{\texorpdfstring{ $\omega,u,\Delta$}{omega, u, Delta}}
%%%%%%%%%%%%%%%%%%%%%%%%%%%%%%%%%%%%%%%%%%%%%%%%%%%%%

The position, momentum and celestial bases are related by Fourier and Mellin transforms.
We will normalize the Fourier transform so that 
\be\label{FT}
\mathcal{F}[f](u)=\int_{-\infty}^{+\infty} d\omega e^{i\omega u} f(\omega)\equiv\hat{f}(u)
\ee
with inverse
\be\label{iFT}
\mathcal{F}^{-1}[\hat{f}](\omega)=\frac{1}{2\pi i}\int_{-\infty}^{+\infty} du e^{-i\omega u} \hat{f}(u).
\ee
The Fourier transform is well defined and invertible for $f,\hat{f}\in L^1(\mathbb{R})$. 
The Mellin transform 
\begin{equation}\label{Mellin}
\mathcal{M}[f](\Delta)=\int_0^\infty d\omega \omega^{\Delta-1}f(\omega)\equiv \varphi(\Delta)\,,
\end{equation}
has inverse
\begin{equation}\label{InverseMellin}
\mathcal{M}^{-1}[\varphi](\omega)=\frac{1}{2\pi i}\int_{c-i\infty}^{c+i\infty} d\Delta \, \omega^{-\Delta}\varphi(\Delta)=f(\omega)
\end{equation}
and converges for functions $f(\omega)$ such that 
\begin{equation}
\int_0^\infty d\omega \omega^{k-1}|f(\omega)|<\infty\,,
\end{equation}
for some $k>0$, while the inverse transform requires $c>k$. 

Now let's consider the massless $\mathcal{S}$-matrix.   It is a function of on-shell external momenta which we can parametrize in terms of an energy $\omega_i$, a direction $z_i$, and a sign $\epsilon_i$ indicating incoming~($-$) and outgoing~($+$)
\be
p_i=\epsilon_i\omega_i(1+z_i\bz_i,z_i+\bz_i,i(\bz_i-z_i),1-z_i\bz_i).
\ee
If we formally transform the external states in the scattering amplitude $\mathcal{A}(\omega_i,z_i,\bz_i)$ using~\eqref{FT} or~\eqref{Mellin} then we can create an equivalent object that is a function of $(u_i,z_i,\bz_i)$ or $(\Delta_i,z_i,\bz_i)$, respectively. By looking at how the single particle states
\be\label{extstate}
|\omega_i,z_i,\bz_i\rangle=a^\dagger(\omega_i,z_i,\bz_i)|0\rangle
\ee
transform under Poincar\'e one finds that $u$ is a (null) time coordinate while $\Delta$ is a boost weight (aka a Rindler energy). For a complete set of states we need $u\in\mathbb{R}$.  Similarly a complete basis for finite energy states is captured by $\Delta = 1+i\lambda$ for $\lambda\in\mathbb{R}$~\cite{Pasterski:2017ylz}.

Simply viewed as an integral transformation, we can translate statements about consistent $\mathcal{S}$-matrix elements between any of these bases. We see that this is already holographic in the sense that in the $\mathcal{S}$-matrix program we are interested in probing the imprint of consistency conditions of the bulk quantum theory on this function of on-shell data. The goal of this section is to illustrate that it is also holographic in a sense closer to the AdS/CFT dictionary, whereby we can look at scattering as a correlation function of operators positioned on the conformal boundary of the spacetime. 

\subsection{From Boundary to Bulk}
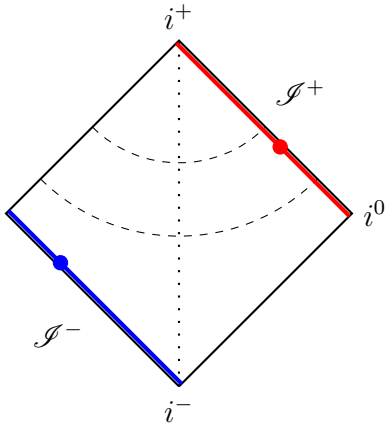
\begin{figure}[th!]
\centering
\begin{tikzpicture}[scale=2.3]
\definecolor{darkgreen}{rgb}{.0, 0.5, .1};
\draw[thick,loosely dotted] (0,0) --(0,2);
\draw[dash pattern=on 3pt off 3pt] (-1+.5,1+.5) to [bend right=45] (1-.5,1+.5);
\draw[dash pattern=on 3pt off 3pt] (-1+.2,1+.2) to [bend right=45] (1-.2,1+.2);
\draw[thick](0,0) --(1,1) node[right] {$i^0$} --(0,2)node[above] {$i^+$} --(-1,1) --(0,0)  node[below] {$i^-$} ;
\draw[ultra thick,blue] (0+.015,0+.015) -- (-1+.015,1+.015);
\draw[ultra thick,red] (0-.015,2-.015) -- (1-.015,1-.015);
\draw[fill,red] (1-.4-.015,1+.4-.015) circle (0.1em);
\draw[fill,blue] (-.7+.015,.7+.015) circle (0.1em);
\node at (1/2+.2,3/2+.2) {$\mathscr{I}^+$};
\node at (-1/2-.2,1/2-.2) {$\mathscr{I}^-$};
\end{tikzpicture}
\caption{When the saddle-point approximation is valid, the boundary operators correspond to  operators inserted on $\mathscr{I}^\pm$.  For the momentum and celestial bases they are smeared along the null generators, as illustrated  by the antipodally placed blue (past) and red (future) contours.  The celestial basis views scattering in terms of constant-Rindler time cuts~\cite{Pasterski:2022jzc} (dotted), in contrast to the constant-$u$ cuts considered in~\cite{Laddha:2020kvp} (dashed).
\label{in_out_states}}
\end{figure}

Before tying these two pictures together, let us take a step back from the quantum amplitude and look at the classical scattering problem.  Consider the Penrose diagram in figure~\ref{in_out_states}.  The conformal boundary includes null components $\mathscr{I}^\pm$ where massless excitations enter and exit.  We will focus on future null infinity here.  In this case it is convenient to use retarded Bondi coordinates $(u,r,z,\bz)$. These are related to the Cartesian coordinates $X^\mu$ by the transformation
\begin{equation}\label{eq:co}
 X^0=u+r\,, \quad X^i=r \hat{X}^i(z,\bz)\,, \quad \hat{X}^i(z,\bz)=\frac{1}{1+z \bar z}(z+\bz,-i(z-\bz),1-z\bz)\,
\end{equation}
for Minkowski space, in which case the line element becomes
\begin{equation}\label{eq:gBondi}
ds^2=-du^2-2du dr+2r^2 \gamma_{z \bar z} dz d\bar z \quad \text{with}\quad \gamma_{z \bar z}=\frac{2}{(1+z \bar z)^2}\,.
\end{equation}
Future null infinity $\mathscr{I}^+$ is then reached by taking $r$~large while keeping $(u,z,\bz)$ fixed.

The free data of the radiative fields are specified as a function of $(u,z,\bz)$. The constraint equations impose restrictions on the $u$ evolution of the leading large-$r$ behavior of the non-radiative metric components, while the remaining equations of motion take us into the bulk. Let's look at this in more detail. For solutions in the standard radiative phase space we can write the linearized perturbations in harmonic gauge as the following large~$r$ expansion~\cite{Himwich:2019dug}
\be\label{Ahmode}
A_{\mu}=\sum_n r^{-n} A_{\mu}^{(n)} +r^{-n}\log r~ \bar{A}_{\mu}^{(n)}\,, \quad h_{\mu\nu}=\sum_n r^{-n} h_{\mu\nu}^{(n)} +r^{-n}\log r~ \bar{h}_{\mu\nu}^{(n)}\,.
\ee
The source-free Maxwell and linearized Einstein equations along with the harmonic gauge conditions amount to 
\be\label{Aheoms}
\Box A_{\mu}=0,~~~\nabla^\mu A_{\mu}=0\,, \quad \Box h_{\mu\nu}=0,~~~\nabla^\mu h_{\mu\nu}=0.
\ee
The order by order expansions of these equations in terms of the modes~\eqref{Ahmode} can be found in appendix~\ref{larger}. We can consistently impose the following leading falloffs\footnote{For source-free solutions we use the radial gauge fixing $u A_{u}+r A_{r} =0$ and $u h_{u\mu}+r h_{r\mu} =0$. These are weaker falloffs than in~\cite{Himwich:2019dug} because of our choice of residual gauge fixing. In what follows we can drop the log terms since they will not enter our analysis below. }
\be\begin{array}{c}
A_{u}\sim\mathcal{O}(r^{-1}),~~~A_{r}\sim\mathcal{O}(r^{-2}),~~~A_{C}\sim\mathcal{O}(1)\,,
\end{array}\ee
and 
\be\begin{array}{c}
h_{uu}\sim\mathcal{O}(r^{-1}),~~~h_{ur}\sim\mathcal{O}(r^{-2}),~~~h_{rr}\sim\mathcal{O}(r^{-3}),\\ [.5em]
h_{uA}\sim \mathcal{O}(1),~~~h_{rA}\sim \mathcal{O}(r^{-1}),~~~h_{z\bz}\sim \mathcal{O}(r^{0}),~~~h_{zz}\sim \mathcal{O}(r)\,.
\end{array}\ee The free data is given by $A_{z}^{(0)}(u,z,\bz)$ and its complex conjugate $A_{\bz}^{(0)}(u,z,\bz)$ for gauge theory, and  $h_{zz}^{(-1)}(u,z,\bz)$ and its complex conjugate $h_{\bz\bz}^{(-1)}(u,z,\bz)$ for gravity, in terms of which the other components and subleading orders are determined by~\eqref{Aheoms} and our residual gauge fixing.

In the same way, we can go from the creation and annihilation operators for the single particle states to a local bulk operator in the free theory by demanding that the operator obey the Heisenberg equations of motion. In the standard momentum space basis for scattering states we use the following mode expansions for the bulk operators in harmonic gauge, namely for spin-1
 \begin{equation}\label{eq:Aexp}
    \hat{A}_\mu(X)=e\int \frac{d^3k}{(2\pi)^3}\frac{1}{2k^0}\left[a_\mu e^{ik\cdot X}+a_\mu^\dagger e^{-ik\cdot X}\right]\,,
\end{equation}
and for spin-2
\begin{equation}\label{eq:hexp}
    \hat{h}_{\mu\nu}(X)=\kappa\int \frac{d^3k}{(2\pi)^3}\frac{1}{2k^0}\left[a_{\mu\nu} e^{i k\cdot X}+a_{\mu\nu}^\dagger e^{-ik\cdot X}\right]\,,
\end{equation}
where $\kappa=\sqrt{32\pi G}$. Here we have defined
\be\label{eq:apolspin12}
a_\mu=\sum_{\alpha\in\pm} \epsilon_\mu^{\alpha *}a_\alpha,~~~a_{\mu\nu}=\sum_{\alpha\in\pm} \epsilon_{\mu\nu}^{\alpha *}a_\alpha\,,
\ee
where we take $\epsilon_{\mu\nu}^{\alpha*}=\epsilon_\mu^{\alpha*} \epsilon_\nu^{\alpha*}$ and $\alpha=+(-)$ is the helicity.

%%%%%%%%%%%%%%%%%%%%%%%%%%%%%%%%%%%%%%%%%%%%%%%%%%%%%
\subsection{From Bulk to Boundary}
%%%%%%%%%%%%%%%%%%%%%%%%%%%%%%%%%%%%%%%%%%%%%%%%%%%%%

We would now like to connect the bulk and boundary pictures and show that, even in the momentum space case, we can get to the external state data $\mathcal{A}(\omega_i,z_i,\bz_i)$ from a boundary limit of the bulk operators~\eqref{eq:Aexp} and~\eqref{eq:hexp}. For the momentum, we use the parametrization  $k^\mu=\omega q^\mu$ where
\be\label{qmu}
q^\mu=(1+w\bw,w+\bw,i(\bw-w),1-w\bw)\,,
\ee
in terms of which the polarization vectors become $\sqrt{2}\epsilon_w^\mu=\p_w q^\mu$ and $\sqrt{2}\epsilon_\bw^\mu=\p_\bw q^\mu$.  We can then perform the large-$r$ expansion of~\eqref{eq:Aexp} and~\eqref{eq:hexp} using the saddle point approximation\footnote{Here $\theta$ is the angle between $\hat q=q^i/q^0$ and $\hat X^i$.}
\be\label{eq:saddlepoint}
\lim_{r\rightarrow\infty}\sin\theta e^{i\omega q^0r(1-\cos\theta)}=\frac{i}{\omega q^0 r}\delta(\theta)+\mathcal{O}((\omega q^0 r)^{-2})\,,
\ee
whereby the point $(w,\bw)$ on the celestial sphere gets identified with the spacetime direction $(z,\bz)$ in the large-$r$ limit. From this we find~\cite{He:2014laa}
 \begin{equation}\label{eq:hatA}
   \lim_{r\to \infty}  \hat{A}_z=\frac{-ie}{\sqrt{2}(2\pi)^2} \int_0^\infty d\omega \left[a_+(\omega,z,\bz)e^{-i\omega (1+z\bz) u}-a^\dagger_-(\omega,z,\bz)e^{i\omega (1+z\bz) u}\right]\,,
 \end{equation}
 and
 \begin{equation}\label{eq:hatH}
  \hat{C}_{zz}=\lim_{r\to \infty}  \frac{1}{r} \hat h_{zz}= \frac{-i\kappa}{(2\pi)^2} \frac{1}{(1+z\bz)} \int_0^\infty d\omega \left[a_+(\omega,z,\bz)e^{-i\omega (1+z\bz) u}-a^\dagger_-(\omega,z,\bz)e^{i\omega (1+z\bz) u}\right]\,,
 \end{equation}
 implying that the boundary limit of the field operators are $u$-Fourier transforms of the creation and annihilation operators. Similar expressions are obtained for the opposite helicity expressions $\hat{A}_\bz$ and $\hat{C}_{\bz\bz}$. 

 Acting on the out vacuum state selects the annihilation operator terms. Similar antipodally related expressions near past null infinity select the creation operator terms. If we now apply the inverse Fourier transform~\eqref{iFT} we can get back to the momentum operators 
 \be\label{eq:extrapolateomega}
\langle 0|a_+(\omega) \propto\lim_{r\rightarrow \infty} \int du e^{i\omega (1+z\bz)u} \langle 0|\hat{A}_z, \quad \langle 0|a_+(\omega) \propto\lim_{r\rightarrow \infty} \frac{1}{r}\int du e^{i\omega (1+z\bz)u}\langle 0|\hat{h}_{zz}.
\ee
We see that as an alternative to the LSZ prescription~\cite{Pasterski:2021dqe}  we can  view massless $\mathcal{S}$-matrix elements as correlators of operators supported on null rays of the conformal boundary.  This is illustrated by the red and blue lines in figure~\ref{in_out_states}. We see that the field operators at the boundary create radiative single particle states whose momenta are directed towards that point on the night sky.  Moreover, rather than interpreting the Fourier transformed amplitudes as an impact parameter representation of scattering, we can attach a spacetime interpretation that identifies $\mathcal{A}(u_i,z_i,\bz_i)$ with operators at points on $\mathscr{I}^+$ (see also~\cite{Banerjee:2018gce,Banerjee:2019prz}).

\subsection{The Extrapolate Dictionary}
\label{sec:ExtraDict}

Treating null infinity as the boundary 3-manifold in the position representation is closest to an AdS$_4$/CFT$_3$ analog. We are viewing the bulk in terms of correlators at the boundary, in the same way that the $\mathcal{S}$-matrix program gives a stereoscopic view of consistent bulk physics.  This lends itself to the view of flat-space holography advocated in~\cite{Laddha:2020kvp} related to earlier BMS-field theory and Carrollian approaches (see ex.~\cite{Dappiaggi:2005ci,Ciambelli:2018wre}).  In contrast to the AdS$_4$ case, we run into the issue of the time direction being null, as well as the question of how to connect {\it in} to {\it out} states, since there are two null boundary components. However, we can consider different ways of slicing up the bulk and boundary.  The celestial approach, alternatively, tries to make sense of the theory as a quotient of the boundary by the null generators.  This can be thought of as cutting the spacetime into slices of constant (spacelike) Rindler `time' as opposed to retarded time~$u$ or advanced time~$v$~\cite{Pasterski:2022jzc}. 

In any of these bases we can define an extrapolate map that prepares an operator that lives in this abstract 3-dimensional space $\mathrm b\times \mathbb{CP}^1$ for $\mathrm b\in\{\omega,u,\Delta\}$ in terms of a boundary limit of the bulk quantum field operators (we focus again on future null infinity). Given\begin{enumerate}
    \item  a spin-$s$ bulk operator $\hat{O}^s(X)$,
       \item a wavefunction $\Phi^s_R(X;\mathrm b,w,\bw)$ satisfying the free spin-$s$ equations of motion and transforming under appropriate bulk ($s$) and boundary ($R$) representations of Poincar\'e, and 
    \item an inner product $(\cdot,\cdot)_\Sigma$ on the single particle wavefunctions defined on a Cauchy slice $\Sigma$, which can be constructed from the symplectic product via $\Omega(\Phi,\Phi')=i(\Phi,{\Phi'}^*)_\Sigma$,
\end{enumerate}
we can construct the boundary operator
\begin{equation}\label{eq:3Dmu}
   \O^{\epsilon}_R(\mathrm b,w,\bw)= (\hat O^s(X),\Phi^s_R(X_{-\epsilon};\mathrm b,w,\bw)^*)_\Sigma,
\end{equation}
where the incoming and outgoing modes are selected by the $i\epsilon$ prescription $X^0_\epsilon=X^0\mp i\epsilon$.
This effectively intertwines the 3D and 4D representations so that this object transforms under the specified boundary representation of Poincar\'e, by construction.

For the $\mathcal S$-matrix we use plane wave solutions. By taking a limit of the Cauchy slice to past and future null infinity $\Sigma\mapsto\mathscr{I}^\pm$ we can reduce this to a boundary limit of the bulk operators, which we expect to still make sense when we go beyond the perturbative quantum field theory limit of our quantum gravity theory.  Using the same saddle point approximation as in the last section, the bulk and boundary $\mathbb{CP}^1$ coordinates get localized to the same point on the celestial sphere.  The inner product~\eqref{eq:3Dmu} serves to project out the polarization tensors, so that the operator $\O^{-}_\ell(\omega,w,\bw)$ is just the annihilation operator $a_\ell(\omega,w,\bw)$.  This is particularly useful for the half-integer spin case~\cite{Pano:2021ewd}.  We can then use the transforms~\eqref{FT} and~\eqref{Mellin} to go to the other bases. We study this construction in more detail for the celestial basis in section~\ref{sec:CelestialTriangle}.

%%%%%%%%%%%%%%%%%%%%%%%%%%%%%%%%%%%%%%%%%%%%%%%%%%%%%%%%%%
\section{The Soft Limit of Scattering}
\label{sec:Soft}
%%%%%%%%%%%%%%%%%%%%%%%%%%%%%%%%%%%%%%%%%%%%%%%%%%%%%%%%%%
In this section we look at features of the low-energy limit of scattering in the momentum, celestial, and position bases, in turn.

%%%%%%%%%%%%%%%%%%%%%%%%%%%%%%%%%%%%%%%%%%%%%%%%%%%%%%%%%%
\subsection{\texorpdfstring{$A(\omega)$}{A(omega)}}
%%%%%%%%%%%%%%%%%%%%%%%%%%%%%%%%%%%%%%%%%%%%%%%%%%%%%%%%%%

Let us start with the $\mathcal{S}$-matrix in the standard momentum basis and consider how the scattering amplitude behaves as we take one external leg corresponding to a massless gauge boson soft. At tree level, a scattering process involving $n$ `hard' particles has the following  Laurent expansion in the energy $\omega$ of the additional soft emission
\be\label{eq:softtheoremA}
\mathcal{A}_{n+1}(k^\mu=\omega q^\mu)\sim \omega^{-1}\mathcal{A}^{(-1)}_n+\omega^0 \mathcal{A}^{(0)}_n+\omega \mathcal{A}^{(1)}_n+\O(\omega^2)\,.
\ee
The leading and subleading terms take a universal form in terms of  the $n$-point amplitude we get upon removing the soft emission.  Namely, we can write
\be
\omega^{j}\mathcal{A}^{(j)}_n=\sum_k S_k^{(j+1)} \mathcal{A}_{n},
\ee
where $\mathcal{A}_{n}$ is the $n$-point amplitude without the soft particle and the $S_k^{(i)}$ are `soft factors.' This factorization property of scattering amplitudes when the energy of an external massless particle is taken to zero is referred to as a `soft theorem'. We have the following leading~\cite{Weinberg:1965nx} and (sub)-subleading~\cite{Low:1954kd,Cachazo:2014fwa} soft theorems in gauge theory
\be\label{softgauge}
S_k^{(0)\pm}=eQ_k\frac{p_{k\mu} \varepsilon^{\pm\mu}}{p_k\cdot q}\,,~~~S_k^{(1)\pm}=-ieQ_k\frac{q_\mu \varepsilon^\pm_\nu J_k^{\mu\nu}}{p_k\cdot q}\,,
\ee
and gravity 
\be\label{softgrav}
S_{k}^{(0)\pm}=\frac{\kappa}{2}\frac{p_{k\mu}p_{k\nu} \varepsilon^{\pm\mu\nu}}{p_k\cdot q}\,,~~~S_k^{(1)\pm}=-i\frac{\kappa}{2}\frac{p_{k\mu}\varepsilon^{\pm\mu\nu}q^\lambda J_{k\lambda\nu}}{p_k\cdot q}\,,~~~S_k^{(2)\pm}=-\frac{\kappa}{4}\frac{\varepsilon^\pm_{\mu\nu}q_{\rho}q_{\lambda}J_k^{\rho\mu}J_k^{\lambda\nu}}{p_k\cdot q}\,.
\ee
In what follows we will set $\kappa=2$. At 1-loop the soft expansion gets modified to
\be\label{eq:logsofttheorem}
\mathcal{A}_{n+1}\sim \omega^{-1}\mathcal{A}^{(-1)}_n+\log\omega \bar{\mathcal{A}}^{(0)}_n+\omega^0 \mathcal{A}^{(0)}_n+\omega \log \omega \bar{\mathcal{A}}^{(1)}_n+\omega \mathcal{A}^{(1)}_n+\mathcal{O}(\omega^2\log \omega)\,.
\ee
We can see an example of why this effects how one extracts the soft theorems by replacing $\omega$ with the dimensionless variable $\Lambda_{IR}^{-1}\omega$ in the loop-level logs and noting how this mixes with the terms that were there at tree level. Resumming virtual soft exchanges leads to a  vanishing of amplitudes without additional real soft emissions when we remove the IR cutoff in 4D.  While the textbook procedure for handling this issue is to turn to inclusive quantities, one can also interpret this vanishing as a result of violating charge conservation for the corresponding asymptotic symmetry~\cite{Kapec:2017tkm}.  This viewpoint lends itself to more flexibility in how one decides to dress the {\it in} and {\it out} states with additional soft radiation. For instance the dressed amplitudes of~\cite{Chung:1965zza,Kulish:1970ut} are in a sector with vanishing asymptotic symmetry charge.

The asymptotic symmetry interpretation has also led to renewed interest in extracting subleading terms in the soft expansion.  An infinite tower of soft theorems was studied at tree level in~\cite{Hamada:2018vrw}, where they were interpreted as Ward identities for an infinite tower of gauge transformations.  Within the MHV sector, a similar infinite tower was found to persist at loop level~\cite{He:2014bga}.  At higher loop level one finds higher powers of $\log\omega$; however, there is a nice hierarchy regarding whereby the corrections for terms that scale with a fixed power of $\omega$ terminate at a fixed loop order. The leading $i=0$ soft factors~\cite{Weinberg:1965nx} are universal to all loop order.  The subleading $i=1$ soft graviton theorem~\cite{Cachazo:2014fwa} receives a 1-loop exact correction. Meanwhile the sub-subleading $i=2$ soft graviton theorem only gets corrections up to 2-loop order~\cite{Bern:2014oka}. This is consistent with the probe-scatterer approximations~\cite{Laddha:2018myi,Laddha:2018vbn,Sahoo:2018lxl,Ghosh:2021hsk} where at $\ell$-loop the leading log correction is $\omega^{\ell-1}\log^\ell\omega$. We will now turn to the conformal primary basis where this hierarchy of multi-logs is most neatly organized.

%%%%%%%%%%%%%%%%%%%%%%%%%%%%%%%%%%%%%%%%%%%%%%%%%%%%%%%%%%
\subsection{\texorpdfstring{$A(\Delta)$}{A(Delta)}}\label{sec:softADelta}

%%%%%%%%%%%%%%%%%%%%%%%%%%%%%%%%%%%%%%%%%%%%%%%%%%%%%%%%%%
When we go to the celestial basis, we integrate over all energy scales.  While naively this precludes the notion of an energetically soft particle ($\omega \to 0$), universal factorization properties actually do persist in this basis, but are organized in a novel way in terms of conformally soft \cite{Donnay:2018neh} limits $(\Delta \to1-\mathbb{Z}_{\geq 0})$.\footnote{Or $\Delta \to \frac{1}{2}-\mathbb{Z}_{\geq 0} $ for fermionic soft theorems.} After Mellin transforming~\eqref{eq:softtheoremA}, the celestial correlator~\eqref{eq:celestialamplitude} has residues at \mbox{$\Delta=-1,0,1,...$} which correspond to the celestial analogue of the tree-level soft theorems~\cite{Cheung:2016iub,Fan:2019emx,Nandan:2019jas,Pate:2019mfs,Adamo:2019ipt,Puhm:2019zbl,Guevara:2019ypd}. 
More explicitly, assuming the amplitude is sufficiently damped at high energies, we can use the relation
\be
\lim\limits_{\epsilon\rightarrow0}\frac{\epsilon}{2}\omega^{\epsilon-1}=\delta(\omega)
\ee
to extract the residues~\cite{Pate:2019mfs}
\be\label{alim}
\lim\limits_{\Delta\rightarrow-n}(\Delta+n)\int_0^\infty d\omega \omega^{\Delta-1}\sum_k\omega^k A^{(k)}=A^{(n)}
\ee
from the tree-level expansion~\eqref{eq:softtheoremA}. For gravity, these conformally soft gravitons with \mbox{$\Delta=-n$} with $n=-2,-1,0,1,...$ generate the $w_{1+\infty}$ algebra of the single helicity sector that has become a very active topic as of late~\cite{Strominger:2021lvk,Adamo:2021lrv,Jiang:2021csc,Jiang:2021ovh,Himwich:2021dau}.  The implications of this infinite-dimensional celestial symmetry algebra for gravity in asymptotically flat spacetimes is an interesting problem. 

From our discussion of log corrections to $\mathcal{A}(\omega)$, we expect this procedure to be modified at loop level.  This is easiest to see with an explicit cutoff separating IR and UV behavior. Noting that
\be\label{eq:omegaintegral}
\int_0^{\omega_*} d\omega\, \omega^{\Delta-1+m}  = \frac{\omega_*^{\Delta+m}}{\Delta+m}\,,
\ee
we again see that simple poles at $\Delta=-m$ for $m=-1,0,1,...$ pick out the conformal analog of the ordinary soft theorems of tree-level amplitudes, namely the Mellin transforms of $\A^{(m)}_n$. Now we can easily incorporate the log corrections following a procedure analogous to that of~\cite{Arkani-Hamed:2020gyp}, but for the external energy scale. For the 1-loop log corrections we have 
\be\label{eq:omegalogomegaintegral}
\int_0^{\omega_*} d\omega\, \omega^{\Delta-1+m} \log \omega =  \frac{\p}{\p \Delta} \int_0^{\omega_*} d\omega\, \omega^{\Delta-1+m} 
= -\frac{\omega_*^{\Delta+m}}{(\Delta+m)^2}+\frac{\omega_*^{\Delta+m} \log \omega_*}{\Delta+m}\,,
\ee
where ${\rm Re}\Delta>-m$.   We see that 1-loop corrections in~\eqref{eq:logsofttheorem} are picked out by double poles at $\Delta=-m$ but now starting from $m=0,1,...$. We can similarly pull down higher powers of $\log\omega$ by taking further $\p_\Delta$ derivatives. At loop level~$\ell$ we expect logarithmic corrections of the form $\omega^{\ell-1} (\log \omega)^\ell$ in the energetically soft expansion of momentum space amplitudes to be picked out by poles of $\ell^{th}$ order in celestial amplitudes. Moreover we see from the scatterer-probe limit computations of~\cite{Laddha:2018rle,Laddha:2018myi,Sahoo:2018lxl,Ghosh:2021hsk} that there will be poles up to order $2-k$  at $\Delta=k$ for $k=\{1,0,-1,-2,...\}$. 

This shows that at loop level we want to modify the limit in~\eqref{alim} to a contour integral that extracts the residue.  While this appears to generically modify the definition of the $w_{1+\infty}$ generators at loop level, this need not be the case for all theories.  For example, the $w_{1+\infty}$ symmetry was recently shown to take the same form at loop level in self-dual gravity~\cite{Ball:2021tmb}. While the fate of celestial symmetry generators at loop level is very interesting, the main takeaway in what follows is that we need to be careful how we take limits in $\Delta$. From the point of view of the extrapolate dictionary, this connects to an order of limits issue between taking $r\rightarrow\infty$ versus $\Delta$ conformally soft.

In the extrapolate dictionary,  taking the large-$r$ limit of radiative data lands us on the saddle-point approximation which we can use to define the asymptotic states by acting with the operators~\eqref{eq:2Dopintro}.
However, analytically continuing the data that survives in the large-$r$ limit of radiative solutions to conformally soft values $\Delta\in \mathbb{Z}$ (for $s\in \mathbb{Z}$)  is not the same as analytically continuing the full bulk conformal primary wavefunctions to conformally soft values of $\Delta$~\cite{Donnay:2018neh}. We see symptoms of this already in the helicity redundancy/relations of the $\Delta=1$ conformally soft limit in the soft charges~\cite{Donnay:2020guq} and the dressing operators~\cite{Arkani-Hamed:2020gyp,Pasterski:2021dqe} which include linear combinations of both helicities. We dive further into this issue in section~\ref{sec:CelestialTriangle}, and this order of limits will play an important role in how we define our subleading soft charges in section~\ref{sec:Goldilocks}. For now we will close our general discussion of soft limits by turning to the position basis, where our understanding of going from the boundary to the bulk and vice versa will be most clear.

%%%%%%%%%%%%%%%%%%%%%%%%%%%%%%%%%%%%%%%%%%%%%%%%%%%%%%%%%%
\subsection{\texorpdfstring{$A(u)$}{A(u)}}
%%%%%%%%%%%%%%%%%%%%%%%%%%%%%%%%%%%%%%%%%%%%%%%%%%%%%%%%%%
For every soft theorem there should be a memory effect corresponding to the classical expectation value of the operator appearing in the soft charge~\cite{Strominger:2014pwa,1502.06120PSZ,Pasterski:2015zua,Strominger:2017zoo}.\footnote{See also~\cite{Miller:2021hty} for a nice recent review of memory effects as Bremsstrahlung focusing on electromagnetism.}  In the position basis, the semi-classical interpretation of soft physics and memory effects is most apparent. Even the logarithmic corrections to the subleading soft theorem can be expected from this semi-classical perspective, once we take into account the deflection of the external scattering states due to long-range interactions~\cite{Laddha:2018myi}.

Focusing on the $u$-behavior associated to the Fourier transform of the terms in the expansion~\eqref{eq:logsofttheorem} \cite{Laddha:2018rle,Sahoo:2021ctw,Ghosh:2021hsk} we see that the tree level computations give
\be\label{uprof}
\omega^{n-1}\stackrel{\F}{\longleftrightarrow} \p_u^n \theta(u)
\ee
while the logs give extended tail behavior since
\be
\log |\omega|  \stackrel{\F}{\longleftrightarrow} -\frac{1}{2|u|}-\gamma_E \delta(u)
\ee
where $\gamma_E$ is the Euler-Mascheroni constant. These tails can affect the convergence of the BMS fluxes, and we will run into this issue when we define the subleading symmetry generators. Moreover, the $u$-boundary conditions are important for the antipodal matching of the canonical charges near $\mathscr{I}^+_-$ and $\mathscr{I}^-_+$. 

The position basis is also easiest to go into the bulk. With this $u$-behavior of the radiative data, we can use the equations of motion to propagate to subleading orders in $r$.  The structure of the recursions in appendix~\ref{larger} show that in harmonic gauge we expect to integrate in $u$ for every order in $r$.  While this is sensitive to the residual gauge fixing, it tells us that the tree level soft theorem corresponding to $\omega^{n-1}$ (or $\Delta=1-n$) leads to a shift in the metric or gauge field suppressed by $r^{-n}$ compared to radiative order.\footnote{We similarly see that the tail terms coming at loop-level will grow like $\log u$ at subleading order in $r^{-1}$.} Such subleading shifts in the metric match the tower of memory effects discussed in~\cite{Compere:2019odm}, which have been speculated to be connected to multipole moments~\cite{Seraj:2016jxi,Compere:2017wrj}.\footnote{Note that the charge conservation laws involve both a hard and soft part and  the boundary conditions for integrating by parts on the sphere therefore do not imply any additional global symmetries.} 
While this interpretation is under some debate~\cite{Mao:2020vgh} and alternate proposals are able to capture the $w_{1+\infty}$ generators without augmenting the phase space~\cite{Freidel:2021ytz} we will be able to make a connection between the multipole memory and what $w_{1+\infty}$ generators are measuring in section~\ref{sec:Discussion}.  To connect to the proposal in~\cite{Compere:2019odm}, we will need to understand the way that residual gauge symmetries can appear in the conformal primaries, which we resolve in section~\ref{sec:Goldilocks}.

%%%%%%%%%%%%%%%%%%%%%%%%%%%%%%%%%%%%%%%%%%%%%%%%%%%%%%%%%%
\section{The Celestial Triangle}
\label{sec:CelestialTriangle}
%%%%%%%%%%%%%%%%%%%%%%%%%%%%%%%%%%%%%%%%%%%%%%%%%%%%%%%%%%

We now discuss the Infrared Triangle from a purely celestial perspective. The aspects that we can lift to any basis are: 1.~the identification of two low energy modes related by the symplectic paring and 2.~a corresponding Ward identity for their insertions.  We review the conformal primary wavefunctions that pick out the memory and Goldstone operators for the cases with a known large gauge symmetry in section~\ref{GM} and show that the decoupling of primary descendants implies conservation laws for the full tower in section~\ref{3diamond}. Before doing so, we revisit the extrapolate dictionary for standard conformal primary basis, without restricting to a radiative spectrum in section~\ref{sec:softext}. This is important for how we interpret the celestial currents in the bulk, since the conformally soft and large-$r$ limits do not commute.

%%%%%%%%%%%%%%%%%%%%%%%%%%%%%%%%%%%%%%%%%%%%%%%%%%%%%
\subsection{Extrapolating Conformally Soft Modes}\label{sec:softext}
%%%%%%%%%%%%%%%%%%%%%%%%%%%%%%%%%%%%%%%%%%%%%%%%%%%%%

We will be considering the operators $\O^{\pm}_{\Delta,J}(w,\bw)$ defined via~\eqref{eq:3Dmu} for $\mathrm b=\Delta$, that is
\be
\label{eq:2Dop}
\mathcal{O}^{\pm}_{\Delta,J}(w,\bw)=i(\hat{O}^s(X^\mu),\Phi_{\Delta,J}(X_\mp^\mu;w,\bw)^*)_{\Sigma}.
\ee
 These operators are interpreted as conformal primaries in 2D celestial CFT, inheriting their conformal transformation properties from the wavefunctions $\Phi_{\Delta,J}(X^\mu_\mp;w,\bw)$. Here we will review the bulk spin-$s$ fields that transform as 2D conformal primaries under the $SL(2,\mathbb{C})$ Lorentz group with conformal dimension $\Delta$ and spin $J$. They are gauge equivalent to the Mellin transform of plane waves that we studied in section~\ref{sec:3Bases}.  Here we will also want to take the Cauchy slice $\Sigma$ to null infinity, but we will need to make sure that we consider how this limit works when $\Delta$ is non-radiative, but rather conformally soft.

%%%%%%%%%%%%%%%%%%%%%%%%%%%%%%%%%%%%%%%%%%%
\subsubsection{Conformal Primary Wavefunctions}
%%%%%%%%%%%%%%%%%%%%%%%%%%%%%%%%%%%%%%%%%%%

Focusing on massless particles with four momentum $k^\mu=\omega q^\mu$ the Mellin transform in $\omega$ takes a plane wave to a scalar conformal primary wavefunction
\be\label{scalar1}
\phi^\pm_\Delta=\int_0^\infty d\omega \omega^{\Delta-1}e^{\pm i\omega q\cdot X_\pm}=\frac{\Gamma(\Delta)}{(\pm i)^\Delta}\frac{1}{(-q\cdot X_\pm)^\Delta}\,,
\ee
where the regulator $X^\mu_\pm\equiv X^\mu \mp i\varepsilon(1,0,0,0)$ makes the integral converge. We will suppress the $\pm$ superscripts when unnecessary henceforth. To describe conformal primary wavefunctions with spin it will be convenient to introduce the spacetime dependent polarization vectors~\cite{Pasterski:2020pdk}
\be\label{mKS}
m^\mu=\epsilon^\mu_w+\frac{\epsilon_w\cdot X}{(-q\cdot X)} q^\mu\,, ~~~\bar{m}^\mu=\epsilon^\mu_\bw +\frac{\epsilon_\bw\cdot X}{(-q\cdot X)} q^\mu\,,
\ee
which satisfy $m\cdot\bar{m}=1$ and transform with $\Delta=0$ and $J=\pm1$, respectively. 
Starting with the conformal primary scalar $\Phi_{\Delta,J=0}\equiv \varphi_\Delta$\footnote{The change in normalization from~\eqref{scalar1} is to match the $\Delta=1$ vector primary and the $\Delta=1,0$ metric primaries onto the celestial currents generating large gauge symmetry and BMS supertranslations/superrotations, respectively.
}
\begin{equation}\label{varphi}
    \varphi_\Delta=\frac{1}{(- q\cdot X)^\Delta}\,,
\end{equation}
we can construct the conformal primary vector $\Phi_{\Delta,J=\pm1}\equiv A_{\Delta,\pm1;\mu}$ and metric $\Phi_{\Delta,J=\pm2}\equiv h_{\Delta,\pm2;\mu\nu}$ via dressing with the spacetime dependent normalization vectors
\be\begin{array}{ll}\label{CPWs}
     A_{{\Delta},+1;\mu}=m_\mu \varphi_{\Delta} \,,&~~~ A_{{\Delta},-1;\mu}=\bar{m}_\mu \varphi_{\Delta}\,, \\    h_{{\Delta},+2;\mu\nu}=m_\mu m_\nu \varphi_{\Delta} \,,&~~~  h_{{\Delta},-2;\mu\nu}=\bar{m}_\mu \bar{m}_\nu \varphi_{\Delta}\,.
\end{array}\ee
These radiative conformal primary wavefunctions satisfy the equations of motion for massless spin~$s$ particles in the vacuum and transform under SL(2,$\mathbb{C}$) as 4D tensors with bulk spin~$s$ and as 2D conformal primaries with conformal dimension $\Delta$ and 2D spin $J=\pm s$. A shadow transform takes the conformal primary wavefunctions~\eqref{varphi}-\eqref{CPWs} to wavefunctions with reflected conformal dimension $2-\Delta$ and flipped SL(2,$\mathbb{C}$) spin. The scalar shadow primary wavefunction is given by
\begin{equation}\label{SHvarphi}
    \widetilde{\varphi}_{\Delta}=(-X^2)^{\Delta-1}\varphi_{\Delta}\,, 
\end{equation}
from which we obtain the vector and metric shadow primary wavefunctions\footnote{Note that we omitted the minus sign in the vector primary that arises from matching this 4D shadow transform to the normalization of the 2D shadow transform used below~\cite{Pasterski:2021fjn} for convenience.}
\be\begin{array}{ll}\label{SHCPWs}
    \widetilde{A}_{\Delta,-1;\mu}=\bar{m}_\mu\tvarphi_\Delta\,, &~~~\widetilde{A}_{\Delta,+1;\mu}=m_\mu\tvarphi_\Delta\,,\\
    \widetilde{h}_{\Delta,-2;\mu\nu}\,=\bar{m}_\mu\bar{m}_\nu\tvarphi_\Delta\,, &~~~ \widetilde{h}_{\Delta,+2;\mu\nu}\,={m}_\mu {m}_\nu\tvarphi_\Delta\,.
\end{array}\ee
The radiative conformal primary wavefunctions~\eqref{CPWs} and~\eqref{SHCPWs} satisfy the harmonic (Lorentz) and radial gauge conditions
\begin{equation}
    \nabla^\mu A_\mu=0\,, \quad  \nabla^\mu h_{\mu\nu}=0\,, \quad \quad  X^\mu A_\mu=0\,, \quad  X^\mu h_{\mu\nu}=0\,.
\end{equation}
The metric primaries are also traceless. When $\Delta \in 1+i\mathbb{R}$ is on the principal series of the SL$(2,\mathbb{C})$ Lorentz group the above conformal primary wavefunctions form a complete $\delta$-function normalizable basis~\cite{Pasterski:2017kqt}. The (conformally) soft sector of scattering is instead captured by analytically continuing to $\Delta \in \mathbb{Z}$.

%%%%%%%%%%%%%%%%%%%%%%%%%%%%%%%%%%%%%%%%%%%%%%%%%%%%%%%%%%
\subsubsection{Conformal Primaries Near Null Infinity}
%%%%%%%%%%%%%%%%%%%%%%%%%%%%%%%%%%%%%%%%%%%%%%%%%%%%%%%%%%

The relation between the $\omega$ and $u$ primaries in the extrapolate dictionary of section~\ref{sec:3Bases} can be seen by Mellin transforming the saddle point approximation of the plane wave~\eqref{eq:saddlepoint}:
\begin{equation}\badat{3}\label{eq:sp}
  \lim_{r\rightarrow\infty} \int_0^\infty d\omega \omega^{\Delta-1}e^{\pm i\omega q\cdot X-\varepsilon\omega q^0}
  &=r^{-1}\frac{(u\mp i \varepsilon)^{1-\Delta}\Gamma(\Delta-1)}{(\pm i)^{\Delta}(1+z\bz)^{\Delta-2}} \pi \delta^{(2)}(z-w)+\O(r^{-2})\,.
\eadat\end{equation}
However, our assumption in section~\ref{sec:3Bases} that the wavefunctions obey the radiative fall-off conditions is no longer guaranteed when continuing the finite energy wavefunctions off the principal series~\cite{Pasterski:2017kqt}.  We will now examine the large-$r$ expansion for primaries of general conformal dimension~$\Delta$. 

The Mellin transformed plane wave~\eqref{eq:sp} implies that we have the following contact terms for the scalar primary 
\begin{equation}\label{eq:varphicontact}
     \lim_{r\to \infty}\varphi_\Delta\Big|_{z= w}=r^{-1}\frac{u^{1-\Delta}}{\Delta-1} (1+z\bz)^{2-\Delta}\pi \delta^{(2)}(z-w)+\O(r^{-2})\,.
\end{equation}
Meanwhile series expanding~\eqref{varphi} away from $z=w$ at large-$r$ gives 
\begin{equation}\label{eq:varphinoncontact}
    \lim_{r\to \infty}\varphi_\Delta\Big|_{z\neq w}=\left(\frac{1+z\bz}{2(z-w)(\bz-\bw)}\right)^\Delta r^{-\Delta}+\O(r^{-\Delta-1})\,.
\end{equation}
 We note that the range for which the $\omega$ integral in~\eqref{eq:sp} converges, Re$(\Delta)>1$, is precisely where the saddle point term dominates the large-$r$ expansion. For Re$(\Delta)< 1$, the series expansion~\eqref{eq:varphinoncontact} is leading compared to the contact term~\eqref{eq:varphicontact}. For Re$(\Delta)= 1$ the above expressions contribute at the same order; however, away from the conformally soft limit of $\Delta=1$, the non-contact piece is rapidly oscillating in $r$.

Repeating the same exercise for the shadow transformed scalar primary, we find that the role of the contact and non-contact terms is reversed. Using~\eqref{SHvarphi} with $-X^2=u(2r+u)$ we find
\begin{equation}\label{SHphicontact}
   \lim_{r\to \infty}  \tvarphi_\Delta\Big|_{z= w}=r^{\Delta-2}\frac{2^{\Delta-1}}{\Delta-1}(1+z\bz)^{2-\Delta}\pi\delta^{(2)}(z-w)+\O(r^{\Delta-3})\,,
\end{equation}
while
\begin{equation}
    \lim_{r\to \infty} \tvarphi_\Delta\Big|_{z\neq w}=r^{-1}u^{-1+\Delta} \frac{1}{2}\left(\frac{1+z\bz}{(z-w)(\bz-\bw)}\right)^\Delta +\O(r^{-2})\,.
\end{equation}
We see that at generic points ($z\neq w$) it is the shadow primary which has the standard radiative falloffs. We also note that the pole at $\Delta=1$ in~\eqref{eq:varphicontact} gets canceled by the normalization of the shadow transform~\eqref{2dShadowTransform}.

For $s>0$ conformal (shadow) primaries, special care is required to capture all contact terms that contribute at a given order in the large~$r$ expansion near null infinity. First note that the spacetime-dependent polarization vectors $\{m,\bar m\}$ can be expressed as
\begin{equation}
    m_u=-r\frac{\sqrt{2}(1+z\bw)}{1+z\bz}(\bz-\bw)\varphi_1\,, \quad m_r=u\frac{\sqrt{2}(1+z\bw)}{1+z\bz}(\bz-\bw)\varphi_1\,,
\end{equation}
obeying the radial gauge relation $m_r=-\frac{u}{r} m_u$, while 
\be\label{mzmzb}
m_{z}=-r(2r+u)\frac{\sqrt{2}(\bz-\bw)^2}{(1+z\bz)^2}\varphi_{1}\,, \quad m_{\bz}=ru\frac{\sqrt{2}(1+z\bw)^2}{(1+z\bz)^2}\varphi_{1}\,.
\ee
Similar expressions hold for $\bar{m}$. As in the $s=0$ case, the contact term from the saddle point approximation for $s=1,2$ primary wavefunctions will dominate for ${\rm Re}(\Delta)>1$. At ${\rm Re}(\Delta)=1$ both terms are of the same order.

%%%%%%%%%%%%%%%%%%%%%%%%%%%%%%%%%%%%%%%%%%%%%%%%%%%%%%%%%%
\subsubsection{An Order-of-Limits}
%%%%%%%%%%%%%%%%%%%%%%%%%%%%%%%%%%%%%%%%%%%%%%%%%%%%%%%%%%

 In the radiative case ($\Delta\in 1+i\mathbb{R}$), the extrapolate dictionary~\eqref{eq:3Dmu} indeed reduces to what we expect from Mellin transforming the saddle point~\eqref{eq:sp}.  Namely the 2D celestial operators that are supported on null rays of the conformal boundary (as in figure~\ref{in_out_states}). The operators that create outgoing modes with positive 2D spin (positive helicity) are given by~\cite{Pasterski:2021dqe}
\be\label{eq:extrapolateDelta}
\mathcal{O}^{+}_{\Delta,1}\propto\lim_{r\rightarrow \infty} \int du u_+^{-\Delta} \hat{A}_z({u},r,z,\bz)\,, \qquad \mathcal{O}^{+}_{\Delta,2}\propto\lim_{r\rightarrow \infty} \frac{1}{r}\int du u_+^{-\Delta} \hat{h}_{zz}({u},r,z,\bz) \,,
\ee
where following the notation for $X_\pm$ we have $u_+=u-i\varepsilon$. Analogous statements hold for the opposite helicity modes, as well as for the $\mathcal{O}^-_{\Delta,J}$ selecting creation operators on $\mathscr{I}^-$.

If we start from the radiative scattering amplitudes, evaluate the transform on the principal series, and then analytically continue to $\Delta\in\mathbb{C}$ we can continue to write operators of this form.  In contrast, if we use the large-$r$ limit expansion of the bulk conformal primary wavefunctions we see that we are outside the standard phase space.  In particular the inner products~\cite{Ashtekar:1987tt,Crnkovic:1986ex,Lee:1990nz,Wald:1999wa}
 \begin{equation}\label{eq:IPspin1}
 (A, A')_\Sigma =- i \int d\Sigma^\rho \,\left[ A^{\nu} {F'}_{\rho\nu}^{*} 
 -{A'}^{* \nu} F_{\rho\nu} 
 \right] ,
 \end{equation}
for spin-1 and 
\begin{equation}\label{eq:IPspin2}
( h,h')_\Sigma=-i\int d\Sigma^\rho \Big[ h^{\mu \nu} \nabla_\rho h'^{*}_{\,\,\mu \nu}-2h^{\mu \nu} \nabla_\mu h'^{*}_{\,\,\rho\nu} - (h \leftrightarrow h'^{*})\Big],
\end{equation}
for spin-2 do not converge. We will run into this issue in full force in section~\ref{sec:Goldilocks}, though we have already encountered it when dealing with Diff$(S^2)$ transformations in~\cite{Donnay:2020guq}. From our discussion in section~\ref{sec:softADelta} of conformally soft limits in the $\Delta$ basis, we expect the amplitudes to be singular as we limit to negative integer values of $\Delta$.  We see that both the $r$ behavior of the conformally soft modes and the $u$ behavior of the amplitudes point us towards using the order of limits implicit in the derivation of the `soft theorem = Ward identity' relation.  Namely we take large-$r$ before we take $\Delta$ conformally soft. However, as in~\cite{Donnay:2020guq} we will see that this can be reconciled with the brute force approach of including overleading modes upon renormalizing the symplectic product.

%%%%%%%%%%%%%%%%%%%%%%%%%%%%%%%%%%%%%%%%%%%%%%%%%%%%%%%%%%
\subsection{Conformal Goldstones and their Memories}\label{GM}
%%%%%%%%%%%%%%%%%%%%%%%%%%%%%%%%%%%%%%%%%%%%%%%%%%%%%%%%%%
We now turn to the question of identifying the Goldstone and memory modes in the celestial basis. Energetically and conformally soft theorems have been related to asymptotic symmetries in~\cite{He:2014laa,Kapec:2014opa,He:2014cra,Avery:2015iix,Lysov:2015jrs} and~\cite{Kapec:2016jld,Cheung:2016iub,Donnay:2018neh,Donnay:2020guq,Pano:2021ewd}. These are the residual gauge transformations that are consistent with the conformal primary gauge fixing. By rewriting the 2D operator~\eqref{eq:2Dop} in terms of the symplectic product
\be\label{eq:2Dopsympprod}
\mathcal{O}^{\pm}_{\Delta,J}(w,\bw)=\Omega(\hat{O}^s(X^\mu),\Phi_{\Delta,J}(X_\pm^\mu;w,\bw))\,,
\ee
it is clear that for the values of $\Delta$ where $\Phi_{\Delta,J}$ is pure gauge, the operator $\mathcal{O}_{\Delta,J}$ computed in the linearized theory will reduce to the soft part of the canonical charge for that gauge transformation. This is the memory operator. The symplectic pairing relates modes of dimension $\Delta$ and $2-\Delta$, so that the Goldstone wavefunction picks out the memory operator and vice versa.   The Goldstone operator decouples from amplitudes because of gauge invariance while the memory operator is correspondingly constrained to be related to matter contributions.   Indeed, the operator $\O_{\Delta=1,J=\pm1}$ corresponds to the conformally soft photon current associated to large gauge symmetry, while the $SL(2,\mathbb{C})$ descendant of $\O_{\Delta=1,J=\pm2}$ corresponds to the BMS supertranslation current; meanwhile $\O_{\Delta=0,J=\pm2}$ gives rise to the shadow of the 2D stress tensor generating superrotation symmetry.

The spectrum of Goldstone primary wavefunctions is summarized in table~\ref{tab:Goldstone}. Their shadows are similarly Goldstone modes.  The corresponding memory modes are related by $\Delta\mapsto 2-\Delta$.  We review the construction for the leading soft theorems in section~\ref{sec:leadmem} and add the subleading soft graviton's analog in section~\ref{sec:submem}. 
 
\vspace{1em}
\begin{table}[bh!]
\renewcommand*{\arraystretch}{1.3}
\centering
\begin{tabular}{|c|c|cc|}
\hline
  & \multicolumn{1}{c|}{${A}^{\Delta}_{\mu}$} &  \multicolumn{2}{c|}{${h}^{\Delta}_{\mu\nu}$}\\
  \hline
 $\Delta$ &1  &  1&  0 \\
 symmetry & large $U(1)$  & supertranslation &  Diff$(S^2)$ superrotation \\
\hline
\end{tabular}
\caption{Conformal Goldstone modes for bosonic particles with spin $s\le2$. 
}
 \label{tab:Goldstone}
\end{table}

%%%%%%%%%%%%%%%%%%%%%%%%%%%%%%%%%%%%%%%%%%%%%%%%%%%%%%%%%%
\subsubsection{Celestial Electromagnetic and Displacement Memory Modes}\label{sec:leadmem}
%%%%%%%%%%%%%%%%%%%%%%%%%%%%%%%%%%%%%%%%%%%%%%%%%%%%%%%%%%
The Goldstone modes of the spontaneously broken large gauge and supertranslation symmetry are given by the $\Delta=1$ pure gauge conformal primary wavefunctions~\cite{Donnay:2018neh}
\begin{equation}
     A^{\rm G}_{1,+1;\mu} =m_\mu \varphi_1\,, \quad h^{\rm G}_{1,+2;\mu\nu} =m_\mu m_\nu \varphi_1\,,
\end{equation}
with $m_\mu \mapsto \bm_\mu$ for the opposite spin.
These Goldstone modes are symplectically paired with the conformally soft photon and graviton modes constructed in~\cite{Donnay:2018neh,Pasterski:2020pdk}
\begin{equation}
    A^{\rm CS}_{1,-1;\mu}=\bm_\mu \varphi^{\rm CS}\,, \quad h^{\rm CS}_{1,-2;\mu\nu}=\bm_\mu \bm_\nu \varphi^{\rm CS}\,,
\end{equation}
where\footnote{We have dropped here the logarithmic term in~\cite{Donnay:2018neh}, as in~\cite{Arkani-Hamed:2020gyp}, since it is not needed for this discussion.}
\begin{equation}
    \varphi^{\rm CS}=\theta(X^2) \varphi_1\,.
\end{equation}
Indeed one finds that the symplectic product evaluates to~\cite{Donnay:2018neh}
\begin{equation}
       \Omega(A^{\rm CS}_{1,-1},A^{'{\rm G}}_{1,+1})=(2\pi)^2\delta^{(2)}(w-w')\,, \quad  \Omega(h^{\rm CS}_{1,-2},h^{'{\rm G}}_{1,+2}) =(2\pi)^2\delta^{(2)}(w-w')\,.
\end{equation}
The conformally soft modes $A^{\rm CS}$ and $h^{\rm CS}$ correspond to a shift at null infinity which can be seen from $\lim_{r\to \infty}\theta(X^2)\propto\theta(-u)$ which motivates their interpretation as being relation to electromagnetic~\cite{Bieri:2013hqa,Pasterski:2015zua,Susskind:2015hpa} and gravitational displacement~\cite{1974SvA....18...17Z,Braginsky:1986ia,gravmem3,Strominger:2014pwa} memory. 

%%%%%%%%%%%%%%%%%%%%%%%%%%%%%%%%%%%%%%%%%%%%%%%%%%%%%%%%%%
\subsubsection{Celestial Spin Memory Mode}\label{sec:submem}
%%%%%%%%%%%%%%%%%%%%%%%%%%%%%%%%%%%%%%%%%%%%%%%%%%%%%%%%%%

In gravity there is a further Goldstone mode in the conformal basis given by the $\Delta=0$ pure gauge conformal primary wavefunction or its $\Delta=2$ shadow~\cite{Donnay:2020guq}
\begin{equation}
    h^{\rm G}_{0,-2;\mu\nu} =\bm_\mu \bm_\nu\,, \quad \th^{\rm G}_{2,+2;\mu\nu}=-X^2 m_\mu m_\nu \varphi_2\,,
\end{equation}
with $m_\mu \mapsto \bm_\mu$ for the opposite spin. These generate Diff($S^2$) and Virasoro superrotation symmetry, and the $\Delta=2$ shadow mode can be shown~\cite{Donnay:2020guq} to correspond to the 2D stress tensor for 4D gravity~\cite{Kapec:2016jld}. The angular components of these Goldstone modes are given by
\begin{equation}
\badat{3}
    h^{\rm G}_{0,-2;zz}&=r\bar C^0_{zz}+\O(r^0)\,, ~~~& \th^{\rm G}_{2,+2;zz}&=r \widetilde{C}^2_{zz}+\O(r^0)\,,\\
    h^{\rm G}_{0,-2;\bz\bz}&=r^2 \bar q^0_{\bz\bz}+r \bar C^0_{\bz\bz}+\O(r^0)\,,~~~ &   \th^{\rm G}_{2,+2;\bz\bz}&=r^2 \widetilde{q}^2_{\bz\bz}+r \widetilde{C}^2_{\bz\bz}+\O(r^0)\,,
\eadat
\end{equation}
where
\begin{equation}
    \bar C^0_{zz}=-u D_z^3 \bar Y^z=u 2\pi \delta^{(2)}(z-w)\,,\quad \widetilde{C}^2_{zz}=-u D_z^3 Y^z=\frac{u}{(z-w)^4}\,,
\end{equation}
while the overleading terms $\propto \bar q^0_{\bz\bz},\widetilde{\bar{q}}^2_{\bz\bz}$ as well as the terms $\propto \bar C^0_{\bz\bz},\widetilde{C}^2_{\bz\bz}$ will not be important here as they do not contribute to the renormalized symplectic product~\cite{Donnay:2020guq}.

We propose that spin memory~\cite{1502.06120PSZ} arises from the $\Delta=2$ conformal primary wavefunction or its $\Delta=0$ shadow 
\begin{equation}
    h_{2,+2;\mu\nu} =m_\mu m_\nu\varphi_2\,, \quad \th_{0,-2;\mu\nu}=-\frac{1}{X^2} \bm_\mu \bm_\nu\,.
\end{equation}
Their angular components are given by
\begin{equation}\label{h2exp}
\badat{2}
    h_{2,+2;zz}&=\O(r^{0}), ~~~&  \th_{0,-2;zz}&=\O(r^{0})\,,\\
    h_{2,+2;\bz\bz}&=r C^2_{\bz\bz}+\O(r^{0})\,, ~~~&\th_{0,-2;\bz\bz}&=r \bar{\widetilde{C}}^0_{\bz\bz}+\O(r^0)\,,
\eadat
\end{equation}
where
\begin{equation}
    C^2_{\bz\bz}=\frac{1}{u} \frac{2\pi}{3}\frac{\delta^{(2)}(z-w)}{(1+z\bz)^{2}}\,, \quad \bar{\widetilde{C}}^0_{\bz\bz}=\frac{1}{u} \frac{(z-w)^2}{(\bz-\bw)^2} \frac{1}{(1+z\bz)^2}\,.
\end{equation}
This implies that the following Bondi news at future null infinity
\begin{equation}
\badat{2}
& N^{2,\pm}_{\bz\bz}=-\frac{1}{(u\mp i\varepsilon)^2}\frac{2\pi}{3} \frac{\delta^{(2)}(z-w)}{(1+z\bz)^2}\,, \quad \bar{\tN}^{0,\pm}_{\bz\bz}=-\frac{1}{(u\mp i\varepsilon)^2}\frac{(z-w)^2}{(\bz-\bw)^2}\frac{1}{(1+z\bz)^2}\,,
\eadat
\end{equation}
where we reinstated the $\pm i\varepsilon$ regulator.
We define the following admixture of incoming and outgoing modes as a memory mode
\begin{equation}\label{eq:Nreg}
 N^{\rm M}_{AB}=\lim_{\varepsilon \to 0} \frac{1}{2\pi i} (N^{+}_{AB}-N^{-}_{AB})\,,
\end{equation}
whose Bondi news takes the following form
\begin{equation}
   N^{2,{\rm M}}_{\bz\bz}=\p_u \delta(u) \frac{2\pi}{3}\frac{\delta^{(2)}(z-w)}{(1+z\bz)^2}\,, \quad \bar{\tN}^{0,{\rm M}}_{\bz\bz}=\p_u \delta(u)\frac{(z-w)^2}{(\bz-\bw)^2}\frac{1}{(1+z\bz)^2}\,.
\end{equation}
We see that the conformally soft modes $h^{\rm M}_{2,\pm2}$ $\th^{\rm M}_{0,\pm2}$ correspond to radiation at null infinity measured by the spin memory effect~\cite{1502.06120PSZ}.  Namely 
\be
\int du u N_{\bz\bz}^{2,\rm M}=-\frac{2\pi}{3}\frac{\delta^{(2)}(z-w)}{(1+z\bz)^2}
\ee 
is finite for the regulated $\Delta=2$ profile we have constructed. This $u$-profile also matches what we anticipated in~\eqref{uprof} which is a nice cross check for when we try to generalize to higher multipoles in section~\ref{sec:Discussion}.

Let us now turn to the pairing between these memory modes and the superrotation Goldstone mode. The renormalized~\cite{Compere:2018ylh,Donnay:2020guq} symplectic pairing between the Diff($S^2$) superrotation Goldstone mode $h^{\rm G}_{0,-2}$ and the regulated memory-like mode $h^{{\rm M}}_{2,+2}$ is 
\begin{equation}
\badat{3}
    \Omega(h^{{\rm M}}_{2,+2},h^{'{\rm G}}_{0,-2})&=2\int du d^2z \gamma^{z\bz} u D_z^3 \bar Y^z(w',\bw') N^{2,\rm M}_{\bz\bz}(w,\bw)\\
    &=\frac{4\pi^2}{3}\delta^{(2)}(w-w')\,.
\eadat
\end{equation}
This is equal to the pairing $\Omega(\th^{{\rm M}}_{0,-2},\th^{'\rm G}_{2,+2})$ between the Virasoro superrotation Goldstone mode $\th^{\rm G}_{2,+2}$ and the regulated memory-like mode $\th^{{\rm M}}_{0,-2}$  since the shadow transform squares to one. The additional pairings
\begin{equation}\label{eq:dualstresstensorIP}
      \Omega(\widetilde{h}^{\rm M}_{0,-2},{h}^{'{\rm G}}_{0,-2})=2\pi \frac{(w-w')^2}{(\bw-\bw')^2}\,,~~~ \Omega({h}^{\rm M}_{2,+2},\widetilde{h}^{'{\rm G}}_{2,+2})=\frac{2\pi}{3}\frac{1}{(w-w')^4}\,,
\end{equation}
are also related by a shadow transform to the one computed above. Note that the {\it dual stress tensor} of~\cite{Ball:2019atb} was constructed from the inner product between the $\Delta=2$ superrotation Goldstone mode and the primary $h_{\Delta,J}$ with $\Delta=2+i\varepsilon$ where $\varepsilon$ was treated as a continuous parameter (i.e. for modes on a shifted copy of the principal series). On hyperbolic slices the result of this different choice of regulation is still proportional to~\eqref{eq:dualstresstensorIP}, but with an extra factor of $\delta(\varepsilon)$ while over a complete Cauchy slice there is an extra factor of $\varepsilon \delta(\varepsilon)$. In contrast to~\cite{Ball:2019atb} we are able to identify a finite symplectic pairing starting with the renormalized inner product from~\cite{Donnay:2020guq} that removes the overleading in~$r$ piece and the regulator prescription~\eqref{eq:Nreg} that mixes incoming and outgoing modes.

%%%%%%%%%%%%%%%%%%%%%%%%%%%%%%%%%%%%%%%%%%%%%%%%%%%%%%%%%%
\subsection{Three Types of Celestial Diamonds}\label{3diamond}
%%%%%%%%%%%%%%%%%%%%%%%%%%%%%%%%%%%%%%%%%%%%%%%%%%%%%%%%%%

While soft factorization of amplitudes persist to more subleading orders, there no longer exists an obvious asymptotic symmetry interpretation, in the standard sense, for the more subleading soft theorems. The subleading soft photon and sub-subleading soft graviton theorems have instead been argued by~\cite{Campiglia:2016hvg,Campiglia:2016efb,Campiglia:2016jdj} to be related to divergent in~$r$ large gauge transformations. In the following section we will show that in the conformal primary basis these are related to `Goldilocks modes' which are not Goldstone or pure gauge modes per se but from which we can nevertheless construct asymptotic pure gauge modes and define an analogue of soft charges.  

\begin{figure}[t]
\centering
\vspace{-0.5em}
\begin{tikzpicture}[scale=1.6,every node/.style={scale=1}]
\definecolor{darkgreen}{rgb}{.0, 0.5, .1};
\draw[thick](0,2)node[above]{$\mathcal{O}_{h, \bar{h}} $} ;
\draw[thick,->](0,2)--node[above left]{$\p_\bw^{ \bar \n}$} (-1+.05,1+.05);
\draw[thick,->](0,2)--node[above right]{$\p_w^{  \n}$} (2-.05,0+.05);
\draw[thick,->] (-1+.05,1-.05)node[left]{$ \p_\bw^{\bar \n} \mathcal{O}_{h, \bar{h}} ~ $} --node[below left]{$\p_w^{ \n}$} (1-.05,-1+.05) ;
\draw[thick,->] (2-.05,0-.05)node[right]{$~ \p_w^{ \n} \mathcal{O}_{h, \bar{h}} $} --node[below right]{$\p_\bw^{ \bar \n}$} (1.05,-1+.05) ;
\draw[thick](1,-1.2)node[below]{$\p_w^{\n} \p_\bw^{\bar \n} \mathcal{O}_{h, \bar{h}}  $};
\filldraw[black] (0,2) circle (2pt) ;
\filldraw[black] (2,0) circle (2pt) ;
\filldraw[black] (-1,1) circle (2pt) ;
\filldraw[black] (1,-1) circle (2pt) ;
\end{tikzpicture}
\caption{Celestial diamond illustrating the nested submodule structure that appears at special values of the conformal dimension such that $h= \frac{1-\n}{2}, \bar{h}=\frac{1-\bar \n}{2}$ for $k,\bar{k}\in\mathbb{Z}_>$. }
\label{Nested_submodules}
\end{figure}
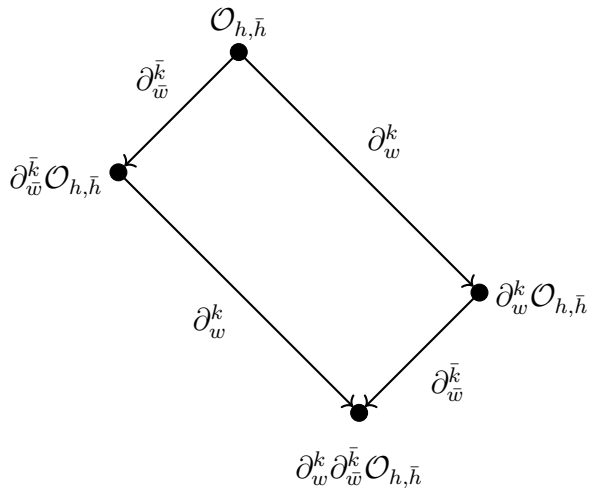

As discussed in the introduction, the advantage of the boost basis is that it really sidesteps these phase space issues. Namely, so long as we can apply radial quantization techniques to CCFT (as motivated by the extraction of symmetry algebras from OPEs~\cite{Guevara:2021abz} and the state operator correspondence proposal of~\cite{Crawley:2021ivb}), we would expect that primary descendants decouple.  This is because we are essentially dimensionally reducing null infinity to a $S^2$ cross-section and are left with a continuous spectrum for each bulk field. Once we are free to analytically continue off the principal series we can choose to limit towards values of the dimension for which there are primary descendants.   In equations, the operator $\mathcal{O}_{\Delta,J}(0,0)$ creates a state $|h,\bar{h}\rangle$ with 
\be
h=\frac{1}{2}(\Delta+J),~~~\bar h=\frac{1}{2}(\Delta-J).
\ee
If we then take $\Delta$ such that
\be
\label{condition_diamond_hhb}
h=\frac{1-\n}{2}
\, , \qquad\qquad
\bar{h}=\frac{1-\bar{\n}}{2}\, ,
\ee
for $k,\bar{k}\in\mathbb{Z}_>$ the derivative $\p_w^k\mathcal{O}_{\Delta,J}$ ($\p_\bw^{\bar k}\mathcal{O}_{\Delta,J}$)  will decouple in correlation functions away from other operator insertions, giving a conservation law. In the case where both conditions in~\eqref{condition_diamond_hhb} are met, we have a nested structure of primary descendants, illustrated in figure~\ref{Nested_submodules}. 
A classification of these primary descendants was performed in~\cite{Pasterski:2021fjn}. Let us summarize the three cases here.

\paragraph{$\boldsymbol{1-s<\Delta<1+s}$}
Each of the soft theorems with known asymptotic symmetry interpretations lie at the left or right corners of a set of nested primary descendants termed `celestial diamonds'~\cite{Pasterski:2021fjn,Pasterski:2021dqe}.  The operators at the bottom of these diamonds give contact terms in correlation functions at the locations of other operators, which act as sources for these generalized currents.  As discussed in the previous subsection, $\mathcal{O}_{\Delta,J}$ give the soft charge for a corresponding asymptotic symmetry whenever the wavefunction $\Phi_{\Delta,J}$ is pure gauge.  This is the case for $1-s < \Delta \le 1 $. The remaining modes at $1 \le \Delta < 1+s $ in this range are the symplectically paired memory modes.  These ranges overlap at $\Delta=1$ for the integer spin case precisely because one needs to keep track of the conformally soft modes studied in~\cite{Donnay:2018neh}.

\paragraph{$\boldsymbol{\Delta<1-s}$}  
 For $\Delta=1-s-n$ the descendant wavefunction at level $n$ is naively zero. While this would seem too trivial to give any interpretation as a sourced generalized current, the algebra of the graviton modes themselves is anything but~\cite{Guevara:2021abz,Strominger:2021lvk}! In appendix~\ref{app:diamondLT} we resolve this tension by observing that if we continue  $\Delta=-n+\epsilon$  and expand to $\mathcal{O}(\epsilon)$, we can capture light transform and shadow relations among the wavefunctions at each corner. This is important for reconciling how the memory operators expected at $2-\Delta$ are related to these diamonds which, in turn, is necessary if we want to make a connection to the proposal of~\cite{Compere:2017wrj} regarding subleading shifts and multipole moments. We will turn to this tantalizing prospect in section~\ref{sec:Discussion}.

\paragraph{$\boldsymbol{\Delta=1-s}$ } 
The case where $\Delta=1-s$ corresponds to a degenerate celestial diamond where a radiative primary descends to its shadow mode at $\Delta'=1+s$. These correspond to the subleading-most soft theorems of~\cite{Low:1954kd,Cachazo:2014fwa}, which imply (at least at tree level) that other celestial operators will source this generalized current and that there should be a non-trivial conservation law despite the fact that the corresponding wavefunctions $\Phi_{1-s,\pm s}$ are not pure gauge. We will turn to these Goldilocks modes in the next section.

\vspace{1em}

With this classification in mind, let us return to the questions that started this subsection.  Namely, why are there extra soft theorems? do we really need to add extra large gauge transformations to our phase space? and why can't we see these gauge transformations in the conformal primary basis despite the fact that the representation theory (even more clearly) points to the existence of generalized currents?  One way to make sense of this in terms of the operator definition~\eqref{eq:2Dopsympprod} is to recall that when we augment or remove directions in our phase space we do so in pairs. This phrasing lets us extend the story to cases where we want to impose a constraint on the physical falloffs, but a gauge interpretation is less clear. In particular from our conformal primary wavefunctions, we see that the Mellin-transformed wavefunctions at the values of the conformal dimension that select the $w_{1+\infty}$ generators~\cite{Strominger:2021lvk} are overleading in $r$. These are the Goldstone-like modes because (like their gauge analogs) we would want to remove these directions from our phase space to get a well defined symplectic form on the physical (gauge invariant) phase space.\footnote{While small gauge transformations give flat directions of the symplectic form, large-gauge transformations can contribute boundary terms, hence the non-trivial pairings we saw in~\cite{Donnay:2018neh,Ball:2019atb} and section~\ref{GM}.} Recall that the diamonds involving operators of dimension $\Delta$ and $2-\Delta$ overlap due to the shadow relations.  The memory modes paired with this tower of Goldstone-like modes are within the radiative phase space.

%%%%%%%%%%%%%%%%%%%%%%%%%%%%%%%%%%%%%%%%%%%%%%%%%%%%%%%%%%
\section{Goldilocks Modes}
\label{sec:Goldilocks}

%%%%%%%%%%%%%%%%%%%%%%%%%%%%%%%%%%%%%%%%%%%%%%%%%%%%%%%%%%

In this section we identify the appropriate Goldstone-like modes and construct the soft charge operators that select the subleading soft photon and sub-subleading soft graviton. Namely, we will evaluate~\eqref{eq:2Dopsympprod} for the conformal dimensions where we know the corresponding conformally soft factorizations occur.\footnote{This subleading soft symmetry has also been the focus of the recent interesting work~\cite{Freidel:2021dfs}.} These modes present two challenges:  First, as discussed above, they have hitherto eluded a gauge symmetry interpretation in the conformal primary basis.  Second, the corresponding wavefunctions have boundary falloffs that would typically not be included in the physical phase space.  

For each of these subleading-most soft theorems, the Mellin-transformed primary will be overleading and outside the ordinary radiative phase space, while it descends to its shadow mode which will be asymptotically locally flat. To deal with this second issue, we will consider two options in turn: 1.~comparing the $r$-before-$\Delta$ order of limits proposal to the expected isomorphism between celestial currents and soft theorems, and 2.~renormalizing the symplectic form to evaluate the inner product between these overleading celestial primaries and radiative solutions.  We will see how the operators constructed in this manner are isomorphic to the expected soft charges. Similarly to what was observed in~\cite{Campiglia:2016hvg,Campiglia:2016efb} the overleading terms before renormalization are proportional to soft charges for the leading symmetries.  By taking suitable linear combinations of the two helicity sectors  we are also able to identify the overleading large gauge transformations that were attached to these soft theorems in~\cite{Campiglia:2016hvg,Campiglia:2016efb}. Finally we examine the associated memory modes.

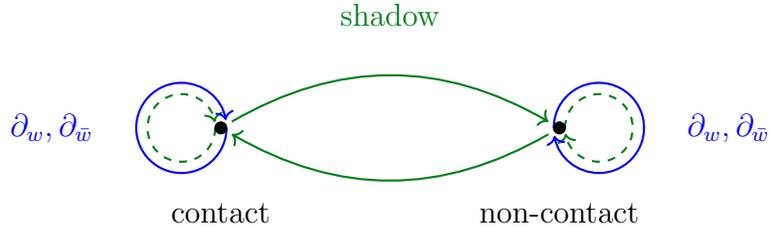
\begin{figure}[t]
\centering
\vspace{-0.5em}
\begin{tikzpicture}[scale=1.5]
 \definecolor{darkgreen}{rgb}{.0, 0.5, .1};
\node (A) at (0,0) {};
\node (B) at (3,0) {};
\node[] at (0,-.75)  {contact};
\node[] at (3,-.75)  {non-contact};
\draw[thick, darkgreen,->] (A) to [bend left] (B);
\draw[thick, darkgreen,->] (B) to [bend left] (A);
\draw[thick, dashed, darkgreen,->] ($(A)+(-.05,0)$) arc (0:-350:3mm);
\draw[thick, blue,->] ($(A)+(.05,0)$)  arc (0:-350:4mm);
\draw[thick, dashed, darkgreen,->] ($(B)+(+.05,0)$) arc (0:-350:-3mm);
\draw[thick, blue,->] ($(B)+(-.05,0)$)  arc (0:-350:-4mm);
\draw[fill] (A) circle (1.5pt);
\draw[fill] (B) circle (1.5pt) ;
\node[blue] at (-1.5,0)  {$\p_w,\p_\bw$};
\node[darkgreen] at (1.5,1)  {shadow};
\node[blue] at (4.5,0)  {$\p_w,\p_\bw$};
\node[] at (4.5,0)  {};
\end{tikzpicture}
\caption{Mapping between contact and non-contact terms under descendants (blue) and shadows (green). Both operations commute with the large-$r$ expansion. The dashed maps between terms of the same type will be less relevant for what follows. 
}
\label{fig:contactmixing}
\end{figure}

Before proceeding with the spin-1 and spin-2 cases, in turn, we discuss one nice feature we take advantage of for the $\Delta=1-s$ modes. In contrast to~\cite{Donnay:2018neh}, where we had to evaluate the symplectic product~\eqref{eq:2Dopsympprod} separately for all conformally soft modes (since missing a contact term in one expansion can have a drastic effect on its shadow), we can focus solely on the $\Delta=0,-1$ modes.  The results for the $\Delta=2,3$ shadows follow without needing a non-local smearing. This is because for these particular conformal dimensions, the degenerate celestial diamonds imply that these modes descend to their own shadows - see Figures~\ref{fig:subleadingphoton} and~\ref{fig:subsubleadinggraviton} below. Namely, the shadow operation 
 \begin{equation}\label{2dShadowTransform}
 \widetilde{\O}_{\tDelta,\tJ}(w,\bw)=\frac{k_{\Delta,J}}{2\pi} \int d^2w' \frac{\O_{\Delta,J}(w',\bw')}{(w'-w)^{2-\Delta-J}(\bw'-\bw)^{2-\Delta+J}}\,,
 \end{equation} 
 where $k_{\Delta,J}=\Delta-1+|J|$, amounts to a Green's function for the corresponding descendancy relation.  This is also true at the level of the full bulk wave function.  In particular, it will be useful to keep in mind that the operations of descending and shadowing commute with our large $r$ expansions and mix contact and non-contact terms in the manner illustrated in figure~\ref{fig:contactmixing}.

%%%%%%%%%%%%%%%%%%%%%%%%%%%%%%%%%%%%%%%%%%%%%%%%%%%%%%%%%%
\subsection{Subleading Conformally Soft Photon}
%%%%%%%%%%%%%%%%%%%%%%%%%%%%%%%%%%%%%%%%%%%%%%%%%%%%%%%%%%
In~\cite{Himwich:2019dug}, it was shown that the subleading soft charge for QED~\cite{Lysov:2014csa} could be presented as an additional current in the celestial CFT.  Based on the Goldstone mode accounting in~\cite{Pasterski:2017kqt}, we do not expect to find a subleading Goldstone mode for gauge theory amongst our conformal primary wavefunctions.  However, we do expect to be able to construct a soft charge operator of the correct conformal dimension.  Since soft theorems are gauge invariant, it should be possible to identify the relevant modes in the conformal basis. Here we argue that the subleading conformally soft photon can be associated to a conformal primary wavefunction $\Phi_{\Delta,J=\pm1}$ with conformal dimension $\Delta=0$:
\begin{equation}
 A_{0,+1;\mu}=m_{\mu} \,,\quad  A_{0,-1;\mu}=\bm_{\mu} \,,
\end{equation}
or, equivalently, its shadow with conformal dimension $\Delta=2$:
\begin{equation}
 \tA_{2,+1;\mu}=-X^2m_{\mu}\varphi_{2} \,,\quad  \tA_{2,-1;\mu}=-X^2\bm_{\mu}\varphi_{2} \,.
\end{equation}
One can indeed check that neither of the two primaries above is pure gauge.

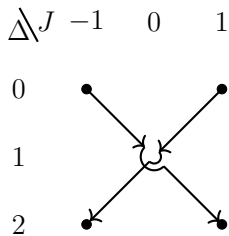
\begin{figure}[htb]
    \centering
   \begin{tikzpicture}[scale=.9,every node/.style={scale=.9}]
\definecolor{darkgreen}{rgb}{.0, 0.5, .1};
\draw[thick] (0+.1414/2,1+.1414/2) arc (45:-135:.1);
\node at (-2,0) {$2$};
\node at (-2,1) {$1$};
\node at (-2,2) {$0$};
\node at (-1.6,3) {$J$};
\node at (-2,2.87) {$\tiny{\Delta}$};
\node at (-1,3) {$-1$};
\node at (0,3) {$0$};
\node at (1,3) {$1$};
\filldraw[black] (1,0) circle (2pt) ;
\filldraw[black] (-1,0) circle (2pt) ;
\filldraw[black] (1,2) circle (2pt) ;
\filldraw[black] (-1,2) circle (2pt) ;
\draw[->,thick] (1,2) --  (0+.07,1+.07);
\draw[->,thick]  (0-.07,1-.07) -- (-1+.05,.05);
\draw[thick] (0-.1414,1+.1414) arc (135:315:.2);
\draw[->,thick] (-1,2) --  (0-.1414,1+.1414);
\draw[->,thick] (0+.1414,1-.1414) -- (1-.03,0.03);
\draw[thick] (-2+.01,3+.2) -- (-2+.28,3-.3);
\end{tikzpicture}
    \caption{Celestial `Diamonds' for the Subleading Soft Photon.}
    \label{fig:subleadingphoton}
\end{figure}

This fits with the result that celestial amplitudes obey a subleading {\it conformally soft} factorization theorem when the conformal dimension of one of the external spin-1 wavefunctions is taken to $\Delta=0$~\cite{Adamo:2019ipt}. 
In the following we will focus on the negative helicity wavefunction $A_{0,-1}$. The positive spin wavefunction $\tA_{2,+1}$ is related to it by a shadow transform.
We will furthermore take advantage of the fact that these conformally soft modes are related by the descendancy relation~\cite{Pasterski:2021fjn}
\begin{equation}\label{adesc}
    \frac{1}{2!}\p_w^2 A_{0,-1}=\tA_{2,+1}\,,\quad  \frac{1}{2!}\p_\bw^2 A_{0,+1}=\tA_{2,-1}\,,
\end{equation}
illustrated in figure~\ref{fig:subleadingphoton}. 

%%%%%%%%%%%%%%%%%%%%%%%%%%%%%%%%%%%%%%%%%%%%%%%%%%%%%%%%%%
\subsubsection{Goldstone-like Primaries}
Let us now keep track of the leading expansion for the Goldilocks photon primary that appear in our final expressions for the soft charges.  The contact terms arise from the saddle point approximation discussed above.  Whenever we have overleading modes, we will need to keep track of more terms in the expansion in order to evaluate the inner product with radiative order wavefunctions.

%%%%%%%%%%%%%%%%%%%%%%%%%%%%%%%%%%%%%%%%%%%%%%%%%%%%%%%%%%
\paragraph{$\Delta=0$ Primary}
%%%%%%%%%%%%%%%%%%%%%%%%%%%%%%%%%%%%%%%%%%%%%%%%%%%%%%%%%%
The conformal spin-1 primary has the following angular components
\begin{equation}
    A_{\Delta,-1;z}=ru\hat{\bm}_z \varphi_{\Delta+1}\,, \quad A_{\Delta,-1;\bz}=r(2r+u)\hat{\bm}_\bz \varphi_{\Delta+1}\,,
\end{equation}
where we have defined
\be \label{mhat}
\hat{\bm}_z=\frac{\sqrt{2}(1+w\bz)^2}{(1+z\bz)^2}\,, \quad \hat{\bm}_\bz=-\frac{\sqrt{2}(z-w)^2}{(1+z\bz)^2} \,.
\ee
Near null infinity this yields the following expansion for $\Delta=0$
\begin{equation}\label{A0Scri}
\badat{2}
   A_{0,-1;z}&=-\lim_{\Delta \to 0} \frac{u^{1-\Delta}}{\Delta} D_z^2 \bar{y}^z+u \bar{W}_{z}+\O(r^{-1})\, \virg
   A_{0,-1;\bz}&=r \bar{y}_{\bz}+u \bar{W}_{\bz}+\O(r^{-1})\,.
\eadat
\end{equation}
Here
\begin{equation}
    \bar{y}^z=-\frac{1}{\sqrt{2}}(1+z\bz)\frac{z-w}{\bz-\bw} \quad \Rightarrow \quad D_z^2 \bar{y}^z=-\sqrt{2}\pi (1+z\bz)\delta^{(2)}(z-w)\,,
\end{equation}
is precisely the kernel for the soft charge that was identified in \cite{Campiglia:2019wxe}.
The subleading terms in these expansions include the following (time-independent) non-contact terms
\begin{equation}
    \bar W_{z}=\frac{1}{\sqrt{2}}\frac{(1+w\bz)^2}{(z-w)(\bz-\bw)(1+z\bz)}\,, \quad \bar W_{\bz}=\frac{1}{\sqrt{2}}\frac{(1+z\bw)(1+w\bz)}{(\bz-\bw)^2(1+z\bz)}\,, %\quad  \bar V=-\frac{1+w\bz}{\sqrt{2}(\bz-\bw)}\,,
\end{equation}
where we note $ \bar W_{\bz}=\frac{1}{2}D^2 \bar y_{\bz}$.  Meanwhile the temporal $A_u$ component starts at $\mathcal{O}(r^0)$ while the radial one is fixed by the gauge condition $A_{r}=-\frac{u}{r}A_u$.

\paragraph{$\Delta=2$ Shadow Primary}
%%%%%%%%%%%%%%%%%%%%%%%%%%%%%%%%%%%%%%%%%%%%%%%%%%%%%%%%%%
From~\eqref{adesc} the terms in this expansion can be written as descendants of the $\Delta=0$ case.  We will fix our notation, nonetheless. Th spin-1 shadow primary has the following angular components
\begin{equation}
    \tA_{\Delta,+1;z}=-ru^{-1}\hat m_z \tvarphi_{\Delta+1}\,, \quad \tA_{\Delta,+1;\bz}=-r(2r+u)^{-1}\hat m_\bz \tvarphi_{\Delta+1}\,,
\end{equation}
where $\hat{m}_z=(\hat{\bm}_\bz)^*$, $\hat{m}_\bz=(\hat{\bm}_z)^*$. For $\Delta=2$ we get the following expansions near null infinity
\begin{equation}\label{tA2zzbScri}
\badat{2}
   \tA_{2,+1;z}
 &=\lim_{\Delta \to 2}\frac{u^{\Delta-1}}{2-\Delta}  D_z^2 \widetilde{y}^z
   +u \tW_{z}+\O(r^{-1})\virg
  \tA_{2,+1;\bz}&= r \ty_{\bz}+u \tW_{\bz}+\O(r^{-1})\,.
\eadat
\end{equation}
Here
\begin{equation}
   \ty_{\bz}=\frac{1}{2!}\p_w^2 \bar{y}_{\bz} \,,\quad  \tW_{z}=\frac{1}{2!}\p_w^2 \bar{W}_{z}\,, \quad \tW_{\bz}=\frac{1}{2!}\p_w^2 \bar{W}_{\bz}\,, 
\end{equation}
matching the expected descendancy relations~\eqref{adesc}. Now, all tilde quantities in~\eqref{tA2zzbScri} are (time-independent) contact terms, except for 
\begin{equation}\label{tWz}
    \tW_z=-\frac{1}{\sqrt{2}} \frac{(1+z\bz)}{(z-w)^3(\bz-\bw)}\,,
\end{equation}
which we recognize as the shadow of the $\Delta\rightarrow0$ residue of the saddle point term in $A_{\Delta,-1;z}$. 

\subsubsection{Subleading Soft Charge in Gauge Theory}\label{subchargeA}
%%%%%%%%%%%%%%%%%%%%%%%%%%%%%%%%%%%%%%%%%%%%%%%%%%%%%%%%%%
Because the $\Delta=2$ shadow mode descends from the $\Delta=0$ primary, we need only evaluate \eqref{eq:IPspin1} for $A=\hat{A}$ and $A'=A_{0,-1}$ to see the form of both operators.  Our definition of the celestial operators~\eqref{eq:2Dop} is precisely so that this evaluates to $\mathcal{O}_{\Delta\to 0,J=-1}$. 

\paragraph{Leading saddle-point contribution}
Let us start by evaluating the leading contribution as $\Delta\rightarrow0$ of the term that arises from the saddle-point approximation.  Plugging the contact term of~\eqref{A0Scri} into~\eqref{eq:IPspin1} we find the contribution 
\begin{equation}\label{eq:Asaddlepoint}
\badat{2}
\mathcal{O}_{\Delta\rightarrow 0,-1}&\ni \lim_{\Delta \to 0} \frac{1}{\Delta}\int du d^2z \, u^{1-\Delta} D_z^2 \bar y^z\overset{\leftrightarrow}{\p}_u \hat{A}^{(0)}_\bz,\\
&\simeq \lim_{\Delta \to 0} \frac{\Delta-1}{\Delta}2\sqrt 2 \pi (1+w\bw)\int du  \, u^{-\Delta} \hat{A}^{(0)}_\bw\,.
\eadat
\end{equation}
Here we integrated by parts and renormalized by dropping a boundary that would vanish for $\mathrm{Re}\Delta>1$, which we recall is the regime where the saddle point dominates the large-$r$ expansion.  The ${\Delta}^{-1}$ in~\eqref{eq:Asaddlepoint} cancels the residue of a $\Gamma(\Delta)$ that appears when we relate this integral to the Mellin transformed mode operators
\be\label{mellinmode}
a_{\Delta,\pm s}(w,\bw)=\int_0^\infty d\omega \omega^{\Delta-1}a_\pm(\omega,w,\bw)\,,\quad a^\dagger_{\Delta,\mp s}(w,\bw)=\int_0^\infty d\omega \omega^{\Delta-1}a^\dagger_\pm(\omega,w,\bw)\,,
\ee
so that the subleading soft photon charge is thus determined by
\begin{equation}\label{eq:subsoftphoton}
    \mathcal{O}_{0,-1}\ni ie[a_{0,-1}+a^\dagger_{0,-1}]\,,
\end{equation}
matching our expectations from~\cite{Pasterski:2021dqe}.  Inserted in celestial amplitudes~\eqref{eq:subsoftphoton} gives rise to the subleading conformally soft photon theorem~\cite{Adamo:2019ipt,Guevara:2019ypd}.

\paragraph{$\mathcal O(r)$ symplectic product} 
We will now turn to non-contact terms and their contribution to the symplectic product $\Omega=\int_{\scri^+}\omega$.  Stripping off the volume element, we can write the presymplectic form $\omega=\omega^\rho (d^3 x)_\rho$ in terms of the following components
\begin{equation}\label{omegapre}
\begin{array}{ll}
\omega^{u(0)}&=-{A}_\bz^{(-1)}\hat A^{(0)}_z\,
,\\
\omega^{r(-1)}&= {A}_\bz^{(-1)}\p_u \hat A^{(0)}_z\,,\\
\omega^{r(0)}
&={A}_{\bz}^{(0)}\p_u \hat A_z^{(0)}- \hat A_z^{(0)}\p_u{A}_{\bz}^{(0)}+{A}_{z}^{(0)}\p_u \hat A_\bz^{(0)}- \hat A_\bz^{(0)}\p_u{A}_{z}^{(0)}\\
&~~~ +{A}_\bz^{(-1)}\hat A_z^{(0)}-2\gamma^{z\bz}D_z {A}_\bz^{(-1)}(D_z \hat A_\bz^{(0)}+D_\bz \hat A_z^{(0)})+ {A}_\bz^{(-1)} \partial_u \hat A_z^{(1)}.
\end{array}
\end{equation}
Here we have dropped the subscripts on the mode $A_{0,-1}$ to simplify our notation and used $A_u^{(0)}=D^C A_{C}^{(-1)}$, $u\p_u \hat A_u^{(1)}=-D^C \hat A_C^{(0)}$, as well as integration by parts. As expected, the presymplectic form diverges in $r$. 

Because the normal of null infinity is $n=\p_u-\frac{1}{2}\p_r$ the only combination we will care about is $-(\frac{1}{2}\omega^{u}+\omega^{r})$. Before doing any renormalization, we can observe that the $\mathcal{O}(r)$ mode is proportional to the soft charge for large gauge transformations, namely
\be
\int du [n_\mu \omega^{\mu}]^{(-1)}\propto \int du\p_u \hat{A}_{z}^{(0)},
\ee
up to an integration by parts in $u$. The fact that the over-leading terms are proportional to an operator with known Ward identities matches what was found in the charge for the over-leading gauge transformation proposed in~\cite{Campiglia:2016hvg}, which we will turn to below. One can renormalize away the $\omega^{r(-1)}$ divergent term, in a similar fashion as what we did in~\cite{Donnay:2020guq} for the subleading soft graviton. The renormalization procedure (see e.g. \cite{Iyer:1994ys,Compere:2018ylh,Compere:2020lrt, Fiorucci:2021pha}) amounts to shifting the presymplectic potential 
 $\Theta$, which is related to the symplectic form via 
\be
 \omega[\delta g, \delta'g]=\delta \Theta[\delta'g]-\delta'\Theta[\delta g]\,,
 \ee
by a spacetime co-dimension 2 form $\mathcal Y$, i.e. $\Theta \to \Theta + d\mathcal Y$.
Subtracting $d\mathcal Y$ induces the change $\omega+\delta\omega \equiv \omega_{ren}$, 
where $\delta \omega^u=-\p_r \mathcal Y^{ur}$ and $\delta \omega^r=-\p_u \mathcal Y^{ru}$. 

Considering here the symplectic form given in \eqref{omegapre} and taking
\be
\mathcal Y^{ru}=-\mathcal Y^{ur}=A_\bz \hat A_z+A_z \hat A_\bz\,,
\ee
we find that $\omega_{ren}^{u(0)}=\omega_{ren}^{r(-1)}=0$ leaving the finite part
\be\begin{array}{ll}
\omega_{ren}^{r(0)}&= \hat A_z^{(0)}[- 2\p_u{A}_{\bz}^{(0)}+D^2 {A}_\bz^{(-1)} ] +\hat A_\bz^{(0)}[-2 \p_u {A}_{z}^{(0)}+2D^\bz D_z {A}_\bz^{(-1)}] \,.\\
\end{array}\ee
Writing this in terms of the free data in~\eqref{A0Scri}, we find (dropping contact terms for consistency) 
\be\begin{array}{ll}
\omega_{ren}^{r(0)}&= \hat A_z^{(0)}[- 2\bar{W}_{\bz}+D^2 \bar{y}_{\bz}]
-2 \hat A_\bz^{(0)}\bar{W}_{z}\,.
\\
\end{array}\ee
Using that $D^2 \bar{y}_{\bz}=2\bar{W}_{\bz}$ and setting $ \Omega[\hat A,A_{0,-1}]=-\delta \Q_{0,-1}$ gives
\begin{equation}\label{qrenA0}
    \Q_{0,-1}=2\int d^2 z du  \bar{W}_{z} \hat A^{(0)}_\bz\,. 
\end{equation}
Judiciously adding another boundary term 
\begin{equation}
    Q^{ren}_{0,-1}=\Q_{0,-1}+\int du \p_u \Q'_{0,-1}\, \virg  \Q'_{0,-1}=-2\int d^2z  \bar{W}_{z} u \hat A^{(0)}_\bz\,,
\end{equation}
we finally get
\be
    Q^{ren}_{0,-1}=-2\int d^2z du \bar{W}_{z} u \p_u \hat A^{(0)}_\bz\,.
    \ee
This converges for radiative solutions with the standard $u$-falloffs~(see e.g. \cite{He:2014cra}). The descendancy relation~\eqref{adesc} lets us straightforwardly evaluate the $Q^{ren}_{2,1}$ using $\tW_{z}=\frac{1}{2!}\p_w^2 \bar{W}_{z}$.  We have
\begin{equation}
     Q^{ren}_{2,1}=-\int d^2z du \tW_{z} u \p_u \hat A^{(0)}_\bz\,,
\end{equation}
with $\tW_{z}$ given in~\eqref{tWz}.
We thus see that the renormalized inner product gives an operator that is determined by -- but not isomorphic to -- the subleading soft photon theorem.

%%%%%%%%%%%%%%%%%%%%%%%%%%%%%%%%%%%%%%%%%%%%%%%%%%%%%%%%%%
%%%%%%%%%%%%%%%%%%%%%%%%%%%%%%%%%%%%%%%%%%%%%%%%%%%%%%%%%%
\subsubsection{Over-leading Symmetry from Mixed Helicity Combinations 
}\label{rsymA}

We would now like to understand to what extent the Goldilocks modes discussed above can be related to the $\mathcal O(r)$ large gauge transformations of Campiglia-Laddha (CL) \cite{Campiglia:2016hvg}. There, it was shown that Low's subleading soft photon theorem could be related to the existence of residual gauge transformations parametrized by solutions to the wave equation $\Box \lambda=0$ that take the following form
\begin{equation}\label{lambda}
\lambda=r \mu(z,\bz) +u(1+D^2/2) \mu(z,\bz)+\cdots.
\end{equation}
Here $\mu(z,\bz)$ is unconstrained and plays the role of free data for over-leading gauge transformations. Charges at null infinity associated to~\eqref{lambda} are divergent but, by projecting out a leading soft photon contribution, they are rendered finite. Moreover, they are equivalent to the charges obtained in~\cite{Lysov:2014csa}, establishing the equivalence of its Ward identity with Low's subleading soft photon theorem.

A generic over-leading gauge transformation can be constructed from by smearing the charge that generates $\mu\propto\delta^{(2)}(z-w)$. Aside from a finite dimensional kernel, such a large gauge transformation will necessarily have both the $A_z$ and $A_\bz$ components be overleading.  In contrast, the large-$r$ expansions of conformal primaries of a given spin only have one or the other component growing in $r$, namely (for the shadow mode)
\be
\widetilde{A}_{2,+1;\bz}=-{\sqrt{2}\pi}r (1+z\bz)^{-1}\delta^{(2)}(z-w)+\cdots
\ee
with $\widetilde{A}_{2,-1;z}$ given by the same expression.
Upon taking linear superpositions of states in these two helicity sectors we can construct a mode that is within what we will call the `CL phase space': radiative solutions augmented by overleading pure gauge transformations~\eqref{lambda}. 

We do not expect the large gauge transformation to be a primary, but we can still organize things in terms of scaling dimension and spin.  A large gauge symmetry should be a spin-0 mode. Naturally we look for descendants of the $\Delta=2$ modes with this spin. Noting that 
\be\badat{3}
\p_\bw \widetilde{A}_{2,+1;\bz}&={\sqrt{2}\pi}r (1+z\bz)^{-1} \p_\bz\delta^{(2)}(z-w)+\cdots\\
&={\sqrt{2}\pi}r\p_\bz\left[ (1+z\bz)^{-1} \delta^{(2)}(z-w)\right]+{\sqrt{2}\pi}r\frac{z}{(1+z\bz)^2}\delta^{(2)}(z-w)+\cdots
\eadat\ee
 while $\p_\bw  \widetilde{A}_{2,+1;z}=\mathcal O(r^0) $ (and similar expressions for the opposite spin) we see that 
\be\label{eq:A2helicitycombo}\badat{3}
\left(\p_\bw+\frac{w}{1+w\bw}\right) \widetilde{A}_{2,+1;\mu}+\left(\p_w+\frac{\bw}{1+w\bw}\right)  \widetilde{A}_{2,-1;\mu}&=\nabla_\mu \left[{\sqrt{2}\pi}r(1+z\bz)^{-1}\delta^{(2)}(z-w)\right]\\
&~~~+~\mathrm{radiative}.
\eadat\ee
This linear combination of primaries and descendants is in the CL phase space. Moreover the descendant terms have the expected spin and matching scaling dimensions. 

We note that to get an isomorphism between the subleading soft theorem and the charge the authors of~\cite{Campiglia:2016hvg} needed to decompose the vector field $Y$ appearing in their parameterization of the charge~\cite{Lysov:2014csa} into two parts
\be\label{Ymutildemu}
Y^A=\frac{1}{2}D^A\mu-i\epsilon^{AB}D_B\tilde{\mu}\,.
\ee
The first contribution is the overleading gauge transformation we have focused on here, while the second contribution is the corresponding magnetic dual. Namely, it corresponds to an overleading large gauge transformation of the dual gauge field.  To compare to our expressions above, we would want to decompose $y^A$ in the same manner. Phrased in this way, we see that the linear combinations in~\eqref{eq:A2helicitycombo} combine to an overleading gauge field of the form $y^A=D^A\mu$.  Meanwhile we can always locally write $y^A$ in the form~\eqref{Ymutildemu} for some complexified parameters $\mu,\tilde{\mu}$. We see that the charges constructed from the single helicity operators generate a combination of overleading large gauge and dual large gauge transformations.

%%%%%%%%%%%%%%%%%%%%%%%%%%%%%%%%%%%%%%%%%%%%%%%%%%%%%%%%%%
%%%%%%%%%%%%%%%%%%%%%%%%%%%%%%%%%%%%%%%%%%%%%%%%%%%%%%%%%%
\subsubsection{Memory}
%%%%%%%%%%%%%%%%%%%%%%%%%%%%%%%%%%%%%%%%%%%%%%%%%%%%%%%%%%
The Goldilocks modes discussed above are symplectically paired with the memory modes, which arise from the spin-one $\Delta=2$ conformal primary wavefunction or its $\Delta=0$ shadow
\begin{equation}
A_{2,+1;\mu}=m_\mu \varphi_2 \virg \tA_{0,-1;\mu}=-\frac{1}{X^2}\bar m_\mu\,. 
\end{equation}
Note that these modes transform respectively as primaries of weights $(h,\bar h)=(\frac{3}{2},\frac{1}{2})$ and $(\frac{1}{2},-\frac{1}{2})$, the latter matching the weights of the subleading soft photon current $J_z^{(1)}$ that was identified in \cite{Himwich:2019dug}. We define the (subleading) memory mode $ A_{2}^{\rm M}$ analogously to~\eqref{eq:Nreg}, so that near $\mathscr I^+$
\begin{equation}
  A^{\rm M}_{2,+1;z}=\O(r^{-1})\,, \quad   A^{\rm M}_{2,+1;\bz}=-\delta(u)(1+z\bz)^{-1} \pi \delta^{(2)}(z-w)+\O(r^{-1})\,.
\end{equation}
 While the Goldilocks wavefunctions are overleading, these memory modes are radiative.  Furthermore, the corrections to the saddle point approximation are sub-radiative.
We would like to compute the pairing of the regulated $\Delta=2$ memory mode with the $\Delta=0$ primary.  The saddle point contribution gives
  \begin{equation}
\badat{3}
 \Omega(A^{\rm M}_{2,+1},A_{0,-1})|_{\rm pole}&=\int d^2z du \,A_{0,-1;z}|_{\rm pole}\overset{\leftrightarrow}{\p}_u A^{\rm M}_{2,+1;\bz}\\
    &=-2\sqrt{2}\pi^2 \delta^{(2)}(w-w')\lim_{\Delta \to 0} \frac{1-\Delta}{\Delta} \int du  u^{-\Delta} \delta(u)\\
     &=-2\sqrt{2}\pi^2 \lim_{\Delta \to 0} \frac{1}{\Delta}\delta^{(2)}(w-w').
\eadat
\end{equation}
While this pairing is divergent, the pole in $\Delta$ is canceled by the normalization of the shadow transform.  Indeed, as discussed above, the non-contact term that survives in the renormalized inner product for $\tDelta=2$ is the shadow of the saddle point contribution from the $\Delta=0$ mode. Namely,
    \begin{equation}\label{2pair}
\badat{3}
   \Omega(A^{\rm M}_{2,+1},\tA_{2,+1})|_{\rm sh~\circ~pole} &=-\sqrt{2}\pi \frac{1}{(w-w')^3(\bw-\bw')}.
\eadat
\end{equation}
Meanwhile this expression descends from the renormalized inner product~\eqref{qrenA0} with the $\Delta=0$ mode
    \begin{equation}
\badat{3}
      \Omega(A^{\rm M}_{2,+1},A_{0,-1})|_{\rm ren}
    &=-\sqrt{2}\pi \frac{1}{(w-w')(\bw-\bw')}.
\eadat
\end{equation}
The pairing~\eqref{2pair} has the expected transformation properties of a conformal 2-point function of primaries with the specified weights and hints at an off diagonal level for the subleading soft photon current.

%%%%%%%%%%%%%%%%%%%%%%%%%%%%%%%%%%%%%%%%%%%%%%%%%%%%%%%%%%
\subsection{Sub-subleading Soft Graviton}
%%%%%%%%%%%%%%%%%%%%%%%%%%%%%%%%%%%%%%%%%%%%%%%%%%%%%%%%%%

At tree level, the gravitational soft theorem extends to sub-subleading order~\cite{Cachazo:2014fwa}. In the conformal basis, this factorization theorem should correspond to a pole at $\Delta=-1$~\cite{Adamo:2019ipt}. Based on the Goldstone mode accounting in~\cite{Pasterski:2017kqt}, we do not expect to find a sub-subleading Goldstone mode for gravity amongst our conformal primary wavefunctions.  However, we argue here that the sub-subleading conformally soft graviton can be associated to a conformal primary wavefunction $\Phi_{\Delta,J=\pm2}$ with conformal dimension $\Delta=-1$:
\begin{equation}
 h_{-1,+2;\mu\nu}=m_{\mu}m_{\nu}\varphi_{-1} \,,\quad  h_{-1,-2;\mu\nu}=\bm_{\mu}\bm_{\nu}\varphi_{-1} \,,
\end{equation}
or, equivalently, its shadow with conformal dimension $\Delta=3$:
\begin{equation}
 \th_{3,+2;\mu\nu}=(-X^2)^2m_{\mu}m_{\nu}\varphi_{3} \,,\quad  \th_{3,-2;\mu\nu}=(-X^2)^2\bm_{\mu}\bm_{\nu}\varphi_{3} \,.
\end{equation}
One can indeed check that neither of the two primaries above is a pure diffeomorphism.

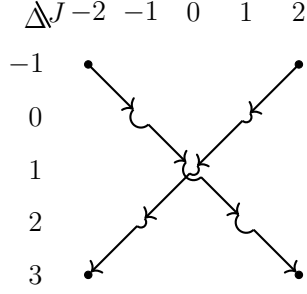
\begin{figure}[htb]
    \centering
\begin{tikzpicture}[scale=.7,every node/.style={scale=0.9}]
\draw[thick] (0+.1414/2,1+.1414/2) arc (45:-135:.1);
\draw[thick] (1+.1414/2,2+.1414/2) arc (45:-135:.1);
\draw[thick] (-1+.1414/2,0+.1414/2) arc (45:-135:.1);
\filldraw[black] (-2,3) circle (2pt) ;
\filldraw[black] (2,3) circle (2pt) ;
\filldraw[black] (2,-1) circle (2pt) ;
\filldraw[black] (-2,-1) circle (2pt) ;
\node at (-3,-1) {$3$};
\node at (-3,0) {$2$};
\node at (-3,1) {$1$};
\node at (-3,2) {$0$};
\node at (-3,3) {$-1~~$};
\node at (-2.6,4) {$J$};
\node at (-3,3.87) {$\tiny{\Delta}$};
\node at (-2,4) {$-2$};
\draw[thick] (-3+.01,4+.2) -- (-3+.28,4-.3);
\node at (-1,4) {$-1$};
\node at (0,4) {$0$};
\node at (1,4) {$1$};
\node at (2,4) {$2$};
\draw[->,thick] (1-.07,2-.07) --  (0+.07,1+.07);
\draw[->,thick]  (0-.07,1-.07) -- (-1+.07,.07);
\draw[->,thick]  (2,3) -- (1+.07,2.07);
\draw[->,thick]  (-1-.07,0-.07) -- (-2+.05,-1+.05);
\draw[thick] (0-.1414,1+.1414) arc (135:315:.2);
\draw[thick] (1-.1414,0+.1414) arc (135:315:.2);
\draw[thick] (-1-.1414,2+.1414) arc (135:315:.2);
\draw[->,thick] (-1+.1414,2-.1414) --  (0-.1414,1+.1414);
\draw[->,thick] (0+.1414,1-.1414) -- (1-.1414,.1414);
\draw[->,thick] (1+.1414,0-.1414) -- (2-.05,-1+.05);
\draw[->,thick] (-2,3) --  (-1-.1414,2+.1414);
\end{tikzpicture}
    \caption{Celestial `Diamonds' for the Sub-subleading Soft Graviton.}
    \label{fig:subsubleadinggraviton}
\end{figure}

In the following we will focus on the negative helicity wavefunction $h_{-1,-2}$. The positive spin wavefunction $\th_{3,+2}$ is related to it by a shadow transform. The descendancy relation is
\begin{equation}\label{hdesc}
\frac{1}{4!}\p_w^4 h_{-1,-2}=\th_{3,+2} \virg  \frac{1}{4!}\p_\bw^4 h_{-1,+2}=\th_{3,-2},
\end{equation}
as illustrated in figure~\ref{fig:subsubleadinggraviton}.

%%%%%%%%%%%%%%%%%%%%%%%%%%%%%%%%%%%%%%%%%%%%%%%%%%%%%%%%%%
\subsubsection{Goldstone-like Primaries}\label{subsec:Gold}
Let us now keep track of the leading expansion for the $\Delta=-1$ primary. As in the spin-one case discussed above, the contact terms arise from the saddle point approximation.
%%%%%%%%%%%%%%%%%%%%%%%%%%%%%%%%%%%%%%%%%%%
\paragraph{$\Delta=-1$ Conformal Primary}
The conformal spin-2 primary has the following angular components
\begin{equation}
    h_{\Delta,-2;zz}=(ru\hat{\bm}_z)^2 \varphi_{\Delta+2}\,, \quad h_{\Delta,-2;\bz \bz}=(r(2r+u)\hat{\bm}_\bz)^2 \varphi_{\Delta+2}\,,
\end{equation}
with the polarization dependent parts defined in \eqref{mhat}. Near null infinity, this yields for $\Delta=-1$
 \begin{equation}\label{hn1scri}
 \badat{2}
   h_{-1,-2;zz}&=r\lim_{\Delta \to -1}\frac{u^{1-\Delta}}{6(\Delta+1)}D_z^4 \bar{Y}^{zz}+r u^2  \bar{W}_{zz}+\O(1)\,, \\
   h_{-1,-2;\bz \bz}&=r^3\bar{Y}_{\bz \bz}+r^2 u \bar{W}_{\bz \bz}+\O(r)\,.
 \eadat
 \end{equation}
 Here
  \begin{equation}
 \bar{Y}^{zz}=(1+z \bz) \frac{(z-w)^3}{(\bz-\bw)}\quad \Rightarrow \quad D_z^4 \bar{Y}^{zz}=12\pi (1+z\bz) \delta^{(2)}(z-w)\,,
\end{equation}
 is precisely the kernel of the soft charge that was identified in \cite{Campiglia:2016efb}. The subleading terms in these expansions include the following (time-independent) non-contact terms 
 \begin{equation}
 \bar{W}_{zz}=\frac{(1+w\bz)^4}{|z-w|^2 (1+z\bz)^3}, \quad \bar{W}_{\bz\bz}=\frac{2(z-w)^2(1+2\bw z+2w\bz-z\bz+w\bw(z\bz-1))}{(\bz-\bw)^2 (1+z\bz)^3}\,.
 \end{equation}

%%%%%%%%%%%%%%%%
\paragraph{$\Delta=3$ Shadow Primary}
%%%%%%%%%%%%%%%%%%%%%%%%%%%%%%%%%%%%%%%%%%%%%%%%%%%%%%%%%%
The  spin-2 shadow primary has the following angular components
\begin{equation}\label{th3}
    \th_{\Delta,+2;zz}=(ru^{-1}\hat m_z)^2 \tvarphi_{\Delta+2}\,, \quad \th_{\Delta,+2;\bz \bz}=(r(2r+u)^{-1}\hat m_\bz)^2 \tvarphi_{\Delta+2}\,.
\end{equation}
For $\Delta=3$ we get, near null infinity, 
 \begin{equation}
 \badat{2}
  \th_{3,+2;zz}&=r\lim_{\Delta \to 3}\frac{u^{\Delta-1}}{6(3-\Delta)}D_z^4 \tY^{zz}+r u^2  \tW_{zz}+\O(1),~~~
   \th_{3,+2;\bz \bz}&=r^3\tY_{\bz \bz}+r^2 u \tW_{\bz \bz}+\O(r)\,.
 \eadat
 \end{equation}
 Here
 \begin{equation}\label{Wdesc}
   \tY_{\bz\bz}=\frac{1}{4!}\p_w^4 \bar{Y}_{\bz\bz} \,, \quad \tW_{zz}=\frac{1}{4!}\p_w^4 \bar{W}_{zz}\,, \quad \tW_{\bz\bz}=\frac{1}{4!}\p_w^4 \bar{W}_{\bz\bz}\,, 
\end{equation}
matching the descendancy relations~\eqref{hdesc}. The expansion of the $\Delta=3$ spin-two shadow primary only gives contact terms except for 
 \begin{equation}
\tW_{zz}=\frac{(1+z\bz)}{(z-w)^5(\bz-\bw)}.  
\end{equation}
We recognize this term as the shadow of the $\Delta\rightarrow-1$ residue of the saddle point term in $h_{\Delta,-2;zz}$. 

%%%%%%%%%%%%%%%%%%%%%%%%%%%%%%%%%%%%%%%%%%%%%%%%%%%%%%%%%%
\subsubsection{Sub-subleading Soft Charge in Gravity}
Because the $\Delta=3$ shadow mode descends from the $\Delta=-1$ primary, we need only evaluate \eqref{eq:IPspin2} for $ h=\hat{h}$ and $h'=h_{-1,-2}$ to see the form of both operators.  Our definition of the celestial operators~\eqref{eq:2Dop} is precisely so that this evaluates to $\mathcal{O}_{\Delta\to -1,J=-2}$.

%%%%%%%%%%%%%%%%%%%%%%%%%%%%%%%%%%%%%%%%%%%%%%%%
\paragraph{Leading saddle-point contribution}
We will start by evaluating the leading contribution as $\Delta\rightarrow-1$ of the term that arises from the saddle-point approximation.  Plugging the contact term of~\eqref{hn1scri} into~\eqref{eq:IPspin2} we find the contribution 
\begin{equation}\label{eq:hsaddlepoint}
\badat{2}
\mathcal{O}_{\Delta\rightarrow -1,-2}&\ni \lim_{\Delta \to -1}\frac{1}{\Delta+1}\int du d^2 z \frac{1}{6} D_z^4 Y^{zz}_{\bw \bw}u^{1-\Delta} \overset{\leftrightarrow}{\p}_u \hat{h}^{(-1)}_{\bz\bz}{\gamma^{z\bz}}\,,\\
&\simeq 4\pi (1+w\bw) \lim_{\Delta \to -1} \frac{\Delta-1}{\Delta+1}\int du  \, u^{-\Delta} \p_u \hat{h}^{(-1)}_{\bw\bw}{\gamma^{w\bw}}\,,
\eadat
\end{equation}
where, as in the spin-one case, we dropped a boundary term that would vanish for $\mathrm{Re}\Delta>1$. We thus find the sub-subleading soft graviton charge
\begin{equation}\label{eq:subsubsoftgraviton}
    \O_{-1,-2}\ni 4i \kappa [a_{-1,-2}+a^\dagger_{-1,-2}]\,.
\end{equation}
Inserted in celestial amplitudes~\eqref{eq:subsubsoftgraviton} gives rise to the sub-subleading conformally soft graviton theorem~\cite{Guevara:2019ypd}.

\paragraph{$\mathcal O(r^2)$ symplectic product}
The non-contact terms give the following contributions to the symplectic form
\begin{equation}
\begin{array}{ll}
\omega^{u(-1)}&=2\gamma^{z\bz}{h}_{\bz\bz}^{(-3)}\hat h^{(-1)}_{zz}\,
,\\
\omega^{u(0)}&=\gamma^{z\bz}{h}_{\bz\bz}^{(-2)}\hat h^{(-1)}_{zz}+3\gamma^{z\bz}h^{(-3)}_{\bz\bz}\hat h_{zz}^{(0)}+2\gamma^{z\bz}h_{\bz\bz}^{(-3)}D_z \hat h_{rz}^{(1)}\,
,\\
\omega^{r(-2)}&= -\gamma^{z\bz}{h}_{\bz\bz}^{(-3)}\overset{\leftrightarrow}{\p}_u \hat h^{(-1)}_{zz}
,\\
\omega^{r(-1)}&=-2\gamma^{z\bz}{h}_{\bz\bz}^{(-3)} \hat h^{(-1)}_{zz}-\gamma^{z\bz}{h}_{\bz\bz}^{(-2)}\overset{\leftrightarrow}{\p}_u \hat h^{(-1)}_{zz}
-\gamma^{z\bz}h_{\bz\bz}^{(-3)}\overset{\leftrightarrow}{\p}_u\hat h_{zz}^{(0)}+2\gamma^{z\bz}h_{\bz\bz}^{(-3)}D_z \hat h_{uz}^{(0)},\\
\end{array}
\end{equation}
where for conciseness we have omitted the lengthy expression for $\omega^{r(0)}$. As expected, the presymplectic form now diverges in $r^2$. We see that the $\mathcal{O}(r^2)$ mode is proportional to the soft charge for supertranslations and the $\mathcal{O}(r)$ contribution is proportional to the soft charge for Diff$(S^2)$ superrotations namely
\be
\int du [n_\mu \omega^{\mu}]^{(-2)}\propto \int du\p_u \hat{h}_{zz}^{(-1)},~~\int du [n_\mu \omega^{\mu}]^{(-1)}\propto \int du u\p_u \hat{h}_{zz}^{(-1)},
\ee
up to an integration by parts in $u$.  Here we have used the equations of motion in appendix~\ref{larger} to rewrite the subleading components of the metric in terms of the radiative free data. Specifically
\be\label{eom2}
\begin{aligned}
[\square \hat{h}_{\bz\bz}]^{(1)} &= 2 \partial_u\hat{h}_{\bz\bz}^{(0)} + [D^2  - 2]\hat{h}_{\bz\bz}^{(-1)}- 4 D_\bz \hat{h}_{u\bz}^{(0)}\,,
\end{aligned}
\ee
and
\begin{equation}
\begin{aligned}%\label{eq:hg1}
\left[\nabla^{\mu}\hat{h}_{\mu z}\right]^{(1)} &=  (u\partial_u  -1) \hat{h}_{uz}^{(0)} + D^z\hat{h}_{zz}^{(-1)}\,,\\
\end{aligned}
\end{equation}
from the linearized Einstein equations and harmonic gauge conditions.  The fact that the overleading terms are proportional to operators with known Ward identities matches what was found in the charge for the overleading diffeomorphism proposed in~\cite{Campiglia:2016efb}, which we will turn to later.

%%%%%%%%%%%%%%%%%%%%%%%%%%%%%%%%%%%%%%%%%%%%%%%%%%%%%%%%%%
\subsubsection{Over-Leading Symmetry from Mixed Helicity Primaries}
As for the subleading spin-1 example~\cite{Campiglia:2016hvg}, there are overleading diffeomorphisms that can be associated to the sub-subleading soft graviton theorem~\cite{Campiglia:2016efb}. They are parametrized by a divergence free vector field
\be\badat{3}\label{xiV}
\xi^A=r X^A(z,\bz)+\dots,~~~D_AX^A=0\,,
\eadat\ee
where the other components are determined by demanding consistency with the harmonic gauge condition $\Box \xi^\mu=0$.   Allowing this residual gauge transformation is not compatible with our conformal primary gauge fixing, explaining why we do not find associated pure gauge primaries at $\Delta=-1$.

We see from the form of $\nabla_{(\mu}\xi_{\nu)}$ that such a large gauge transformation will have an overleading $h_{AB}$ at order $r^3$
\be
\mathcal{L}_{\xi}\eta_{AB}=r^3(\nabla_A X_B+\nabla_B X_A)\,.
\ee
If $X^A$ is a real vector field then, aside from a finite dimensional kernel, both the $h_{zz}$ and $h_{\bz\bz}$ should be overleading at this order. The large-$r$ expansions of our conformal primaries with fixed spin only have one or the other component overleading in $r$. Namely for the shadow mode we have 
\be
\tilde{h}_{3,+2;\bz\bz}=
-2\pi r^3 (1+z\bz)^{-3} \delta^{(2)}(z-w)+\dots
\ee
with $\tilde{h}_{3,-2;zz}$ given by the same expression.
Let us now see if, like the spin-1 case,  we can construct a linear combination of these modes and their descendants that is within the `CL phase space': radiative solutions augmented by overleading pure gauge transformations~\eqref{xiV}. 

We do not expect the large gauge transformation to be a primary, but we can still organize things in terms of scaling dimension and spin. A large diffeomorphism symmetry can in principle be a spin-1 mode.  So we naturally want to start by looking at descendants of the $\Delta=3$ modes with this spin.  Noting that
\be\badat{3}
\p_\bw \tilde{h}_{3,+2;\bz\bz}&=2\pi r^3 (1+z\bz)^{-3} \p_\bz\delta^{(2)}(z-w)+\cdots\\
&=2\pi r^3\p_\bz\left[ (1+z\bz)^{-3} \delta^{(2)}(z-w)\right]+2\pi r^3\frac{3z}{(1+z\bz)^4}\delta^{(2)}(z-w)+\cdots
\eadat\ee
while $\partial_\bw \th_{3,+2;zz}=\mathcal O(r^2)$ 
(and similar expressions for the opposite spin) we see that 
\be
\left(\p_\bw+\frac{w}{1+w\bw}\right) \tilde{h}_{3,+2;\bz\bz}=D_\bz \xi_\bz+\cdots
\ee
for $\xi_\bz\propto r v_\bz\delta^{(2)}(z-w)$ (keeping track of the index structure), and similarly for the complex conjugate.  In order to be trace-free at this order, we need a linear combination for which
\be
\p_z X_\bz+\p_\bz X_z=0\,.
\ee
We can make our lives easier by finding a better parameterization for $X_A$ that automatically satisfies the divergence free condition of~\cite{Campiglia:2016efb}.  Locally we can write
\be
X_\bz=i\p_\bz X\,.
\ee
Iterating our steps above we find
\be\label{eq:h2helicitycombo}
i\left(\p_\bw+\frac{3w}{1+w\bw}\right)\left(\p_\bw+\frac{w}{1+w\bw}\right) \tilde{h}_{3,+2;\mu\nu}+c.c. =2\nabla_{(\mu}\xi'_{\nu)}+\cdots 
\ee
for 
\be\xi'_A=r^3i\p_A X,~~~~X=-2{\pi}(1+z\bz)^{-3} \delta^{(2)}(z-w).
\ee
This combination of primaries and descendants is in the CL phase space. Because this overleading behavior is 2 orders above radiative order, one needs to verify that the order $r^2$ term of this combination is also pure gauge, which is indeed the case.\footnote{Using the shadow relation helps with this because the contact terms for $\th^{(-3)}_{3,+2}$ translate to non-contact terms of $h^{(-3)}_{-1,-2}$. One can see from~\eqref{hn1scri} that the only overleading free data is in $h_{-1,-2;zz}^{(-3)}$.  Namely there is no extra $u$-integration constant introduced at the $r^2$ order. Once we identify a linear combination of the free data that is pure gauge, it will propagate into the bulk as such.  Meanwhile, the conformal primaries have additional radiative order free data hence, like the spin-1 case, the ellipsis includes radiative order terms.} We have thus matched a linear combination of the radiative modes to the set of overleading divergence-free diffeomorphisms that was proposed to correspond to the sub-subleading soft graviton theorem by~\cite{Campiglia:2016efb}.

We note that to get an isomorphism between the sub-subleading soft theorem and the charge the authors of~\cite{Campiglia:2016efb}  needed to decompose the tensor $Y_{AB}$ appearing the the soft charge as follows
\be\label{YAB}
Y_{AB}=(D_A X_B+i\epsilon_{BC} D_A X'^C)^{\rm STF},~~~D_A X^A=D_A X'^A=0
\ee
where STF stands for symmetric trace-free.  To compare to our expressions above, we should identify this $Y_{AB}$ with the one in our expansions of the $\Delta=0,3$ conformal primary metrics. Similar to the electromagnetic case, the additional term in~\eqref{YAB} is identified with a magnetic dual charge. We see that the charges constructed from the single helicity operators generate a combination of overleading and dual transformations.

%%%%%%%%%%%%%%%%%%%%%%%%%%%%%%%%%%%%%%%%%%%%%%%%%%%%%%%%%%
%%%%%%%%%%%%%%%%%%%%%%%%%%%%%%%%%%%%%%%%%%%%%%%%%%%%%%%%%%
\subsubsection{Memory}
The Goldilocks modes discussed above are symplectically paired with the sub-subleading memory modes, which arise from the  spin-two $\Delta=3$ conformal primary wavefunction or its $\Delta=-1$ shadow
\begin{equation}
h_{3,+2;\mu \nu}=m_\mu m_\nu \varphi_3 \virg \th_{-1,-2;\mu \nu}=\frac{1}{X^4}\bar m_\mu \bar m_\nu \varphi_{-1}.
\end{equation}
Note that these modes transform respectively as primaries of weights $(h,\bar h)=(\frac{5}{2},\frac{1}{2})$ and $(-\frac{3}{2},\frac{1}{2})$. We again define the regulated mode $ h^{\rm M}_3$ analogously to~\eqref{eq:Nreg}, yielding
\begin{equation}
   h^{\rm M}_{3,+2;zz}=\O(r^{-2})\,, % vs -1
   \quad   h^{\rm M}_{3,+2;\bz \bz}=\frac{r}{8}\p_u\delta(u)(1+z\bz) \pi \delta^{(2)}(z-w)+\O(r^{0})\,.
\end{equation}
We would like to compute the pairing of the regulated $\Delta=3$ memory mode with the $\Delta=-1$ primary. The  saddle point contribution gives
    \begin{equation}
\badat{3}
    \Omega(h^{\rm M}_{3,+2},h_{-1,-2})|_{\rm pole}&=\int d^2z du\, h_{-1,-2;zz}|_{\rm pole}\overset{\leftrightarrow}{\p}_u h^{\rm M}_{3,+2;\bz\bz}\\
    &=-\frac{\pi^2}{2} \delta^{(2)}(w-w')\lim_{\Delta \to -1} \frac{\Delta(\Delta-1)}{\Delta+1} \int du  u^{-1-\Delta} \delta(u)\\
     &=-{\pi^2}\lim_{\Delta \to -1} \frac{1}{\Delta+1}\delta^{(2)}(w-w').
\eadat
\end{equation}
While this pairing is divergent, the pole in $\Delta$ is canceled by the normalization of the shadow transform.  Indeed, as discussed above, the non-contact term that survives in the renormalized inner product for $\tDelta=3$ is the shadow of the saddle point contribution from the $\Delta=0$ mode. Namely,
    \begin{equation}\label{2pairh}
\badat{3}
    \Omega(h^{\rm M}_{3,+2},\th_{3,+2})|_{\rm sh~\circ~pole} &=-\frac{\pi}{2} \frac{1}{(w-w')^5(\bw-\bw')}.
\eadat
\end{equation}
Meanwhile we expect this expression to descend from the renormalized inner product $ \Omega(h^{\rm M}_{3,+2},h_{-1,-2})|_{\rm ren}$ as in the spin-1 case. The pairing~\eqref{2pairh} has the expected transformation properties of a conformal 2-point function of primaries with the specified weights and hints at an off diagonal level for the sub-subleading soft graviton current.

\vspace{1em}

We have seen that we can reconcile the renormalization of non-contact pieces of the the overleading modes with the charges we get from the saddle point approximation for the the $\Delta=1-s$ modes.  In particular, we saw for the electromagnetic case that the shadow transformation takes us from the pole term to the renormalized term. The residue  saddle point term also gives a well-defined symplectic pairing with the memory modes, and we see that it is sufficient for extracting the interesting physics.  If we kept the contact terms to $\mathcal{O}((\Delta-1-s)^0)$ we would find an additional $\log u$ term in the integral kernels~\eqref{eq:Asaddlepoint} and \eqref{eq:hsaddlepoint}, for which we don't have an associated Ward identity. This corroborates our discussion in section~\ref{sec:softADelta} of the expected analytic behavior in $\Delta$, and we will take the conformally soft limit last. We will now turn to the task of generalizing our story to the full $w_{1+\infty}$ tower, where this prescription will come in handy.

%%%%%%%%%%%%%%%%%%%%%%%%%%%%%%%%%%%%%%%%%%%%%%%%%%%%%%%%%%

%\pagebreak

%%%%%%%%%%%%%%%%%%%%%%%%%%%%%%%%%%%%%%%%%%%%%%%%%%%%%
\section{Memories, Multipoles, and a Conformally Soft Tower}
\label{sec:Discussion}
%%%%%%%%%%%%%%%%%%%%%%%%%%%%%%%%%%%%%%%%%%%%%%%%%%%%%
We will close with a summary of what we have learned about the conformally soft phase space, and see how we can apply these lessons to the remaining tower of conformally soft modes. One of the main takeaways throughout our discussions is to take the conformally soft limit last. This prescription has multiple advantages. First, it lets us make sense of analytically continuing the Mellin transform of a radiative amplitude to conformal dimensions where the wavefunctions would be outside the standard radiative phase space.  Second,  this prescription avoids convergence issues for the $u$-integrals needed to define the corresponding charges. Finally, it extends the picture of overlapping memory and Goldstone-like celestial diamonds to the full conformally soft tower, to which we now turn. Our examinations of the Goldilocks modes generalize as follows.
\begin{itemize}
    \item The $\Delta=1-s-n$ conformal primaries are the Goldstone-like modes that should be used in~\eqref{eq:2Dopintro} to select celestial currents with primary descendants. These are the operators giving conformally soft theorems in~\eqref{eq:celestcurrent}.
    \item The $\Delta=1+s+n$ conformal primaries corresponding to memory modes. These have non-trivial pairings with the Goldstone-like modes of weight $2-\Delta$.
%    \item 
\end{itemize}
As we show in appendix~\ref{app:diamondLT} the (shadowed Mellin, respectively Mellin) conformal primaries at $\Delta=1-s-n+\epsilon$ descend to their own shadows (Mellin, respectively shadowed Mellin)  at $\Delta=1+s+n$ if we keep terms of $\mathcal{O}(\epsilon)$
\be\label{OSHdesc}
\p_w^k\p_\bw^{\bar{k}} \mathcal{O}_{\Delta+\epsilon,J}\propto \epsilon \widetilde{\mathcal O}_{2-\Delta,-J}
\ee
where $\Delta,J$ are related to $k,\bar{k}$ via~\eqref{condition_diamond_hhb}. Because these conformally soft theorems correspond to poles in the Mellin amplitude, these descendants can give finite contributions as $\epsilon\rightarrow0$, and we indeed see that we should not be too quick in taking conformally soft limits.

Let us now tie this together with the conversations about how to interpret these soft theorems that started our story. We will address three aspects of this tower in turn: the role of over-leading gauge transformations for the Goldstone-like modes, the residual gauge dependence of phrasing the memory modes as vacuum transitions, and a multipole interpretation for what the $w_{1+\infty}$ generators couple to.

\paragraph{The Goldstone-like Tower}
We will start with the Goldstone-like modes first.  We see that from the perspective of the bulk wavefunctions, the angular components of the $\Delta=1-s-n$ modes grow like 
\be
A_{-n,-1;\bz}\propto r^{n+1},~~~h_{-1-n,-2;\bz\bz}\propto r^{n+3}
\ee
the same goes for the shadow modes $\tA_{n+2,+1;\bz}$ and $\th_{n+3,+2;\bz\bz}$.  However, the form of the conformal primary wavefunctions~\eqref{CPWs} and~\eqref{SHCPWs} give the Mellin and shadowed Mellin modes rather different functional dependence on the celestial sphere.  The $\Delta=1-s-n$ modes involve a positive power of $(-q\cdot X)$ and therefore vanish when $q$ and $X$ are collinear.  Meanwhile the $\Delta=1+s+n$ modes involve a negative power and will be singular in the same limit.  From the expansion~\eqref{SHphicontact} we see that there is a contact term 
\be
\tA_{n+2,+1;\bz\bz}\sim r^{n+1}(1+z\bz)^{-n-1}\delta^{(2)}(z-w)\,,~~~\th_{n+3,+2;\bz\bz}\sim r^{n+3}(1+z\bz)^{-n-3}\delta^{(2)}(z-w)\,.
\ee
It is perhaps surprising that we can reconcile these different behaviors and write this shadow as a descendant of the Mellin primary by continuing the dimension slightly away from these integer values as in~\eqref{OSHdesc}.  However, with this result, we are at the same starting point as we were when we set out to identify an overleading gauge transformation from mixed helicity combinations of the Goldilocks modes, and the constructions in~\eqref{eq:A2helicitycombo} and~\eqref{eq:h2helicitycombo} generalize. 

Before proceeding to the memory modes, let us point out one more way for us to quickly see the relevance of overleading gauge transformations to the Ward identities for these generalized currents.  As in~\cite{Donnay:2020guq}  we can write the conformal primary wavefunctions in terms of a Mellin representative and a pure gauge part
\begin{equation}\label{ADelta}
 A_{\Delta,J}
=\sqrt 2  \frac{\Delta-1}{\Delta}\epsilon_{J} \varphi_{\Delta}+\nabla\alpha_{\Delta,J}\,,~~~\alpha_{\Delta,J}=\frac{\sqrt{2}\epsilon_J\cdot X}{\Delta (-q\cdot X)^\Delta}\,,
\end{equation}
for spin-1, and
\begin{equation}\label{hDelta}
 h_{\Delta,J}
 =\frac{\Delta-1}{\Delta+1} \epsilon_{J}\epsilon_{J}\varphi_{\Delta}
 +\mathcal{L}_{\zeta_{\Delta,J}}\eta\,, ~~~ 
\zeta_{\Delta,J}=\frac{1}{2(\Delta+1)}\left(\frac{\epsilon_{J} (\epsilon_{J}\cdot X)}{(-q\cdot X)^\Delta}+\frac{1}{2}\frac{q (\epsilon_{J}\cdot X)^2}{(-q\cdot X)^{\Delta+1}}\right)\,,
\end{equation}
for spin-2. For $\Delta=1-s-n$ we see that the angular components of these gauge parameters indeed scale with $r$ in the way we expect for the tower of overleading diffeomorphisms.

\paragraph{The Memory Tower}
Let us now turn to the memory modes.  Starting from the conformal primaries with $\Delta=1+s+n$, we can construct an analog of the M modes in~\eqref{eq:Nreg}.  The leading behavior for these Mellin-representatives comes from the saddle point contribution and is of radiative order.  From~\eqref{eq:varphicontact} and the large-$r$ expansions of the tetrad, we see that the radiative components scale like $u^{1-\Delta}$.
Writing
\be
(u+i\epsilon)^{-m-1}-(u-i\epsilon)^{-m-1}=\frac{(-1)^{m}}{(m-1)!}\p_u^{m} (\frac{1}{u+i\epsilon}-\frac{1}{u-i\epsilon})\propto \p_u^{m}\delta(u)
\ee
for $m\in\mathbb{Z}_{>0}$, we find that the $u$-profile matches the expectation from~\eqref{uprof} coming from the $\omega^{\Delta-2}$ which is the soft theorem selected by $\mathcal{O}_{1-s-n,\pm s}$. Meanwhile the non-contact terms scale like
\be
A_{n+2,-1;\bz}\propto r^{-n-1},~~~h_{n+3,-2;\bz\bz}\propto r^{-n-1}
\ee
again matching the order in $r$ expected for the subleading memory interpretation of~\cite{Compere:2019odm}.

We will now want to understand how to merge these two pictures for the towers of soft theorems.  While the bulk modes match the scaling-in-$r$ expected from the residual gauge interpretation of~\cite{Compere:2019odm}, the celestial perspective is quite different. Indeed the power of the celestial basis comes from the SL$(2,\mathbb{C})$ representations identifying operators that decouple from scattering, with or without a gauge interpretation. This point of view is robust to the choice of residual gauge fixing and, in that sense, can be viewed as a stronger footing for the physicality of these symmetries.  Indeed, the recent work of~\cite{Freidel:2021ytz} is able to identify the $w_{1+\infty}$ generators in gravity without needing to augment the phase space with overleading gauge directions. Here we appear to see aspects of both approaches by comparing the behavior of the results we extract from the saddle-point versus from the full bulk profile. The fact that we are able to match part of the large gauge transformation encourages us to connect these two approaches further.

One point worth clarifying is that in this discussion of vacuum transitions there are actually normally four modes under consideration rather than two.  The two that we care about are the paired Goldstone and memory modes for a given soft theorem.  One of these is a gauge transformation at order $\times~ r^m$ overleading relative to radiative solutions the other is a dynamical shift in the metric data at order $\times~ r^{-m}$ subleading. The extra two modes one can ask about are a dynamical shift at order $\times~ r^m$ and a gauge transformation at order $\times~ r^{-m}$. The former would normally be excluded by boundary conditions while the latter would have vanishing asymptotic symmetry charge on radiative configurations, and thus not be considered part of the asymptotic symmetry group. So long as we label our diamonds by symplectically paired modes there should be no confusion.  However, one should take care about what we mean when we say vacuum-to-vacuum transition, which we will use $\Delta A^{(m)}$ to indicate --  namely the difference between the early and late time values --  versus the mode which implements a change in the vacuum state $A^{(m)}\mapsto A^{(m)}+\delta  A^{(m)}$.  The source of this confusion comes from the fact that these two operators are the same for the leading soft theorem examples. 

Let us briefly review the first few subleading cases, where this degeneracy is broken. Even the word `memory effect' evokes a notion of a final imprint on the metric rather than just a constrained mode, so it is helpful to understand the phrasing in terms of residual vacuum-to-vacuum transitions.  For the subleading soft photon, the dynamical shift occurs at $\mathcal{O}(r^{-1})$~\cite{Himwich:2019dug} in contrast to the $\mathcal{O}(r)$ overleading gauge transformations from~\cite{Campiglia:2016hvg} that we discussed in section~\ref{rsymA}.  Similarly, for the subleading soft graviton we can write the spin memory effect~\cite{1502.06120PSZ} as a shift in a subleading metric component~\cite{Himwich:2019qmj}.  Namely, for an appropriate residual gauge fixing
the late/early $u$-asymptotics of the leading angular metric components are
\be\label{bndy}
h_{\bz\bz,\pm}^{(-1)}=-uD_z^3 \hat{Y}^\bz(\bz)-2D_\bz^2 \hat{f}^\pm(z,\bz),~~~h_{\bz\bz,\pm}^{(0)}=u[D^2-2]D_\bz^2\hat{f}^\pm(z,\bz)+2D_\bz \hat{V}_\bz^\pm(z,\bz).
\ee
Here $\hat{f}^\pm$ and $\hat{Y}^\pm$ label the early and late time supertranslation and superrotation vacua. Meanwhile, under a subleading diffeomorphism of the form
\be
\xi^A=\frac{1}{r^2}V^A(z,\bz),~~~\hat{V}_{\bz}^\pm\mapsto\hat{V}_{\bz}^\pm+V_\bz.
\ee
While in the leading case we only have two modes $\hat{f}^+$ and $\Delta \hat{f}$, when we turn to the subleading soft graviton we encounter the four modes (for a single helicity sector) $\{\hat{V}_\bz^+,\hat{Y}_\bz^+,\Delta\hat{V}_\bz,\Delta \hat{Y}_{\bz}\}$.  
The mode $\Delta \hat{V}_{\bz}$ corresponds to the spin memory effect.\footnote{For comparison with the celestial diamond construction of~\cite{Pasterski:2021fjn,Pasterski:2021dqe} we note that at the linearized level $\Delta \hat{V}_\bz$ matches the top corner of the subleading graviton memory diamond.} It is symplectically paired with the superrotation Goldstone mode $\hat{Y}_\bz^+$. By contrast a dynamical shift in the superrotation vacuum $\Delta\hat{Y}_\bz$ has been associated to snapping cosmic strings~\cite{Strominger:2016wns} (see also~\cite{Compere:2016jwb,Adjei:2019tuj}), and would be paired with the subleading Goldstone mode $\hat{V}_\bz^+$. 

Note that this rewriting of the spin memory as a subleading shift starts from the radiative profile for the celestial stress tensor and uses the equations of motion and residual gauge fixing to propagate into the bulk. As advocated above, by phrasing everything in terms of the radiative order data we can compare statements made in different residual gauges. This is how we identified the corresponding celestial Goldstone-like and memory modes. In particular, the radial gauge fixing for these primaries does not give rise to subleading shifts and is incompatible when there are sources.   Nevertheless it is tempting to still use these modes to describe the gauge fields in the regions near $\mathscr{I}^\pm_{\pm}$ where there are no sources at null infinity.  In that case the generalized primaries $\theta(\pm X^2) \Phi_{\Delta,J}$ have the correct SL$(2,\mathbb{C})$ transformation properties and only violate the source-free equations at the lightcone, which is outside the regions of interest~\cite{Pasterski:2020pdk}.  We see that this gives the correct accounting of modes: namely multiplying by $\theta(u)$ does not add anything for the leading soft theorems, since the conformally soft modes already have this step function profile~\cite{Donnay:2018neh}, while for the other conformally soft modes in the tower we get four modes for the data on $\mathscr{I}^+$ whose non-contact terms have the same radial falloffs as predicted in the residual diffeomorphism discussion.

\paragraph{A Multipole Interpretation}
Let us close with a discussion of what these $w_{1+\infty}$ generators and their gauge theory analogs couple to. We will focus on the electromagnetic case where the connection to multipole moments is easiest to see. We will consider a toy set-up that avoids the issues encountered in~\cite{Campiglia:2016hvg} with regards to being unable to renormalize overleading terms in the expected tower of charges by recognizing them as the charges for more-leading soft theorems. In the end, we can sidestep the question of a large gauge interpretation since the existence of a conformally soft theorem is enough for us to proceed and the form of the soft factor acting on the other operators indicates what feature it is measuring.

We start by noting that the standard electromagnetic memory measures a kinematics-weighted combination of the electric charges for the particles in the initial and final states~\cite{Bieri:2013hqa,Pasterski:2015zua}. Namely the Li\'enard-Wiechert potential of a point charge moving with four momentum $p^\mu= E(1,\beta_{s})$ is
\be
A^\mu=\frac{1}{4\pi\varepsilon_0}\left(\frac{-eQp^\mu}{(n\cdot p)|\vec{r}-\vec{r}_s|}\right)_{t_r}
\ee
where $n^\mu=(1,\hat{n})$ parameterizes the direction where the observer is sitting relative to the source and we will assume $r\gg r_s$. If we expand in large $r$ we see an angle dependence in $F_{ru}^{(2)}$ by nature of the nonzero three-velocity of the charge. Of course for a single massive charge we could always go to its rest frame, but the point is that we can build up an arbitrary angle dependence to $F_{ru}^{(2)}$ by superimposing charges moving in different directions. When we think about decomposing the asymptotic charge for large $U(1)$ gauge transformations 
\be\label{qlambda}
Q^+(\lambda)= \lim\limits_{r\rightarrow\infty} \int d^2z\sqrt{\gamma}~\lambda (r^2 F_{ru})
\ee
into a basis of spherical harmonics for $\lambda$, we see that if we could not build such a general dependence for $F_{ur}^{(2)}$ some of these charges would vanish.

Let us now consider a situation where all of the external states are neutral.  This corresponds to a non-ionizing regime which one could create by scattering, for example, some molecules with intrinsic dipole moments. For simplicity we will only consider electric dipoles, though the generalization to the magnetic case is straightforward. Now in electrostatics we are used to constructing an ideal dipole by taking two oppositely charged monopoles close together \be
A^\mu_{dip}=A^\mu(Q,\vec{x}_s+\vec{d}/2)+A^\mu(-Q,\vec{x}_s-\vec{d}/2)
\ee
in the limit that $|\vec{d}|\rightarrow 0$, $\vec{P}=Q\vec{d}$ finite.  For the scattering configurations we are interested in we want to boost these pairs together.  Now the net monopole charge $F^{(2)}_{ru}$ will vanish and the leading term near spatial infinity will be $F^{(3)}_{ru}$. Similar to the leading case, its angular dependence will be a kinematics-weighted sum/projection of the dipole moments.  This will be measured by the subleading soft photon theorem while the leading soft factor will vanish for this scattering process. Indeed, comparing to the charge from~\cite{Lysov:2014csa} 
\be
Q^+(Y)=\frac{1}{e^2}\int d^2z du \p_u (uD_zY^z(F_{ur}^{(2)}\gamma_{z\bz}+F^{(0)}_{z\bz})+2Y^z\gamma_{z\bz}F^{(2)}_{zr}+c.c.
\ee
the first two terms vanish for this configuration while 
\be
\gamma^{AB} \p_A F_{rB}^{(2)}=-F_{ru}^{(3)}
\ee
for massive scatterers.  

We can iterate this procedure of constructing higher intrinsic multipoles. The electrostatic fields fall of with progressively higher powers of $r^{-1}$.  Their boost images preserve this fall-off structure. Despite our simplified set up, the robust statement is that these intrinsic $\ell$-multipole moments of the external scatterers start contributing to the soft theorems at the corresponding $\ell^{th}$ subleading order and below.  While it is clear from~\eqref{qlambda} that an overleading $\lambda$ would pick out such subleading modes of $F_{ru}$, the equations of motion again tie us back to the radiative data, where we can make statements in any scattering basis.

\section*{Acknowledgements}
We would like to thank  Laurent Freidel, Carlo Heissenberg, Elizabeth Himwich, Prahar Mitra, Yorgo Pano, Andrew Strominger, Emilio Trevisani,  and Herman Verlinde for useful conversations.  LD is supported by the Austrian Science Fund (FWF) START project Y 1447-N. SP is supported by the Sam B. Treiman Fellowship at the Princeton Center for Theoretical Science. AP is supported by the European Research Council (ERC) under the European Union’s Horizon 2020 research and innovation programme (grant agreement No 852386).

%\pagebreak

\appendix{

\section{\texorpdfstring{Large-$r$ Expansions}{Large-r Expansions}}\label{larger}
In this appendix we collect the radial expansions of the equations of motion and gauge fixing conditions for spin-1 and spin-2 wavefunctions.

%%%%%%%%%%%%%%%%%%%%%%%%%%%%%%%%%%%%%%%%%%%%%%%%%%%%%
\subsection{Harmonic gauge field with radial gauge fixing}\label{app:harmA}
\label{app:harm}
%%%%%%%%%%%%%%%%%%%%%%%%%%%%%%%%%%%%%%%%%%%%%%%%%%%%%

Given our ansatz for the radial expansion of the vector field in~\eqref{Ahmode}, a similar decomposition of the harmonic gauge equations $\Box A_\mu=0$ into powers of $r^{-n}$ yields
\begin{equation}\scalemath{0.92}{\label{eq:boxA}
\begin{array}{ll}
[\Box A_u]^{(n)} &= 2(n-2)\partial_u A_u^{(n-1)} + \left[D^2 +(n-2)(n-3) \right]A_u^{(n-2)}
%\\ &~~~
+ (5-2n)\bar{A}_u^{(n-2)} - 2\partial_u\bar{A}_u^{(n-1)}, \\
\left[ \Box A_r \right]^{(n)} &= 2(n-2)\partial_u A_r^{(n-1)} + \left[D^2 +(n-2)(n-3) - 2\right]A_r^{(n-2)}+ 2 A_u^{(n-2)}- 2D^CA_C^{(n-3)} \\
&~~~+ (5-2n)\bar{A}_r^{(n-2)} - 2\partial_u\bar{A}_r^{(n-1)}\,, \\
\left[\Box A_A\right]^{(n-1)} &= 2(n-2)\partial_u A_A^{(n-2)} + \left[D^2 +(n-2)(n-3) -1 \right]A_A^{(n-3)}- 2 \partial_A\left(A_u^{(n-2)} - A_r^{(n-2)}\right) \\
&~~~+ (5-2n)\bar{A}_A^{(n-3)} - 2\partial_u\bar{A}_A^{(n-2)} \,,
\end{array}
}\end{equation}
and at logarithmic order 
\begin{equation}\scalemath{0.92}{\label{eq:boxAlog}
\begin{array}{ll}
\left[\Box\bar{A}_u\right]^{(n)} &= 2(n-2)\partial_u \bar{A}_u^{(n-1)} + \left[D^2 +(n-2)(n-3) \right]\bar{A}_u^{(n-2)}\,, \\
\left[\Box \bar{A}_r\right]^{(n)} &= 2(n-2)\partial_u \bar{A}_r^{(n-1)} + \left[D^2 +(n-2)(n-3) - 2\right]\bar{A}_r^{(n-2)} + 2 \bar{A}_u^{(n-2)} - 2 D^C\bar{A}_C^{(n-3)}\,, \\
\left[\Box \bar{A}_A\right]^{(n-1)}& = 2(n-2)\partial_u \bar{A}_A^{(n-2)} + \left[D^2 +(n-2)(n-3) -1 \right]\bar{A}_A^{(n-3)} - 2 \partial_A\left(\bar{A}_u^{(n-2)} - \bar{A}_r^{(n-2)}\right) \,.
\end{array}}
\end{equation}
Meanwhile the harmonic gauge condition takes the form
\begin{equation}\scalemath{0.92}{
\begin{array}{ll}
[\nabla^\mu A_\mu]^{(n)} &=-\p_u A_r^{(n)} +(n-3) (A_u^{(n-1)}-A_r^{(n-1)})+2D^A A_{A}^{(n-2)}-(\bar{A}_u^{(n-1)}-\bar{A}_r^{(n-1)}), \\
\end{array}}\end{equation}
and at logarithmic order
\begin{equation}\scalemath{0.92}{
\begin{array}{ll}
[\nabla^\mu \bar{A}_\mu]^{(n)} &= -\p_u \bar{A}_r^{(n)} +(n-3) (\bar{A}_u^{(n-1)}-\bar{A}_r^{(n-1)})+2D^A \bar{A}_{A}^{(n-2)}. \\
\end{array}}\end{equation}

%\pagebreak

\noindent This system of equations can be solved recursively in $r$.  We want solutions where the tower is only semi-infinite, starting at a finite order in $r$.  The place where this recursion can break down is where the prefactors of the $\p_u$ terms vanish.  The data that can be freely specified as a function of $u$ is $A_A^{(0)}$ while the subleading modes and other components are determined by the equations of motion and gauge fixing conditions.  Meanwhile, overleading solutions are allowed so long as the $u$ dependence is restricted. The log tower is necessary for sourced solutions and can also be sourced by the overleading terms.  

Additional residual gauge fixing is required to fully specify the recursion relations. For source-free solutions we can impose the radial gauge condition $rA_r+uA_u=0$ which implies
\begin{equation}
 A_r^{(n+1)}=-uA_u^{(n)}\,, \quad \bar{A}_r^{(n+1)}=-u\bar{A}_u^{(n)}\,
\end{equation}
within our radial recursions.

\subsection{Harmonic metric with radial gauge fixing}\label{app:DiffS2}

%%%%%%%%%%%%%%%%%%%%%%%%%%%%%%%%%%%%%%%%%%%%%%%%%%%%%
Given our ansatz for the radial expansion of the metric in~\eqref{Ahmode}, a similar decomposition of the harmonic gauge equations $\Box h_{\mu\nu}=0$ into powers of $r^{-n}$ yields~\cite{Himwich:2019qmj}
\begin{equation}\scalemath{0.92}{\label{eq:boxh}
\begin{array}{ll}
\left[\square h_{uu}\right]^{(n)} &= 2(n-2) \partial_uh_{uu}^{(n-1)} + [D^2 + (n-2)(n-3)]h_{uu}^{(n-2)}
+ (5-2n) \bar{h}_{uu}^{(n-2)} - 2\partial_u\bar{h}_{uu}^{(n-1)}, \\
\left[\square h_{ur}\right]^{(n)} &= 2(n-2) \partial_uh_{ur}^{(n-1)} + [D^2 + (n-2)(n-3) - 2]h_{ur}^{(n-2)}+ 2h_{uu}^{(n-2)}- 2D^Ah_{uA}^{(n-3)} \\
&~~~+ (5-2n) \bar{h}_{ur}^{(n-2)} - 2\partial_u\bar{h}_{ur}^{(n-1)}, \\
\left[\square h_{rr}\right]^{(n)} &= 2(n-2) \partial_uh_{rr}^{(n-1)} + [D^2 + (n-2)(n-3)]h_{rr}^{(n-2)}+ 4\left(h_{ur}^{(n-2)}- h_{rr}^{(n-2)}\right) \\
&~~~ - 4D^Ah_{rA}^{(n-3)}+ 2\gamma^{AB}h_{AB}^{(n-4)} + (5-2n) \bar{h}_{rr}^{(n-2)} - 2\partial_u\bar{h}_{rr}^{(n-1)}, \\
\left[\square h_{uA}\right]^{(n-1)} &= 2(n-2) \partial_uh_{uA}^{(n-2)} + [D^2 + (n-3)(n-2) - 1]h_{uA}^{(n-3)}- 2\partial_A\left(h_{uu}^{(n-2)} - h_{ur}^{(n-2)}\right) \\
&~~~+ (5-2n) \bar{h}_{uA}^{(n-3)} - 2\partial_u\bar{h}_{uA}^{(n-2)}, \\
\left[\square h_{rA}\right]^{(n-1)} &= 2(n-2) \partial_uh_{rA}^{(n-2)} + [D^2 + (n-3)(n-2) - 1]h_{rA}^{(n-3)}- 2\partial_A\left(h_{ur}^{(n-2)} - h_{rr}^{(n-2)}\right) \\
&~~~- 2D^Bh_{AB}^{(n-4)} + 4(h_{uA}^{(n-3)}-h_{rA}^{(n-3)})+ (5-2n) \bar{h}_{rA}^{(n-3)} - 2\partial_u\bar{h}_{rA}^{(n-2)}, \\
\left[\square h_{AB}\right]^{(n-2)} &= 2(n-2) \partial_uh_{AB}^{(n-3)} + [D^2 + (n-2)(n-3) - 2]h_{AB}^{(n-4)}\\
&~~~- 2\left(D_Ah_{uB}^{(n-3)} - D_Ah_{rB}^{(n-3)} + D_Bh_{uA}^{(n-3)} - D_Bh_{rA}^{(n-3)}\right) \\
&~~~+ 2 \gamma_{AB}\left(h_{uu}^{(n-2)} - 2h_{ur}^{(n-2)} + h_{rr}^{(n-2)} \right) + (5-2n)\bar{h}_{AB}^{(n-4)} - 2 \partial_u\bar{h}_{AB}^{(n-3)} ,
\end{array}}
\end{equation}
and at logarithmic order 
\begin{equation}\label{eq:boxhlog}
\scalemath{0.92}{
\begin{array}{ll}
\left[\square \bar{h}_{uu}\right]^{(n)} &= 2(n-2) \partial_u\bar{h}_{uu}^{(n-1)} + [D^2 + (n-2)(n-3)] \bar{h}_{uu}^{(n-2)}, \\
\left[\square \bar{h}_{ur}\right]^{(n)} &= 2(n-2) \partial_u\bar{h}_{ur}^{(n-1)} + [D^2 + (n-2)(n-3) - 2]\bar{h}_{ur}^{(n-2)} + 2\bar{h}_{uu}^{(n-2)} - 2D^A\bar{h}_{uA}^{(n-3)}, \\
\left[\square \bar{h}_{rr}\right]^{(n)} &= 2(n-2) \partial_u\bar{h}_{rr}^{(n-1)} + [D^2 + (n-2)(n-3)]\bar{h}_{rr}^{(n-2)} + 4\left(\bar{h}_{ur}^{(n-2)} - \bar{h}_{rr}^{(n-2)}\right) \\
&~~~- 4D^A\bar{h}_{rA}^{(n-3)} + 2\gamma^{AB}\bar{h}_{AB}^{(n-4)}, \\
\left[\square \bar{h}_{uA}\right]^{(n-1)} &= 2(n-2) \partial_u\bar{h}_{uA}^{(n-2)} + [D^2 + (n-3)(n-2) - 1]\bar{h}_{uA}^{(n-3)} - 2\partial_A\left(\bar{h}_{uu}^{(n-2)} - \bar{h}_{ur}^{(n-2)}\right), \\
\left[\square \bar{h}_{rA}\right]^{(n-1)} &= 2(n-2) \partial_u\bar{h}_{rA}^{(n-2)} + \left[D^2 + (n-3)(n-2) - 1\right]\bar{h}_{rA}^{(n-3)} - 2\partial_A\left(\bar{h}_{ur}^{(n-2)} - \bar{h}_{rr}^{(n-2)}\right) \\
&~~~- 2D^B\bar{h}_{AB}^{(n-4)} + 4(\bar{h}_{uA}^{(n-3)}-\bar{h}_{rA}^{(n-3)}), \\
\left[\square \bar{h}_{AB}\right]^{(n-2)} &= 2(n-2) \partial_u\bar{h}_{AB}^{(n-3)} + [D^2 + (n-2)(n-3) - 2)]\bar{h}_{AB}^{(n-4)} \\
&~~~- 2\left(D_A\bar{h}_{uB}^{(n-3)} - D_A\bar{h}_{rB}^{(n-3)} + D_B\bar{h}_{uA}^{(n-3)} - D_B\bar{h}_{rA}^{(n-3)}\right) \\
&~~~+ 2 \gamma_{AB}\left(\bar{h}_{uu}^{(n-2)} - 2\bar{h}_{ur}^{(n-2)} + \bar{h}_{rr}^{(n-2)} \right)  . 
\end{array}
}
\end{equation}
Meanwhile the  harmonic gauge condition becomes
\begin{equation}\scalemath{0.90}{
\begin{array}{ll}\label{eq:hg1}
\left[\nabla^{\mu}h_{\mu u}\right]^{(n)} &= - \partial_u h_{ur}^{(n)} + (n-3)\left(h_{uu}^{(n-1)} - h_{ur}^{(n-1)}\right) + D^Ah_{uA}^{(n-2)}
\left(\bar{h}_{uu}^{(n-1)} - \bar{h}_{ur}^{(n-1)}\right), \\
\left[\nabla^{\mu}h_{\mu r}\right]^{(n)} &= - \partial_u h_{rr}^{(n)} + (n-3)\left(h_{ur}^{(n-1)} - h_{rr}^{(n-1)}\right) + D^Ah_{rA}^{(n-2)}- \gamma^{AB}h_{AB}^{(n-3)} 
\left(\bar{h}_{ur}^{(n-1)} - \bar{h}_{rr}^{(n-1)}\right), \\
\left[\nabla^{\mu}h_{\mu A}\right]^{(n-1)} &= - \partial_u h_{rA}^{(n-1)} + (n-4)\left(h_{uA}^{(n-2)} - h_{rA}^{(n-2)}\right) + D^Bh_{BA}^{(n-3)}
\left(\bar{h}_{uA}^{(n-2)} - \bar{h}_{rA}^{(n-2)}\right)  ,
\end{array}}
\end{equation}
and
\begin{equation}\scalemath{0.92}{
\begin{array}{ll}\label{eq:hg2}
\left[\nabla^{\mu}\bar{h}_{\mu u}\right]^{(n)} &= - \partial_u \bar{h}_{ur}^{(n)} + (n-3)\left(\bar{h}_{uu}^{(n-1)} - \bar{h}_{ur}^{(n-1)}\right) + D^A\bar{h}_{uA}^{(n-2)}, \\
\left[\nabla^{\mu}\bar{h}_{\mu r}\right]^{(n)} &= - \partial_u \bar{h}_{rr}^{(n)} + (n-3)\left(\bar{h}_{ur}^{(n-1)} - \bar{h}_{rr}^{(n-1)}\right) + D^A\bar{h}_{rA}^{(n-2)} - \gamma^{AB}\bar{h}_{AB}^{(n-3)} ,\\
\left[\nabla^{\mu}\bar{h}_{\mu A}\right]^{(n-1)} &= - \partial_u \bar{h}_{rA}^{(n-1)} + (n-4)\left(\bar{h}_{uA}^{(n-2)} - \bar{h}_{rA}^{(n-2)}\right) + D^B\bar{h}_{BA}^{(n-3)} .
\end{array}}
\end{equation}
The structure of the recursions is similar to the spin-1 case.  Now the free data is $h_{zz}^{(-1)}$ and its complex conjugate $h_{\bz\bz}^{(-1)}$.  Meanwhile, overleading solutions are allowed so long as the $u$ dependence is restricted. The log tower is necessary for sourced solutions and can also be sourced by the overleading terms.  

For source free solutions we can impose the radial gauge condition $rh_{r\mu}+uh_{u\mu}=0$ fixes the radial components at a given order by the temporal components  
\begin{equation}\begin{array}{ll}
 h_{rA}^{(n+1)}=-uh_{uA}^{(n)}\,, \quad &\bar{h}_{rA}^{(n+1)}=-u\bar{h}_{uA}^{(n)}\,\\
 h_{ru}^{(n+1)}=-uh_{uu}^{(n)}\,, \quad &\bar{h}_{ru}^{(n+1)}=-u\bar{h}_{uu}^{(n)}\,\\
 h_{rr}^{(n+2)}=u^2h_{uu}^{(n)}\,, \quad &\bar{h}_{rr}^{(n+2)}=u^2\bar{h}_{uu}^{(n)}\,,
\end{array}\end{equation}
and these relations can be plugged into the equations above.
}

\pagebreak

%\section{Diamonds and Light Transforms}%\label{app:diamondLT}

% \section{Light and Shadow  Transforms in Celestial Diamonds}\label{app:diamondLT}

\section{Light and Shadow Transforms as Descendants}\label{app:diamondLT}

The aim of this appendix is to resolve a tension between the finite truncation of the $\Delta=1-s-n$ for $n>0$ multiplets observed in~\cite{Pasterski:2021fjn} and the role these primaries, their light transforms, and shadows play in the extended celestial symmetry algebras.  In particular, we show that by considering conformal dimensions slightly analytically continued from integer values, the shadow~\cite{Pasterski:2017kqt} and light transformed~\cite{Atanasov:2021cje,Sharma:2021gcz} wavefunctions appear within the same diamond.

When we take $w,\bw$-derivatives of the radiative wavefunctions we showed in~\cite{Pasterski:2021fjn} that we can land on primary descendants with $|J|< s$.  This necessitates the introduction of additional tetrad elements
\be
l^\mu=\frac{q^\mu}{-q\cdot X}\,, ~~~n^\mu=X^\mu+\frac{X^2}{2}l^\mu\,
\ee
with weights $(\Delta,J)=(0,0)$ so that $\{l,n,m,\bar{m}\}$ form an orthonormal tetrad. In~\cite{Pasterski:2021fjn} we showed that descendants of these tetrad elements take the form
\begin{equation}
 \badat{4}\label{tetradesc}
\p_w^n l_\mu&=2^{\frac{n}{2}}n! (\epsilon_w\cdot X)^{n-1}\varphi_n m_\mu\,,\\
\p_w^n n_\mu &=\frac{X^2}{2}\p_w^n l_\mu\,,\\
\p_w^n m_\mu&=(\epsilon_w\cdot X)\partial_w^n l_\mu\,,\\
\p_w^{n} \bar{m}_\mu&=\left\{\begin{array}{ll}
    2^{\frac{1}{2}}\varphi_1\left( v_\mu-(\epsilon_w\cdot X)\bar{m}_\mu\right)  &  n=1\\
   \frac{X^2}{2}(\epsilon_w\cdot X)^{-1}\p_w^nl_\mu\,  & n>1
\end{array}\right.\,
 \eadat
\end{equation}
where we have defined the quantity
\be
 v_\mu=\frac{X^2}{2}l_\mu+n_\mu\,
\ee
and $\epsilon_w=\frac{1}{\sqrt{2}}\p_w q_\mu$ is the positive helicity polarization tensor. Similar expressions hold for the $\p_\bw$ descendants.

Now consider radiative primaries of the form
\be
\Phi_{\Delta,s}=m^s\varphi_\Delta,~~~\Phi_{\Delta,-s}=\bar{m}^s\varphi_\Delta.
\ee
For $\Delta=1-s-n$ these functions are polynomial in $w,\bw$ such that the primaries with 
\be
\Delta_{k,\bar{k}}=1-\frac{k+\bar{k}}{2},~~~J=\frac{\bar{k}-k}{2}=\pm s,
\ee
corresponding to left and right conformal weights~\eqref{condition_diamond_hhb} have vanishing level-$k$ and level-$\bar{k}$ descendants
\be\label{desc0}
\p_w^k \Phi_{\Delta_{k,\bar{k}},s}=\p_\bw^{\bar k}\Phi_{\Delta_{k,\bar{k}},s}=0.
\ee
Because these derivatives commute with $X^2$, the same holds true for the shadow primary wavefunctions with $\varphi_\Delta\mapsto\tvarphi^\Delta$.

If we didn't know the specific form of these wavefunctions, we would not necessarily  assume that the primary descendant wavefunctions identically vanish. For example, this is not the case for the diamonds with $1-s\le \Delta<1+s$.  In this paper we have examined various order of limits issues which naturally advocate for considering the analytic behavior near these special values of $\Delta$ rather than taking strict limits~\cite{Donnay:2020guq,Guevara:2021abz,Arkani-Hamed:2020gyp}. If we consider a small shift in the conformal dimension
\be
\Delta=\Delta_{k,\bar{k}}+\epsilon
\ee
and keep the descendancy relations to $\mathcal{O}(\epsilon)$ we will see that more of the structure we saw for the leading diamonds persists to this semi-infinite tower.  In particular, vertices across the diamond are related by shadow transforms and across one edge are related by an analytic continuation of the (2,2) signature light transform.

We will discuss the general structure of these descendants and then give the specific normalizations for the spin-1 and spin-2 cases of interest here.  Writing
\be
\varphi_{\Delta_{k,\bar{k}}+\epsilon}=\varphi_{\Delta_{k,\bar{k}}}-\epsilon \varphi_{\Delta_{k,\bar{k}}}\log(-q\cdot X)+\cdots
\ee
the descendant we will be interested in comes from the second term. We see from~\eqref{tetradesc} and~\eqref{desc0} that at least one derivative will hit the $\log(-q\cdot X)$ term. Because $\p_w^2 q=\p_\bw^2 q=0$, we see that \be
\p_w^n \log(-q\cdot X)\propto (\epsilon_w\cdot X)^{n}\varphi_n ,
\ee
similarly 
\be
\p_w^n m \propto  m(\epsilon_w\cdot X)^n\varphi_n,
\ee
so all the terms contributing to $\p_w^k \Phi_{\Delta_{k,\bar{k}}+\epsilon,s}$ will be proportional to
\be
\p_w^k \Phi_{\Delta_{k,\bar{k}}+\epsilon,s}\propto \epsilon  m^s (\epsilon_w \cdot X)^{1-s-\Delta_{k,\bar{k}}} \varphi_{1-s}+\mathcal{O}(\epsilon^2),
\ee
where we've used $-s=\Delta_{k,\bar{k}}+k-1$. We will drop the explicit $+~\mathcal{O}(\epsilon^2)$ in what follows taking $\epsilon$ to be implicitly infinitesimal. While this object isn't made from the conformally covariant tetrad elements, it does match the proposed shadow wavefunctions~\cite{Atanasov:2021cje,Sharma:2021gcz} and has correct homogeneity expected for an operator at the weyl reflected weight $h\mapsto 1-h$. We see from~\eqref{tetradesc} that we expect $\p_\bw^{\bar{k}} \Phi_{\Delta_{k,\bar{k}}+\epsilon,s}$ to instead have a sum over multiple %$2s+1$ 
terms involving tensor products of $\{m_\mu,v_\mu,\bar{m}_\mu\}$ depending on if the $m$'s are hit with $\p_\bw$ derivatives 0, 1, or $>1$ times. We note that the tensor factor for the descendant wavefunction will still be symmetric, traceless, and orthogonal to $X$, so we expect 3 independent terms for $s=1$ and 5 for $s=2$. This, again, is consistent with what was found in~\cite{Sharma:2021gcz}.  

Let us now restrict to the spin-1 and spin-2 cases where it is easier to be explicit. We have
\be
\p_w^k (m\varphi_{\Delta_{k,\bar{k}}+\epsilon})=\epsilon(-1)^{k-1}2^{\frac{k}{2}}(k-1)! m (\epsilon_w \cdot X)^{-\Delta_{k,\bar{k}}} 
\ee
and
\be\badat{3}
\p_\bw^{\bar k} (m\varphi_{\Delta_{k,\bar{k}}+\epsilon})&=\Big[(\epsilon_\bw\cdot X)m+\frac{\bar k}{k} \frac{X^2}{2} (\epsilon_\bw\cdot X)^{-1}\bar{m}-\frac{\bar k}{k+1} v \Big]\\
&~~~\times\epsilon (-1)^{\bar{k}-1}2^\frac{\bar{k}}{2}(\bar k-1)!(\epsilon_\bw\cdot X)^{1-\Delta_{k,\bar{k}}}\varphi_{2}
\eadat\ee
for spin-1, where we note $\bar{k}=k+2$ so $(\bar k-1)!$ times the relative $k,\bar{k}$-dependent coefficients gives integers. Meanwhile, for spin-2 we have
\be
\p_w^k (mm\varphi_{\Delta_{k,\bar{k}}+\epsilon})=\epsilon(-1)^{k-1}2^{\frac{k}{2}}(k-1)! mm (\epsilon_w \cdot X)^{-1-\Delta_{k,\bar{k}}} \varphi_{-1}
\ee
and
\be\badat{3}
\p_\bw^{\bar k} (mm\varphi_{\Delta_{k,\bar{k}}+\epsilon})&= \Big[(\epsilon_\bw\cdot X)^2 mm+\frac{\bar k}{k} \left(\frac{X^2}{2}\right)^2(\epsilon_\bw\cdot X)^{-2}\bar m\bar m+ \frac{\bar k}{k+3}(\epsilon_\bw\cdot X)(m v+vm) \\
&~~~ -\frac{\bar k}{k+1}\frac{X^2}{2}(\epsilon_\bw\cdot X)^{-1}(\bar m v+v\bar m)+\frac{\bar k}{k+2}\frac{X^2}{2}( m\bar{m} +\bar m m+\frac{2}{X^2} vv)\Big] \\
&~~~ \times \epsilon (-1)^{\bar{k}-1}2^\frac{\bar{k}}{2}(\bar k-1)!(\epsilon_\bw\cdot X)^{1-\Delta_{k,\bar{k}}}\varphi_{3}.
\eadat\ee
Here $\bar k=k+4$, so multiplying the relative $k,\bar k$-dependent coefficients by $(\bar k-1)!$ gives integers. Beautifully, if we combine these two descendancy relations we land on the shadow transformed wavefunctions 
\be
{\p}_\bw^{\bar{k}}\p_w^k (m\varphi_{\Delta_{k,\bar{k}}+\epsilon})=\epsilon (k-1)! \bar{k}! \bar{m}\tvarphi_{2-\Delta_{k,\bar{k}}}
\ee
and\be
{\p}_\bw^{\bar{k}}\p_w^k (mm\varphi_{\Delta_{k,\bar{k}}+\epsilon})=-\epsilon (k-1)! \bar{k}! \bar{m}\bar{m}\tvarphi_{2-\Delta_{k,\bar{k}}},
\ee
which are proper radiative SL$(2,\mathbb{C})$ primaries that only involve our covariant tetrad elements and not factors of $\epsilon_J\cdot X$~\cite{Pasterski:2020pdk}.

%\pagebreak

Curiously, we see that a $\mathbb{R}^{1,3}$ manifestation of the light transform~\cite{Atanasov:2021cje,Guevara:2021tvr} that appears in the celestial diamonds descending from the tower of conformally soft theorems responsible for the $w_{1+\infty}$ currents of~\cite{Guevara:2021abz,Strominger:2021lvk}.   Related manipulations exploring the relation between descendants and light transforms at the level of operators have recently appeared in~\cite{Freidel:2021ytz}.  In the context of our strictly $(1,3)$ signature discussion here, the fact that the shadows are within the same multiplets once we expand in $\Delta$ helps reconcile what we expect for the full tower with our understanding of the paring between Goldstone(-like) and memory modes for the other diamonds, and avoids a naive over-counting of independent SL$(2,\mathbb{C})$ multiplets.

\bibliographystyle{style}
\bibliography{references}

\providecommand{\href}[2]{#2}\begingroup\raggedright\begin{thebibliography}{10}

\bibitem{Strominger:2017zoo}
A.~Strominger, {\em {Lectures on the Infrared Structure of Gravity and Gauge
  Theory}}.
\newblock {Princeton University Press},
2018
%%CITATION = ARXIV:1703.05448;%%.

\bibitem{Pasterski:2019ceq}
S.~Pasterski, \emph{{Implications of Superrotations}}, Phys. Rept. {\bf 829}
  (2019) 1--35,
\href{http://www.arXiv.org/abs/1905.10052}{{\tt 1905.10052}}
% .

\bibitem{Raclariu:2021zjz}
A.-M. Raclariu, \emph{{Lectures on Celestial Holography}},
\href{http://www.arXiv.org/abs/2107.02075}{{\tt 2107.02075}}
% .

\bibitem{Pasterski:2021rjz}
S.~Pasterski, \emph{{Lectures on celestial amplitudes}}, Eur. Phys. J. C {\bf
  81} (2021), no.~12, 1062,
\href{http://www.arXiv.org/abs/2108.04801}{{\tt 2108.04801}}
% .

\bibitem{Pasterski:2021raf}
S.~Pasterski, M.~Pate  and A.-M. Raclariu, \emph{{Celestial Holography}}, in
  {\em {2022 Snowmass Summer Study}}.
\newblock 11, 2021.
\newblock
\href{http://www.arXiv.org/abs/2111.11392}{{\tt 2111.11392}}.
\newblock
% .

\bibitem{Cheung:2016iub}
C.~Cheung, A.~de~la Fuente  and R.~Sundrum, \emph{{4D scattering amplitudes and
  asymptotic symmetries from 2D CFT}}, JHEP {\bf 01} (2017) 112,
\href{http://www.arXiv.org/abs/1609.00732}{{\tt 1609.00732}}
%%CITATION = ARXIV:1609.00732;%%.

\bibitem{Fan:2019emx}
W.~Fan, A.~Fotopoulos  and T.~R. Taylor, \emph{{Soft Limits of Yang-Mills
  Amplitudes and Conformal Correlators}}, JHEP {\bf 05} (2019) 121,
\href{http://www.arXiv.org/abs/1903.01676}{{\tt 1903.01676}}
%%CITATION = ARXIV:1903.01676;%%.

\bibitem{Fotopoulos:2019tpe}
A.~Fotopoulos and T.~R. Taylor, \emph{{Primary Fields in Celestial CFT}}, JHEP
  {\bf 10} (2019) 167,
\href{http://www.arXiv.org/abs/1906.10149}{{\tt 1906.10149}}
%%CITATION = ARXIV:1906.10149;%%.

\bibitem{Pate:2019mfs}
M.~Pate, A.-M. Raclariu  and A.~Strominger, \emph{{Conformally Soft Theorem in
  Gauge Theory}}, Phys. Rev. {\bf D100} (2019), no.~8, 085017,
\href{http://www.arXiv.org/abs/1904.10831}{{\tt 1904.10831}}
%%CITATION = ARXIV:1904.10831;%%.

\bibitem{Adamo:2019ipt}
T.~Adamo, L.~Mason  and A.~Sharma, \emph{{Celestial amplitudes and conformal
  soft theorems}}, Class. Quant. Grav. {\bf 36} (2019), no.~20, 205018,
\href{http://www.arXiv.org/abs/1905.09224}{{\tt 1905.09224}}
% .

\bibitem{Puhm:2019zbl}
A.~Puhm, \emph{{Conformally Soft Theorem in Gravity}}, JHEP {\bf 09} (2020)
  130,
\href{http://www.arXiv.org/abs/1905.09799}{{\tt 1905.09799}}
% .

\bibitem{Guevara:2019ypd}
A.~Guevara, \emph{{Notes on Conformal Soft Theorems and Recursion Relations in
  Gravity}},
\href{http://www.arXiv.org/abs/1906.07810}{{\tt 1906.07810}}
%%CITATION = ARXIV:1906.07810;%%.

\bibitem{Crawley:2021ivb}
E.~Crawley, N.~Miller, S.~A. Narayanan  and A.~Strominger,
  \emph{{State-operator correspondence in celestial conformal field theory}},
  JHEP {\bf 09} (2021) 132,
\href{http://www.arXiv.org/abs/2105.00331}{{\tt 2105.00331}}
% .

\bibitem{Pasterski:2022lsl}
S.~Pasterski and H.~Verlinde, \emph{{Chaos in Celestial CFT}},
\href{http://www.arXiv.org/abs/2201.01630}{{\tt 2201.01630}}
% .

\bibitem{Pasterski:2022jzc}
S.~Pasterski, \emph{{A Shorter Path to Celestial Currents}},
\href{http://www.arXiv.org/abs/2201.06805}{{\tt 2201.06805}}
% .

\bibitem{Donnay:2020guq}
L.~Donnay, S.~Pasterski  and A.~Puhm, \emph{{Asymptotic Symmetries and
  Celestial CFT}}, JHEP {\bf 09} (2020) 176,
\href{http://www.arXiv.org/abs/2005.08990}{{\tt 2005.08990}}
% .

\bibitem{Pano:2021ewd}
Y.~Pano, S.~Pasterski  and A.~Puhm, \emph{{Conformally Soft Fermions}},
\href{http://www.arXiv.org/abs/2108.11422}{{\tt 2108.11422}}
% .

\bibitem{Donnay:2018neh}
L.~Donnay, A.~Puhm  and A.~Strominger, \emph{{Conformally Soft Photons and
  Gravitons}}, JHEP {\bf 01} (2019) 184,
\href{http://www.arXiv.org/abs/1810.05219}{{\tt 1810.05219}}
%%CITATION = ARXIV:1810.05219;%%.

\bibitem{Campiglia:2016hvg}
M.~Campiglia and A.~Laddha, \emph{{Subleading soft photons and large gauge
  transformations}}, JHEP {\bf 11} (2016) 012,
\href{http://www.arXiv.org/abs/1605.09677}{{\tt 1605.09677}}
%%CITATION = ARXIV:1605.09677;%%.

\bibitem{Campiglia:2016efb}
M.~Campiglia and A.~Laddha, \emph{{Sub-subleading soft gravitons and large
  diffeomorphisms}}, JHEP {\bf 01} (2017) 036,
\href{http://www.arXiv.org/abs/1608.00685}{{\tt 1608.00685}}
%%CITATION = ARXIV:1608.00685;%%.

\bibitem{Hamada:2018vrw}
Y.~Hamada and G.~Shiu, \emph{{Infinite Set of Soft Theorems in Gauge-Gravity
  Theories as Ward-Takahashi Identities}}, Phys. Rev. Lett. {\bf 120} (2018),
  no.~20, 201601,
\href{http://www.arXiv.org/abs/1801.05528}{{\tt 1801.05528}}
% .

\bibitem{Seraj:2016jxi}
A.~Seraj, \emph{{Multipole charge conservation and implications on
  electromagnetic radiation}}, JHEP {\bf 06} (2017) 080,
\href{http://www.arXiv.org/abs/1610.02870}{{\tt 1610.02870}}
% .

\bibitem{Compere:2017wrj}
G.~Comp\`ere, R.~Oliveri  and A.~Seraj, \emph{{Gravitational multipole moments
  from Noether charges}}, JHEP {\bf 05} (2018) 054,
\href{http://www.arXiv.org/abs/1711.08806}{{\tt 1711.08806}}
% .

\bibitem{Compere:2019odm}
G.~Comp\`ere, \emph{{Infinite towers of supertranslation and superrotation
  memories}}, Phys. Rev. Lett. {\bf 123} (2019), no.~2, 021101,
\href{http://www.arXiv.org/abs/1904.00280}{{\tt 1904.00280}}
% .

\bibitem{Guevara:2021abz}
A.~Guevara, E.~Himwich, M.~Pate  and A.~Strominger, \emph{{Holographic Symmetry
  Algebras for Gauge Theory and Gravity}},
\href{http://www.arXiv.org/abs/2103.03961}{{\tt 2103.03961}}
% .

\bibitem{Strominger:2021lvk}
A.~Strominger, \emph{{w(1+infinity) and the Celestial Sphere}},
\href{http://www.arXiv.org/abs/2105.14346}{{\tt 2105.14346}}
% .

\bibitem{Himwich:2021dau}
E.~Himwich, M.~Pate  and K.~Singh, \emph{{Celestial Operator Product Expansions
  and ${\rm w}_{1+\infty}$ Symmetry for All Spins}},
\href{http://www.arXiv.org/abs/2108.07763}{{\tt 2108.07763}}
% .

\bibitem{Pasterski:2021fjn}
S.~Pasterski, A.~Puhm  and E.~Trevisani, \emph{{Celestial diamonds: conformal
  multiplets in celestial CFT}}, JHEP {\bf 11} (2021) 072,
\href{http://www.arXiv.org/abs/2105.03516}{{\tt 2105.03516}}
% .

\bibitem{Pasterski:2021dqe}
S.~Pasterski, A.~Puhm  and E.~Trevisani, \emph{{Revisiting the conformally soft
  sector with celestial diamonds}}, JHEP {\bf 11} (2021) 143,
\href{http://www.arXiv.org/abs/2105.09792}{{\tt 2105.09792}}
% .

\bibitem{Donnay:2021wrk}
L.~Donnay and R.~Ruzziconi, \emph{{BMS flux algebra in celestial holography}},
  JHEP {\bf 11} (2021) 040,
\href{http://www.arXiv.org/abs/2108.11969}{{\tt 2108.11969}}
% .

\bibitem{Pate:2019lpp}
M.~Pate, A.-M. Raclariu, A.~Strominger  and E.~Y. Yuan, \emph{{Celestial
  Operator Products of Gluons and Gravitons}},
\href{http://www.arXiv.org/abs/1910.07424}{{\tt 1910.07424}}
%%CITATION = ARXIV:1910.07424;%%.

\bibitem{Fotopoulos:2019vac}
A.~Fotopoulos, S.~Stieberger, T.~R. Taylor  and B.~Zhu, \emph{{Extended BMS
  Algebra of Celestial CFT}}, JHEP {\bf 03} (2020) 130,
\href{http://www.arXiv.org/abs/1912.10973}{{\tt 1912.10973}}
%%CITATION = ARXIV:1912.10973;%%.

\bibitem{Banerjee:2020kaa}
S.~Banerjee, S.~Ghosh  and R.~Gonzo, \emph{{BMS symmetry of celestial OPE}},
  JHEP {\bf 04} (2020) 130,
\href{http://www.arXiv.org/abs/2002.00975}{{\tt 2002.00975}}
% .

\bibitem{He:2014laa}
T.~He, V.~Lysov, P.~Mitra  and A.~Strominger, \emph{{BMS supertranslations and
  Weinberg's soft graviton theorem}}, JHEP {\bf 05} (2015) 151,
\href{http://www.arXiv.org/abs/1401.7026}{{\tt 1401.7026}}
%%CITATION = ARXIV:1401.7026;%%.

\bibitem{Kapec:2014opa}
D.~Kapec, V.~Lysov, S.~Pasterski  and A.~Strominger, \emph{{Semiclassical
  Virasoro symmetry of the quantum gravity $ \mathcal{S}$-matrix}}, JHEP {\bf
  08} (2014) 058,
\href{http://www.arXiv.org/abs/1406.3312}{{\tt 1406.3312}}
%%CITATION = ARXIV:1406.3312;%%.

\bibitem{Lysov:2014csa}
V.~Lysov, S.~Pasterski  and A.~Strominger, \emph{{Low's Subleading Soft Theorem
  as a Symmetry of QED}}, Phys. Rev. Lett. {\bf 113} (2014), no.~11, 111601,
\href{http://www.arXiv.org/abs/1407.3814}{{\tt 1407.3814}}
%%CITATION = ARXIV:1407.3814;%%.

\bibitem{Laddha:2018myi}
A.~Laddha and A.~Sen, \emph{{Logarithmic Terms in the Soft Expansion in Four
  Dimensions}}, JHEP {\bf 10} (2018) 056,
\href{http://www.arXiv.org/abs/1804.09193}{{\tt 1804.09193}}
% .

\bibitem{Laddha:2018rle}
A.~Laddha and A.~Sen, \emph{{Gravity Waves from Soft Theorem in General
  Dimensions}}, JHEP {\bf 09} (2018) 105,
\href{http://www.arXiv.org/abs/1801.07719}{{\tt 1801.07719}}
% .

\bibitem{Sahoo:2018lxl}
B.~Sahoo and A.~Sen, \emph{{Classical and Quantum Results on Logarithmic Terms
  in the Soft Theorem in Four Dimensions}}, JHEP {\bf 02} (2019) 086,
\href{http://www.arXiv.org/abs/1808.03288}{{\tt 1808.03288}}
% .

\bibitem{Ghosh:2021hsk}
D.~Ghosh and B.~Sahoo, \emph{{Spin Dependent Gravitational Tail Memory in
  $D=4$}},
\href{http://www.arXiv.org/abs/2106.10741}{{\tt 2106.10741}}
% .

\bibitem{Arkani-Hamed:2020gyp}
N.~Arkani-Hamed, M.~Pate, A.-M. Raclariu  and A.~Strominger, \emph{{Celestial
  amplitudes from UV to IR}}, JHEP {\bf 08} (2021) 062,
\href{http://www.arXiv.org/abs/2012.04208}{{\tt 2012.04208}}
% .

\bibitem{Pasterski:2017ylz}
S.~Pasterski, S.-H. Shao  and A.~Strominger, \emph{{Gluon Amplitudes as 2d
  Conformal Correlators}}, Phys. Rev. {\bf D96} (2017), no.~8, 085006,
\href{http://www.arXiv.org/abs/1706.03917}{{\tt 1706.03917}}
%%CITATION = ARXIV:1706.03917;%%.

\bibitem{Himwich:2019dug}
E.~Himwich and A.~Strominger, \emph{{Celestial Current Algebra from Low's
  Subleading Soft Theorem}},
\href{http://www.arXiv.org/abs/1901.01622}{{\tt 1901.01622}}
%%CITATION = ARXIV:1901.01622;%%.

\bibitem{Laddha:2020kvp}
A.~Laddha, S.~G. Prabhu, S.~Raju  and P.~Shrivastava, \emph{{The Holographic
  Nature of Null Infinity}}, SciPost Phys. {\bf 10} (2021), no.~2, 041,
\href{http://www.arXiv.org/abs/2002.02448}{{\tt 2002.02448}}
% .

\bibitem{Banerjee:2018gce}
S.~Banerjee, \emph{{Null Infinity and Unitary Representation of The Poincare
  Group}}, JHEP {\bf 01} (2019) 205,
\href{http://www.arXiv.org/abs/1801.10171}{{\tt 1801.10171}}
% .

\bibitem{Banerjee:2019prz}
S.~Banerjee, S.~Ghosh, P.~Pandey  and A.~P. Saha, \emph{{Modified celestial
  amplitude in Einstein gravity}}, JHEP {\bf 03} (2020) 125,
\href{http://www.arXiv.org/abs/1909.03075}{{\tt 1909.03075}}
% .

\bibitem{Dappiaggi:2005ci}
C.~Dappiaggi, V.~Moretti  and N.~Pinamonti, \emph{{Rigorous steps towards
  holography in asymptotically flat spacetimes}}, Rev. Math. Phys. {\bf 18}
  (2006) 349--416,
\href{http://www.arXiv.org/abs/gr-qc/0506069}{{\tt gr-qc/0506069}}
% .

\bibitem{Ciambelli:2018wre}
L.~Ciambelli, C.~Marteau, A.~C. Petkou, P.~M. Petropoulos  and K.~Siampos,
  \emph{{Flat holography and Carrollian fluids}}, JHEP {\bf 07} (2018) 165,
\href{http://www.arXiv.org/abs/1802.06809}{{\tt 1802.06809}}
% .

\bibitem{Weinberg:1965nx}
S.~Weinberg, \emph{{Infrared photons and gravitons}}, Phys. Rev. {\bf 140}
  (1965)
B516--B524
%%CITATION = PHRVA,140,B516;%%.

\bibitem{Low:1954kd}
F.~E. Low, \emph{{Scattering of light of very low frequency by systems of spin
  1/2}}, Phys. Rev. {\bf 96} (1954)
1428--1432
% .

\bibitem{Cachazo:2014fwa}
F.~Cachazo and A.~Strominger, \emph{{Evidence for a New Soft Graviton
  Theorem}},
\href{http://www.arXiv.org/abs/1404.4091}{{\tt 1404.4091}}
%%CITATION = ARXIV:1404.4091;%%.

\bibitem{Kapec:2017tkm}
D.~Kapec, M.~Perry, A.-M. Raclariu  and A.~Strominger, \emph{{Infrared
  Divergences in QED, Revisited}}, Phys. Rev. D {\bf 96} (2017), no.~8, 085002,
\href{http://www.arXiv.org/abs/1705.04311}{{\tt 1705.04311}}
% .

\bibitem{Chung:1965zza}
V.~Chung, \emph{{Infrared Divergence in Quantum Electrodynamics}}, Phys. Rev.
  {\bf 140} (1965)
B1110--B1122
% .

\bibitem{Kulish:1970ut}
P.~P. Kulish and L.~D. Faddeev, \emph{{Asymptotic conditions and infrared
  divergences in quantum electrodynamics}}, Theor. Math. Phys. {\bf 4} (1970)
745
% .

\bibitem{He:2014bga}
S.~He, Y.-t. Huang  and C.~Wen, \emph{{Loop Corrections to Soft Theorems in
  Gauge Theories and Gravity}}, JHEP {\bf 12} (2014) 115,
\href{http://www.arXiv.org/abs/1405.1410}{{\tt 1405.1410}}
% .

\bibitem{Bern:2014oka}
Z.~Bern, S.~Davies  and J.~Nohle, \emph{{On Loop Corrections to Subleading Soft
  Behavior of Gluons and Gravitons}}, Phys. Rev. D {\bf 90} (2014), no.~8,
  085015,
\href{http://www.arXiv.org/abs/1405.1015}{{\tt 1405.1015}}
% .

\bibitem{Laddha:2018vbn}
A.~Laddha and A.~Sen, \emph{{Observational Signature of the Logarithmic Terms
  in the Soft Graviton Theorem}}, Phys. Rev. D {\bf 100} (2019), no.~2, 024009,
\href{http://www.arXiv.org/abs/1806.01872}{{\tt 1806.01872}}
% .

\bibitem{Nandan:2019jas}
D.~Nandan, A.~Schreiber, A.~Volovich  and M.~Zlotnikov, \emph{{Celestial
  Amplitudes: Conformal Partial Waves and Soft Limits}}, JHEP {\bf 10} (2019)
  018,
\href{http://www.arXiv.org/abs/1904.10940}{{\tt 1904.10940}}
%%CITATION = ARXIV:1904.10940;%%.

\bibitem{Adamo:2021lrv}
T.~Adamo, L.~Mason  and A.~Sharma, \emph{{Celestial $w_{1+\infty}$ symmetries
  from twistor space}},
\href{http://www.arXiv.org/abs/2110.06066}{{\tt 2110.06066}}
% .

\bibitem{Jiang:2021csc}
H.~Jiang, \emph{{Celestial OPEs and $ w_{1+\infty}$ algebra from worldsheet in
  string theory}},
\href{http://www.arXiv.org/abs/2110.04255}{{\tt 2110.04255}}
% .

\bibitem{Jiang:2021ovh}
H.~Jiang, \emph{{Holographic Chiral Algebra: Supersymmetry, Infinite Ward
  Identities, and EFTs}},
\href{http://www.arXiv.org/abs/2108.08799}{{\tt 2108.08799}}
% .

\bibitem{Ball:2021tmb}
A.~Ball, S.~A. Narayanan, J.~Salzer  and A.~Strominger, \emph{{Perturbatively
  exact w$_{1+\infty}$ asymptotic symmetry of quantum self-dual gravity}}, JHEP
  {\bf 01} (2022) 114,
\href{http://www.arXiv.org/abs/2111.10392}{{\tt 2111.10392}}
% .

\bibitem{Strominger:2014pwa}
A.~Strominger and A.~Zhiboedov, \emph{{Gravitational Memory, BMS
  Supertranslations and Soft Theorems}}, JHEP {\bf 01} (2016) 086,
\href{http://www.arXiv.org/abs/1411.5745}{{\tt 1411.5745}}
% .

\bibitem{1502.06120PSZ}
S.~Pasterski, A.~Strominger  and A.~Zhiboedov, \emph{{New Gravitational
  Memories}}, JHEP {\bf 12} (2016) 053,
\href{http://www.arXiv.org/abs/1502.06120}{{\tt 1502.06120}}
%%CITATION = ARXIV:1502.06120;%%.

\bibitem{Pasterski:2015zua}
S.~Pasterski, \emph{{Asymptotic Symmetries and Electromagnetic Memory}}, JHEP
  {\bf 09} (2017) 154,
\href{http://www.arXiv.org/abs/1505.00716}{{\tt 1505.00716}}
%%CITATION = ARXIV:1505.00716;%%.

\bibitem{Miller:2021hty}
N.~Miller, \emph{{From Noether's Theorem to Bremsstrahlung: a pedagogical
  introduction to large gauge transformations and classical soft theorems}},
\href{http://www.arXiv.org/abs/2112.05289}{{\tt 2112.05289}}
% .

\bibitem{Sahoo:2021ctw}
B.~Sahoo and A.~Sen, \emph{{Classical soft graviton theorem rewritten}}, JHEP
  {\bf 01} (2022) 077,
\href{http://www.arXiv.org/abs/2105.08739}{{\tt 2105.08739}}
% .

\bibitem{Mao:2020vgh}
P.~Mao, \emph{{Remarks on infinite towers of gravitational memories}}, JHEP
  {\bf 11} (2020) 102,
\href{http://www.arXiv.org/abs/2008.12109}{{\tt 2008.12109}}
% .

\bibitem{Freidel:2021ytz}
L.~Freidel, D.~Pranzetti  and A.-M. Raclariu, \emph{{Higher spin dynamics in
  gravity and $w_{1 + \infty}$ celestial symmetries}},
\href{http://www.arXiv.org/abs/2112.15573}{{\tt 2112.15573}}
% .

\bibitem{Pasterski:2020pdk}
S.~Pasterski and A.~Puhm, \emph{{Shifting Spin on the Celestial Sphere}},
\href{http://www.arXiv.org/abs/2012.15694}{{\tt 2012.15694}}
% .

\bibitem{Pasterski:2017kqt}
S.~Pasterski and S.-H. Shao, \emph{{Conformal basis for flat space
  amplitudes}}, Phys. Rev. {\bf D96} (2017), no.~6, 065022,
\href{http://www.arXiv.org/abs/1705.01027}{{\tt 1705.01027}}
%%CITATION = ARXIV:1705.01027;%%.

\bibitem{Ashtekar:1987tt}
A.~Ashtekar, {\em {Asymptotic Quantization: Based on 1984 Naples Lectures}}.
\newblock
1987
%%CITATION = INSPIRE-256903;%%.

\bibitem{Crnkovic:1986ex}
C.~{Crnkovic} and E.~{Witten}, \emph{{Covariant description of canonical
  formalism in geometrical theories}}, in {\em Three Hundred Years of
  Gravitation}, pp.~676--684.
\newblock {S.~W. Hawking} and { W. Israel},
1987.
\newblock
% .

\bibitem{Lee:1990nz}
J.~Lee and R.~M. Wald, \emph{{Local symmetries and constraints}}, J. Math.
  Phys. {\bf 31} (1990)
725--743
%%CITATION = JMAPA,31,725;%%.

\bibitem{Wald:1999wa}
R.~M. Wald and A.~Zoupas, \emph{{A General definition of `conserved quantities'
  in general relativity and other theories of gravity}}, Phys. Rev. {\bf D61}
  (2000) 084027,
\href{http://www.arXiv.org/abs/gr-qc/9911095}{{\tt gr-qc/9911095}}
%%CITATION = GR-QC/9911095;%%.

\bibitem{He:2014cra}
T.~He, P.~Mitra, A.~P. Porfyriadis  and A.~Strominger, \emph{{New Symmetries of
  Massless QED}}, JHEP {\bf 10} (2014) 112,
\href{http://www.arXiv.org/abs/1407.3789}{{\tt 1407.3789}}
%%CITATION = ARXIV:1407.3789;%%.

\bibitem{Avery:2015iix}
S.~G. Avery and B.~U. Schwab, \emph{{Residual Local Supersymmetry and the Soft
  Gravitino}}, Phys. Rev. Lett. {\bf 116} (2016), no.~17, 171601,
\href{http://www.arXiv.org/abs/1512.02657}{{\tt 1512.02657}}
% .

\bibitem{Lysov:2015jrs}
V.~Lysov, \emph{{Asymptotic Fermionic Symmetry From Soft Gravitino Theorem}},
\href{http://www.arXiv.org/abs/1512.03015}{{\tt 1512.03015}}
% .

\bibitem{Kapec:2016jld}
D.~Kapec, P.~Mitra, A.-M. Raclariu  and A.~Strominger, \emph{{2D Stress Tensor
  for 4D Gravity}}, Phys. Rev. Lett. {\bf 119} (2017), no.~12, 121601,
\href{http://www.arXiv.org/abs/1609.00282}{{\tt 1609.00282}}
%%CITATION = ARXIV:1609.00282;%%.

\bibitem{Bieri:2013hqa}
L.~Bieri and D.~Garfinkle, \emph{{An electromagnetic analogue of gravitational
  wave memory}}, Class. Quant. Grav. {\bf 30} (2013) 195009,
\href{http://www.arXiv.org/abs/1307.5098}{{\tt 1307.5098}}
%%CITATION = ARXIV:1307.5098;%%.

\bibitem{Susskind:2015hpa}
L.~Susskind, \emph{{Electromagnetic Memory}},
\href{http://www.arXiv.org/abs/1507.02584}{{\tt 1507.02584}}
%%CITATION = ARXIV:1507.02584;%%.

\bibitem{1974SvA....18...17Z}
Y.~B. {Zel'dovich} and A.~G. {Polnarev}, \emph{{Radiation of gravitational
  waves by a cluster of superdense stars}}, Sov. Astron. Lett {\bf 18} (1974)
17
% .

\bibitem{Braginsky:1986ia}
V.~B. Braginsky and L.~P. Grishchuk, \emph{{Kinematic Resonance and Memory
  Effect in Free Mass Gravitational Antennas}}, Sov. Phys. JETP {\bf 62} (1985)
  427--430,
[Zh. Eksp. Teor. Fiz.89,744(1985)]
%%CITATION = SPHJA,62,427;%%.

\bibitem{gravmem3}
V.~B. Braginsky and K.~S. Thorne, \emph{Gravitational-wave bursts with memory
  and experimental prospects}, Nature {\bf 327} (1987), no.~6118,
123--125
% .

\bibitem{Compere:2018ylh}
G.~Comp\`ere, A.~Fiorucci  and R.~Ruzziconi, \emph{{Superboost transitions,
  refraction memory and super-Lorentz charge algebra}}, JHEP {\bf 11} (2018)
  200,
\href{http://www.arXiv.org/abs/1810.00377}{{\tt 1810.00377}}
% .

\bibitem{Ball:2019atb}
A.~Ball, E.~Himwich, S.~A. Narayanan, S.~Pasterski  and A.~Strominger,
  \emph{{Uplifting AdS$_{3}$/CFT$_{2}$ to flat space holography}}, JHEP {\bf
  08} (2019) 168,
\href{http://www.arXiv.org/abs/1905.09809}{{\tt 1905.09809}}
% .

\bibitem{Campiglia:2016jdj}
M.~Campiglia and A.~Laddha, \emph{{Sub-subleading soft gravitons: New
  symmetries of quantum gravity?}}, Phys. Lett. {\bf B764} (2017) 218--221,
\href{http://www.arXiv.org/abs/1605.09094}{{\tt 1605.09094}}
%%CITATION = ARXIV:1605.09094;%%.

\bibitem{Freidel:2021dfs}
L.~Freidel, D.~Pranzetti  and A.-M. Raclariu, \emph{{Sub-subleading Soft
  Graviton Theorem from Asymptotic Einstein's Equations}},
\href{http://www.arXiv.org/abs/2111.15607}{{\tt 2111.15607}}
% .

\bibitem{Campiglia:2019wxe}
M.~Campiglia and A.~Laddha, \emph{{Loop Corrected Soft Photon Theorem as a Ward
  Identity}}, JHEP {\bf 10} (2019) 287,
\href{http://www.arXiv.org/abs/1903.09133}{{\tt 1903.09133}}
%%CITATION = ARXIV:1903.09133;%%.

\bibitem{Iyer:1994ys}
V.~Iyer and R.~M. Wald, \emph{{Some properties of Noether charge and a proposal
  for dynamical black hole entropy}}, Phys. Rev. D {\bf 50} (1994) 846--864,
\href{http://www.arXiv.org/abs/gr-qc/9403028}{{\tt gr-qc/9403028}}
% .

\bibitem{Compere:2020lrt}
G.~Comp\`ere, A.~Fiorucci  and R.~Ruzziconi, \emph{{The $\Lambda$-BMS$_4$
  Charge Algebra}},
\href{http://www.arXiv.org/abs/2004.10769}{{\tt 2004.10769}}
% .

\bibitem{Fiorucci:2021pha}
A.~Fiorucci, {\em {Leaky covariant phase spaces: Theory and application to
  \ensuremath{\Lambda}-BMS symmetry}}.
\newblock PhD thesis, Brussels U., Intl. Solvay Inst., Brussels, 2021.
\newblock
\href{http://www.arXiv.org/abs/2112.07666}{{\tt 2112.07666}}.
\newblock
% .

\bibitem{Himwich:2019qmj}
E.~Himwich, Z.~Mirzaiyan  and S.~Pasterski, \emph{{A Note on the Subleading
  Soft Graviton}},
\href{http://www.arXiv.org/abs/1902.01840}{{\tt 1902.01840}}
%%CITATION = ARXIV:1902.01840;%%.

\bibitem{Strominger:2016wns}
A.~Strominger and A.~Zhiboedov, \emph{{Superrotations and Black Hole Pair
  Creation}}, Class. Quant. Grav. {\bf 34} (2017) 064002,
\href{http://www.arXiv.org/abs/1610.00639}{{\tt 1610.00639}}
%%CITATION = ARXIV:1610.00639;%%.

\bibitem{Compere:2016jwb}
G.~Comp\`ere and J.~Long, \emph{{Vacua of the gravitational field}}, JHEP {\bf
  07} (2016) 137,
\href{http://www.arXiv.org/abs/1601.04958}{{\tt 1601.04958}}
% .

\bibitem{Adjei:2019tuj}
E.~Adjei, W.~Donnelly, V.~Py  and A.~J. Speranza, \emph{{Cosmic footballs from
  superrotations}}, Class. Quant. Grav. {\bf 37} (2020), no.~7, 075020,
\href{http://www.arXiv.org/abs/1910.05435}{{\tt 1910.05435}}
% .

\bibitem{Atanasov:2021cje}
A.~Atanasov, W.~Melton, A.-M. Raclariu  and A.~Strominger, \emph{{Conformal
  block expansion in celestial CFT}}, Phys. Rev. D {\bf 104} (2021), no.~12,
  126033,
\href{http://www.arXiv.org/abs/2104.13432}{{\tt 2104.13432}}
% .

\bibitem{Sharma:2021gcz}
A.~Sharma, \emph{{Ambidextrous light transforms for celestial amplitudes}},
\href{http://www.arXiv.org/abs/2107.06250}{{\tt 2107.06250}}
% .

\bibitem{Guevara:2021tvr}
A.~Guevara, \emph{{Celestial OPE blocks}},
\href{http://www.arXiv.org/abs/2108.12706}{{\tt 2108.12706}}
% .

\end{thebibliography}\endgroup

\end{document}